\documentclass[twoside,12pt]{article}
\usepackage{graphicx}
\usepackage{bm,amstext,amssymb}
\usepackage[colorlinks,citecolor=blue]{hyperref}

\begin{document}

\title{ \vspace{1cm} Neutrino-nucleus reactions and their role for
  supernova dynamics and nucleosynthesis}
\author{K. G. Balasi$^1$, K. Langanke$^{2,3}$,\\ and G.
  Mart\'{\i}nez-Pinedo$^{3,2}$\\
\\
$^1$Demokritos National Center of Scientific Research,\\ Agia Paraskevi
Attikis, 15310 Athens, Greece\\
$^2$GSI Helmholtzzentrum f\"ur Schwerionenforschung,\\ Planckstr. 1,
64291 Darmstadt, Germany\\
$^3$Institut f\"ur Kernphysik, Technische Universit\"at
Darmstadt,\\ Schlossgartenstr. 9, 64289 Darmstadt, Germany}
\maketitle
\begin{abstract}
  The description of nuclear reactions induced by supernova neutrinos
  has witnessed significant progress during the recent years. On one
  hand this progress is due to experimental data which serve as
  important constraints to model calculations, on the other hand it is
  related to advances in nuclear modelling itself and in computer
  hardware.  At the energies and momentum transfers relevant for
  supernova neutrinos neutrino-nucleus cross sections are dominated by
  allowed transitions, however, often with non-negligible
  contributions from (first) forbidden transitions. For several nuclei
  allowed Gamow-Teller strength distributions could be derived from
  charge-exchange reactions and from inelastic electron scattering
  data.  Importantly the diagonalization shell model has been proven
  to accurately describe these data and hence became the appropriate
  tool to calculate the allowed contributions to neutrino-nucleus
  cross sections for supernova neutrinos. Higher multipole
  contributions are usually calculated within the framework of the
  Quasiparticle Random Phase Approximation, which describes the total
  strength and the position of the giant resonances quite well. Both
  are the relevant quantities for a reliable calculation of the
  forbidden contributions to the cross sections.

  The current manuscript reviews the recent progress achieved in
  calculating supernova-relevant neutrino-nucleus cross sections and
  discusses its verification by data. Moreover, the review summarizes
  also the impact which neutrino-nucleus reactions have on the
  dynamics of supernovae and on the associated nucleosynthesis. With
  relevance to the supernova dynamics these include the absorption of
  neutrinos by nuclei (the inverse of nuclear electron capture which
  is the dominating weak-interaction process during collapse),
  inelastic neutrino-nucleus scattering and nuclear de-excitation by
  neutrino-pair emission. For supernova nucleosynthesis we discuss the
  role of neutrino-induced reactions for the recently discovered $\nu
  p$ process, for the r-process and for the neutrino process, for
  which neutrino-nucleus reactions have the largest impact. Finally we
  briefly review neutrino-nucleus reactions important for the
  observation of supernova neutrinos by earthbound detectors.
\end{abstract}

\section{Introduction}

February 1987 was the birth of extrasolar neutrino astronomy when
detectors in Japan and the United States registered neutrinos which
travelled over 50 kpc from the Large Magellanic Cloud to Earth and
gave the first indications that a star in this neighbor galaxy of the
Milky Way had exploded as a
supernova~\cite{Hirata.Kajita.ea:1987,Koshiba:1992}. This
extraordinary scientific event also proved the general expectation
that neutrinos are produced in enormous numbers in supernovae
triggered by the core collapse of massive stars. In fact, about 99$\%$
of the gravitational binding energy released in the cataclystic event
is carried away by neutrinos, clearly overpowering the kinetic energy
associated with the expansion of the supernova and the energy radiated
away as light, despite the fascinating fact that supernovae can shine
as bright as entire galaxies.

Although the neutrinos observed from supernova SN1987A were likely all
electron antineutrinos, identified by the Cerenkov light produced by
the relativistic positrons after a charged-current neutrino reaction
on the protons in the water Cerenkov detectors, the amount of observed
neutrinos and their energy spectrum (of order a few 10s of MeV)
confirmed the general understanding of supernova
dynamics~\cite{Bethe90}.  These observations were supplemented by
detailed studies of the SN1987A lightcurve, which as expected,
followed the sequence of half-lives of radioactive nuclides like
$^{56}$Ni, $^{57}$Ni or $^{44}$Ti, which were copiously produced in
the hot supernova environment~\cite{Seitenzahl.Timmes.Magkotsios:2014}.
 
Core-collapse, or Type II, supernovae are the final fate of massive
stars when at the end of hydrostatic burning their inner core,
composed of nuclei in the iron-nickel mass range, runs out of nuclear
fuel and collapses under its own gravity triggering an explosion
during which most of the star's material, partly processed in the hot
environment in what is called explosive nucleosynthesis, is ejected
into the Interstellar Medium~\cite{Bethe90}.  In the general picture
of core-collapse supernovae and their associated nucleosynthesis,
neutrino reactions on nucleons and nuclei play an important role.
Arguably, however, the most important impact neutrinos have on the
supernova dynamics is the fact that, for the typical supernova
neutrino energy scales, they virtually do not interact with matter for
densities smaller than a few $10^{11}$ g~cm$^{-3}$. This makes neutrinos
the most efficient cooling mechanism during the late hydrostatic
burning stages and the early collapse phase, where neutrinos could be
thought of as free streaming with an appropriate energy loss
correction for nuclear reactions mediated by the weak interaction like
$\beta$ decay or electron capture. This picture does not hold at
higher densities, say in excess of $10^{12}$ g~cm$^{-3}$, the neutrino
interaction with matter is strong enough making a detailed bookkeeping
of neutrinos and their interaction with matter a tedious, but
necessary requirement for supernova modeling. During the collapse
phase when densities larger than $10^{12}$ g~cm$^{-3}$ are reached, it is
the elastic scattering of neutrinos on nuclei which changes the
neutrino transport through the dense matter to a diffusion problem
with time scale larger than the competing collapse time scale. As a
consequence neutrinos are trapped during the final stage of the
collapse. By inelastic scattering on electrons and, to a lesser
extent, on nuclei, neutrinos exchange energy with matter and get
thermalized.  Inelastic neutrino-nucleus scattering plays also an
interesting role in a short episode after core bounce, where it alters
the spectrum of electron neutrinos emitted during the so-called
neutrino burst~\cite{Langanke08}.

The supernova explosion is triggered by a shock wave which, when
passing outwards through the Fe-Ni-core, dissociates the heavy nuclei
into free nucleons. The interaction of neutrinos, produced by the hot
matter of the freshly-born neutron star in the center, with the free
protons and neutrons behind the shock are an effective addtional
energy source which, together with effects like convection and
plasma instabilities, are required for successful explosions, as modern
multi-dimensional supernova simulations show~\cite{janka12,Burrows:2013}.

It is also the competition of the various interactions of electron
neutrinos with neutrons and of anti-electron neutrinos with protons,
which determines the proton-to-neutron ratio of the matter ejected
from the surface of the nascent proto-neutron star crucially
influencing the subsequent nucleosynthesis of the ejected matter in
this neutrino-driven wind model. Simulations indicate that there are
periods where the neutrino-driven wind matter is proton-rich (with an
electron-to-nucleon ratio $Y_e>0.5$) and where it is neutron-rich
($Y_e <0.5$). In the former case the nucleosynthesis occuring when the
ejected matter reaches cooler regions at larger distances from the
neutron star surface gives rize to the recently discovered $\nu p$
process. For several years the ejection of neutron-rich matter in the
neutrino-driven wind has been considered as the favorite site for the
astrophysical r-process. However, modern supernova simulations predict
astrophysical conditions which allow for a 'weak
r-process'~\cite{gmp14}, which can produce elements up to the second
peak around mass number $A \sim 130$, which are, however, not
sufficient for the production of r-process nuclides at the
gold-platinum peak and the transactinides.  As the nucleosynthesis in
the neutrino-driven wind scenario occurs in the presence of enormous
neutrino fluxes, neutrino-induced reactions on nuclei might play an
important role in these processes. Particularly relevant are here
neutrino-induced reactions with particle emission (neutron, protons,
$\alpha$) in the final channel as they implicitly alter the nuclear
abundance distributions.  Neutrino-induced particle emission of nuclei
is also the key process in the neutrino nucleosynthesis
process~\cite{Woosley90} which 
is responsible for the production of selected nuclides generated by
particle spallation of more abundant nuclides in charged- or
neutral-current reactions in the outer shells of the star.

Obviously neutrino-induced reactions by the various target nuclei are
the means for detection of supernova neutrinos. After the observation
of anti-electron neutrinos from SN1987A, it is the aim of operational
and future detectors to individually detect the various neutrino types
and proof the differences in the spectra of $\nu_e$, ${\bar \nu_e}$
and $\nu_x$ (usually used to combine $\nu_\mu$ and $\nu_\tau$ neutrinos
and their anti-particles which are expected to have quite similar
spectra) predicted by supernova models.

Neutrino reactions on nuclei can be described in perturbation theory.
For the energies and momentum transfers relevant for supernova
neutrinos the task reduces to a nuclear structure problem where mainly
allowed and first-forbidden transitions are of importance. Fermi
transitions, of relevance for elastic neutrino scattering and for
$(\nu_e,e^-)$ reactions, can only occur to Isobaric Analog States. Due
to progress in computer hardware and impressive advances in shell
model developments Gamow-Teller (GT) strength distributions can now be
calculated for up to medium-mass nuclei taking the relevant nuclear
degrees of freedom into account. This theoretical progress has been
accompanied by novel experimental techniques which allow the
determination of the GT$_-$ strength (in which a neutron is changed
into a proton as it is relevant for $(\nu_e,e^-)$ reactions) by
$(p,n)$~\cite{Gaarde81} and $(^3\text{He},t)$~\cite{Daito98,Zegers06}
charge-exchange reactions and of the GT$_+$ strength (in which a
proton is changed into a neutron as in the $(\bar{\nu}_e,e^+)$
reaction or in electron capture) by $(n,p)$~\cite{Vetterli92} and,
with significantly improved resolution, by the
$(d,^2\text{He})$~\cite{Frekers05} and
$(t,^3\text{He})$~\cite{cole.akimune.ea:2006} reactions at intermediate
energies. The GT$_0$ strength, important for inelastic
neutrino-nucleus scattering, can be studied in (p,p') reactions, but
it is also constrained by precision M1 data obtained in inelastic
electron scattering for spherical nuclei where the orbital
contribution to the M1 transition is small and the spin part is
dominated up to a constant factor by the same operator as the GT$_0$
transitions.  Importantly, the shell model studies reproduce the
measured GT and M1 strength distributions quite well, except for a
constant normalization factor and hence it is the method of choice to
describe the Gamow-Teller contributions to the neutrino-nucleus cross
sections.  The calculation of first-forbidden (dipole) transitions
require multi-shell studies which, due to computer limitations, have
only been possible for light nuclei yet.  Usually dipole contributions
to the neutrino cross sections are derived within the Quasi-Particle
Random Phase Approximation (QRPA) which can handle model spaces with a
larger number of single particle orbitals than the shell model space,
however, for the price of considering only (2 particle -- 2 hole)
correlations among nucleons. The QRPA is well suited to describe the
energy centroid and total strength of giant resonances and hence gives
a fair description of neutrino-nucleus cross sections if the neutrino
energy is sufficiently large so that the cross sections are dominated
by the contributions from giant resonances.

Neutrino-induced spallation reactions are usually described in a
two-step process, calculating the neutrino-induced excitation spectrum
within the framework of the shell model and QRPA, followed by a
determination of the decay probability into the various particle (or
$\gamma$) channels using the statistical model.  Exploiting this
two-step strategy differential cross sections have been derived for
nearly the entire nuclear chart which can now been routinely used in
nucleosynthesis studies.
 
In this article we would like to review recent developments in the
description of neutrino-nucleus reactions. Hereby we will focus solely
on neutrino energies up to a few 10s of MeV, as they are relevant for
supernovae. The description of reactions induced by higher-energetic
neutrinos as they are important for atmospheric, accelerator or
astrophysical neutrinos from other sources require different
theoretical approaches; as outlined e.g. in Refs. \cite{Learned00,Becker08}.  
Our review can
be viewed as an extension and update of the article by Kolbe {\it et
  al.} \cite{Kolbe03}.  Hence we will only briefly repeat some of the
material presented there, but focus mainly on the research performed
later. Our review is organized as follows: in the next section we will
briefly describe the current understanding of core-collapse
supernovae, paying special attention to the role of neutrinos. In
particular we will discuss the neutrino spectra during the collapse
and arising from the cooling of the proto-neutron star. In Section 3
we summarize the basic cross section formulae for neutrino-nucleus
reactions and discuss the models, mainly shell model and QRPA, used to
derive the necessary nuclear input. We will also compare strength
functions obtained with these models with experimental
data. Furthermore we discuss in Section 3 on the basis of the
statistical model the derivation of the decay probabilities into
particle and $\gamma$ channels as required to calculate partial cross
sections for neutrino-induced spallation reactions and summarize some
general results. Sections 4 and 5 are concerned with the description
of specific neutrino-nucleus reactions and their role for the
supernova dynamic and nucleosynthesis, respectively. Finally we
summarize in Section 6 the progress achieved to describe
neutrino-nucleus reactions of importance for operational and future
supernova detectors.

\section{Brief description of core-collapse supernovae}

During their long lifetimes of millions and billions of years stars
generate the energy necessary to achieve hydrostatic equilibrium by
nuclear reactions in their interior
\cite{BBFH,Wallerstein97,Wiescher01a}.  The first nuclear reaction
source which a star tabs is hydrogen burning, the fusion of 4 proton
nuclei to $^4$He. Hydrogen burning occurs, in stars like our Sun, in a
reaction sequence called pp-chains beginning with the formation of a
deuteron by the fusion of two protons mediated by the weak
interaction. In more massive stars than the Sun hydrogen burning
occurs at slightly higher temperatures making hydrogen burning via the
CNO cycle energetically more efficient than by the pp chains
\cite{Adelberger98,Adelberger11}.  It is interesting to note that
hydrogen burning produces a nucleus ($^4$He) with an identical amount
of protons and neutrons (i.e. $Y_e=0.5$) from matter with an
unbalanced proton-to-neutron ratio basically inherited from the Big
Bang. It is an important feature of hydrostatic burning and nuclear
stability that this value of $Y_e=0.5$ is kept during the various
burning stages.

The ashes of hydrogen burning, $^4$He, can be burnt once the
temperature of the star's interior has increased sufficiently by core
contraction to enable the fusion of three $^4$He nuclei to
$^{12}$C. This triple-alpha reaction and the subsequent
$^{12}$C($\alpha,\gamma$)$^{16}$O reaction constitute helium burning.
Both together determine the carbon/oxygen ratio in the Universe. Hence
helium burning is essential for life as we know it.

In sufficiently massive stars the sequence of core contraction
associated with a rise of the temperature repeats itself igniting
successively the ashes of the previous burning phase. Thus following
helium burning massive stars undergo in their centers the sequence of
carbon, neon, oxygen and silicon burning until an 'iron core' is
produced as the product of silicon core burning.  At this phase of its
hydrostatic burning, a massive star consists of concentric shells that
are the remnants of its previous burning phases (hydrogen, helium,
carbon, neon, oxygen, silicon) surrounding the iron core.

The temperature in the iron core (a few $10^9$ K) is sufficiently high
to establish an equilibrium of reactions mediated by the strong and
electromagnetic interaction and their inverse.  Under such conditions
the matter composition is given by Nuclear Statistical Equilibrium
(NSE).  Importantly once such an equilibrium is achieved nuclear
reactions (by strong and electromagnetic forces) cease as energy
sources. Hence the star has lost its nuclear energy source in the
center and the iron core becomes eventually unstable under its own
gravity.  We note that the NSE matter composition depends on the
astrophysical parameters (temperature and density). For the following
discussion it is important that it also depends on the
proton-to-neutron ratio, i.e. on the $Y_e$ value, which can only be
changed by reactions mediated by the weak interaction. Initially the
ashes of silicon burning consists roughly of the same amount of
protons and neutrons ($Y_e \approx 0.5$) favoring, in NSE, nuclei in
the vicinity of the double-magic nucleus $^{56}$Ni. Under (early) core
conditions reactions mediated by the weak interaction are not in
equilibrium with their inverse. This, as we will discuss next, will
change the $Y_e$ value during the collapse and relatedly the matter
composition.
  
While the mass of the core is dominated by the nuclear component, the
pressure is given by the electrons, which under core conditions, form
a degenerate ultrarelativistic Fermi gas; both components are connected by the
electron-to-nucleon ratio $Y_e$.  Once the core density reaches values
of order $10^9$ g~cm$^{-3}$ or higher, the electron Fermi energy grows to
values in excess of MeV, the core matter can efficiently lower its
free energy by electron captures on nuclei. This process has three
important consequences:

1) At densities below $10^{11}$ g~cm$^{-3}$ the neutrinos produced by the
capture process can leave the star unhindered carrying away energy.
This is a very efficient cooling mechanism which, as noted in the
fundamental paper by Bethe, Brown, Applegate and Lattimer (BBAL)
\cite{BBAL} keeps the entropy of the matter low. As a consequence
heavy nuclei survive during the collapse.

2) Electron capture reduces the number of electrons (i.e. $Y_e$ gets
smaller). As a result the pressure which the matter can stem against
the gravitational collapse is reduced and the collapse is accelerated.

3) As electron captures change a proton in the nucleus into a neutron,
the matter composition is driven towards more neutron-rich (and
heavier) nuclei.  With increasing neutron excess, nuclei become
unstable against $\beta$ decay. Under core conditions these decays
are, however, hindered due to Pauli blocking by electrons and at
sufficiently high densities (say in excess of $10^{10}$ g~cm$^{-3}$) the
phase space for beta decays closes and electron capture dominates by a
large factor driving the core composition to neutron-rich,
$\beta$-unstable nuclei

It should be mentioned that $\beta$ decay is temporarily competitive
with electron capture after silicon depletion and during silicon shell
burning \cite{Heger01a,Heger01b}.  Besides counteracting electron
captures by increasing the $Y_e$ value, $\beta$ decays are also an
important neutrino source and thus add to the cooling of the core and
the reduction of the entropy \cite{Heger01a}. Finally, we remark that
it is quite fortuitous that for pre-supernova simulations $\beta$ decay
rates are only required for nuclei for which large-scale shell model
calculations can be performed \cite{Langanke00}.

Exploiting progress in nuclear modelling electron capture rates at
supernova conditions have recently been calculated using an
appropriately chosen hierarchy of approaches \cite{Juodagalvis10}.  At
densities up to about $10^{10}$ g~cm$^{-3}$ supernova weak-interaction
rates are dominated by nuclei in the mass range $A=45-65$ ($pf$-shell
nuclei). At such densities the electron chemical potential is of the
same order as the nuclear $Q$ values requiring a detailed reproduction
of the GT$_+$ distribution due to the strong phase space sensitivity
of the rates. As we will discuss in the next Section, modern shell
model calculations are capable to describe GT$_+$ distributions rather
well. Shell model diagonalization in the $pf$ model space has
therefore been used to calculate the weak-interaction rates for the
nuclei with $A\sim 50-65$ \cite{Caurier99,Langanke00,ADNDT}.  These
rates explicitly consider that, due to the finite temperature in the
star, electron capture can also occur from excited states, which can
have different GT$_+$ distributions than the ground state.  With
continuing electron captures and at higher densities (above $10^{10}$
g~cm$^{-3}$) nuclei become relevant for the capture rates which have
proton numbers $Z<40$ and neutron numbers $N>40$; i.e. in the
Independent Particle Model (IPM) the valence protons are in the $pf$
shell, while the valence neutrons occupy already the next major shell
($sdg$ shell).  For such nuclei the IPM predicts a vanishing GT$_+$
strength as the GT operator acts only in spin-isospin space and all
GT$_+$ transitions of the valence protons would be blocked due to the
complete occupation of $pf$ orbitals by neutrons. Hence electron
captures on such nuclei are only possible due to correlations of
nucleons across the $N=40$ shell gap. Such cross-shell correlations
are a rather slowly converging process and usually require
multi-particle-multi-hole excitations \cite{Caurier03}. Taking this
into account, the electron capture rates for heavier nuclei have been
calculated within a hybrid approach. At first, the Shell Monte Carlo
Approach (SMMC) \cite{Johnson92,Koonin97} has been used to determine
occupation numbers of proton and neutron orbitals by performing SMMC
calculations for individual nuclei at finite temperature within the
complete $pf$-$sdg$ shell and taking the appropriate nuclear
correlations into account. Using these finite-temperature occupation
numbers electron captures have been calculated using the RPA
approach. The latter step is justified as at the conditions under
which the nuclei with $A>65$ are relevant the electron chemical
potential is noticeably larger than the respective nuclear $Q$ values
making the capture rates mainly sensitive to the centroid and total
value of the transition strength, but not to its detailed
structure~\cite{Juodagalvis10}.  Forbidden transitions contribute
increasingly to the capture rates with growing density and, relatedly,
electron chemical potential.  Inspired by the IPM supernova
simulations have for a long time assumed that electron captures on
heavy nuclei (with $N>38$ \cite{Bruenn85}) vanish and that the capture
occurs solely on free protons (which have the larger individual
capture rate than heavy neutron-rich nuclei due to the smaller $Q$
value, but are much less abundant in the core caused by the low
entropy). More recent supernova simulations have incorporated the
modern shell model rates and find that the capture on heavy nuclei
dominates during the entire collapse
\cite{Langanke03,Hix03,Janka07}. By measuring the GT$_+$ strength for
$^{76}$Se (with 34 protons and 42 neutron) it has been experimentally
demonstrated that the strength is indeed unblocked by nuclear
correlations across the $N=40$ gap \cite{Grewe08}.  Data from transfer
reaction measurements \cite{Kay08}, in agreement with large-scale
shell model calculations \cite{Zhi11}, indicate about 4 neutron holes
in the $pf$ shell in the $^{76}$Se ground state, unblocking GT$_+$
transitions of valence protons.

\begin{figure}
  \begin{center}
    \includegraphics[width=0.5\linewidth]{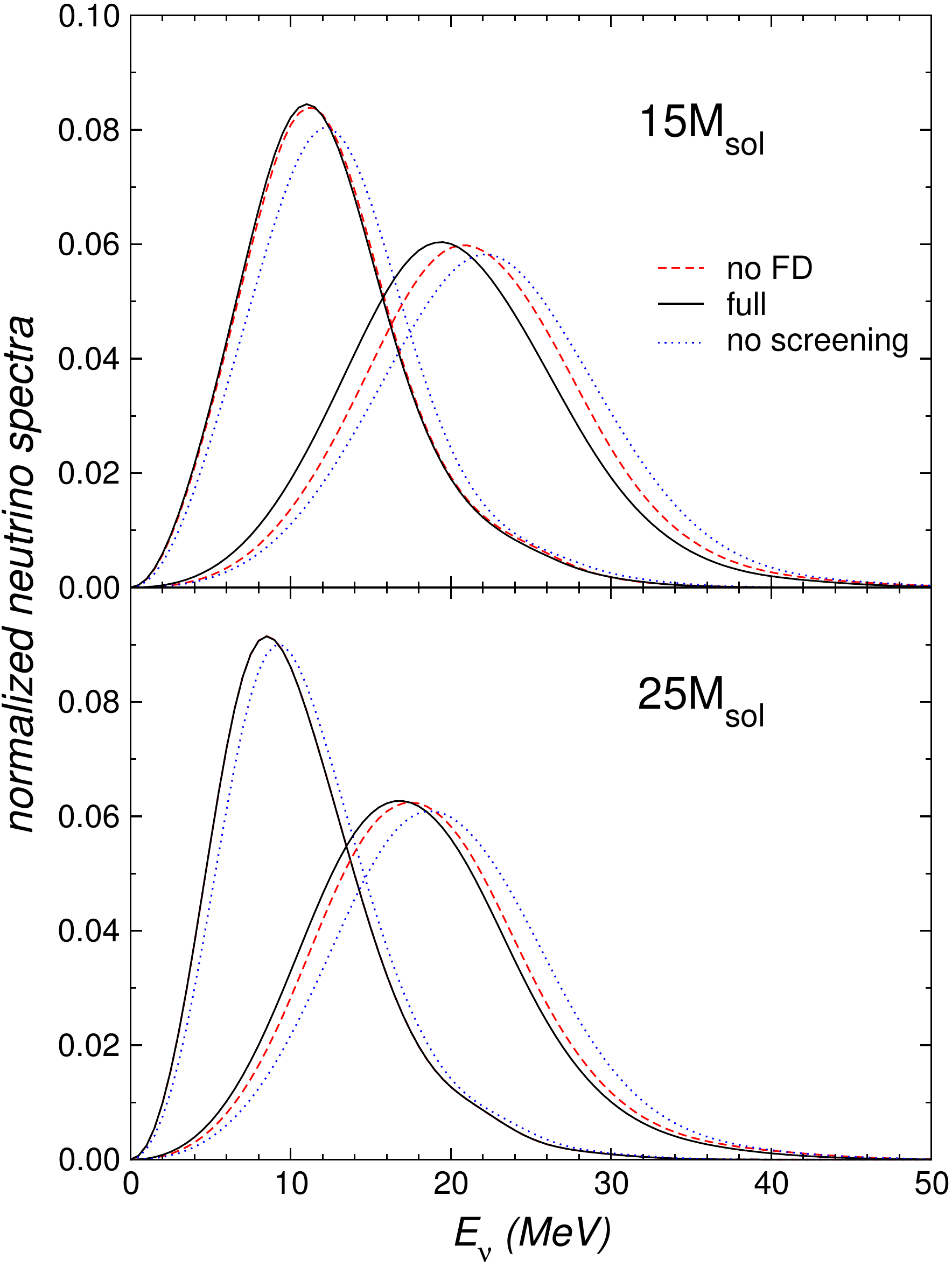}
    \caption{Neutrino spectra produced by electron captures on nuclei
      at two different densities ($3.76 \times 10^{11}$~g~cm$^{-3}$ and
      $2.47 \times 10^{12}$~g~cm$^{-3}$) during the collapse of a star
      with 15~M$_\odot$ and at $1.13 \times 10^{11}$~g~cm$^{-3}$ and $1.54
      \times 10^{12}$~g~cm$^{-3}$ for a 25~M$_\odot$ star.  For the solid
      line an NSE distribution has been assumed for the matter
      composition; the dashed line shows the spectrum if heavy nuclei
      (with $A> 100$) are neglected in the distribution. The dotted
      curve shows the spectra if screening corrections due to the
      astrophysical environment are not included.  (from
      \cite{Juodagalvis10}) \label{fig:neutrino-spectra}}
  \end{center}
\end{figure}

Electron capture on nuclei is the main neutrino source during
collapse.  Fig. \ref{fig:neutrino-spectra} shows the respective
neutrino spectra produced at different moments during the core
collapse of a 15~M$_\odot$ and a 25~M$_\odot$ star, respectively. The
spectra have been calculated from the individual electron captures of
about 3000 nuclei, averaged over an appropriate NSE distribution. The
collapse snapshots are chosen to support the following
discussion on neutrino trapping and thermalization. Due to the finite
temperature of the environment and the presence of many nuclei the
neutrino spectra are quite wide. At densities of a few $10^{11}$
g~cm$^{-3}$ the spectrum is centered around neutrino energies of order 10
MeV.  Due to the increase of electron chemical potential with density
($\sim \rho^{1/3}$) the neutrino spectra are shifted to higher energy
with growing density. At $\rho > 10^{12}$ g~cm$^{-3}$ the centroid is at
neutrino energies $E_\nu \approx 20$ MeV.  Screening due to the
astrophysical environment effectively reduces the chemical electron
potential and enhances the nuclear $Q$ value. These effects decrease
the electron capture rates slightly and shift the neutrino spectra to
lower energies (by about 2 MeV as is shown in
Fig. \ref{fig:neutrino-spectra}).

An important change in the physics of the collapse occurs, when
neutrinos start to get trapped in the core at densities
$\rho_{\text{trap}} \approx 4 \times 10^{11}$~g~cm$^{-3}$. The main
process is elastic scattering of neutrinos on nuclei.  The
respective total cross section for this process is given by
\cite{Bethe90}

\begin{equation}
  \sigma_{\text{el}} (E_\nu)  =
  \frac{G_F^2}{4 \pi} \left[ N - (1-4 \sin^2 \theta_W) Z \right]^2 E_\nu^2
  \label{eq:elastic}
\end{equation}

where $E_\nu$ is the energy of the incident neutrino; $G_F$ and
$\theta_W$ are the Fermi coupling constant and the Weinberg angle
($\sin^2 \theta_W \approx 0.231$), respectively. As $(1-4 \sin^2
\theta_W) \approx 0.08$, elastic neutrino scattering occurs mainly as
a coherent process on the neutrons in the nuclei. In
Eq. \ref{eq:elastic} it has been assumed that the elastic form factors
for protons and neutrons can be replaced by $F_{Z,N} (q^2) = F_{Z,N}
(q^2 =0) =1$ which is justified at the momentum transfers involved.
For odd-A and odd-odd nuclei with ground state angular momenta $J \neq
0$ and at finite temperature there can also be a GT contribution to
the cross section. However, due to the strong fragmentation of the GT
strength this is usually negligibly small. Furthermore under supernova
conditions even-even nuclei dominate the nuclear composition.  As is
demonstrated in \cite{Bethe90} neutrinos with energies about 20 MeV
will have a mean free path of order 0.5~km at densities $10^{12}$
g~cm$^{-3}$ which is significantly smaller than the core radius at the
same density. As a consequence neutrinos scatter quite often in the
core and their propagation has to be described by a diffusion
process. Indeed, the diffusion time scale at $\rho = 10^{12}$ g~cm$^{-3}$
is longer than the competing collapse time scale and neutrinos are
effectively trapped during the final phase of the collapse.

\begin{figure}
  \includegraphics[width=0.33\linewidth]{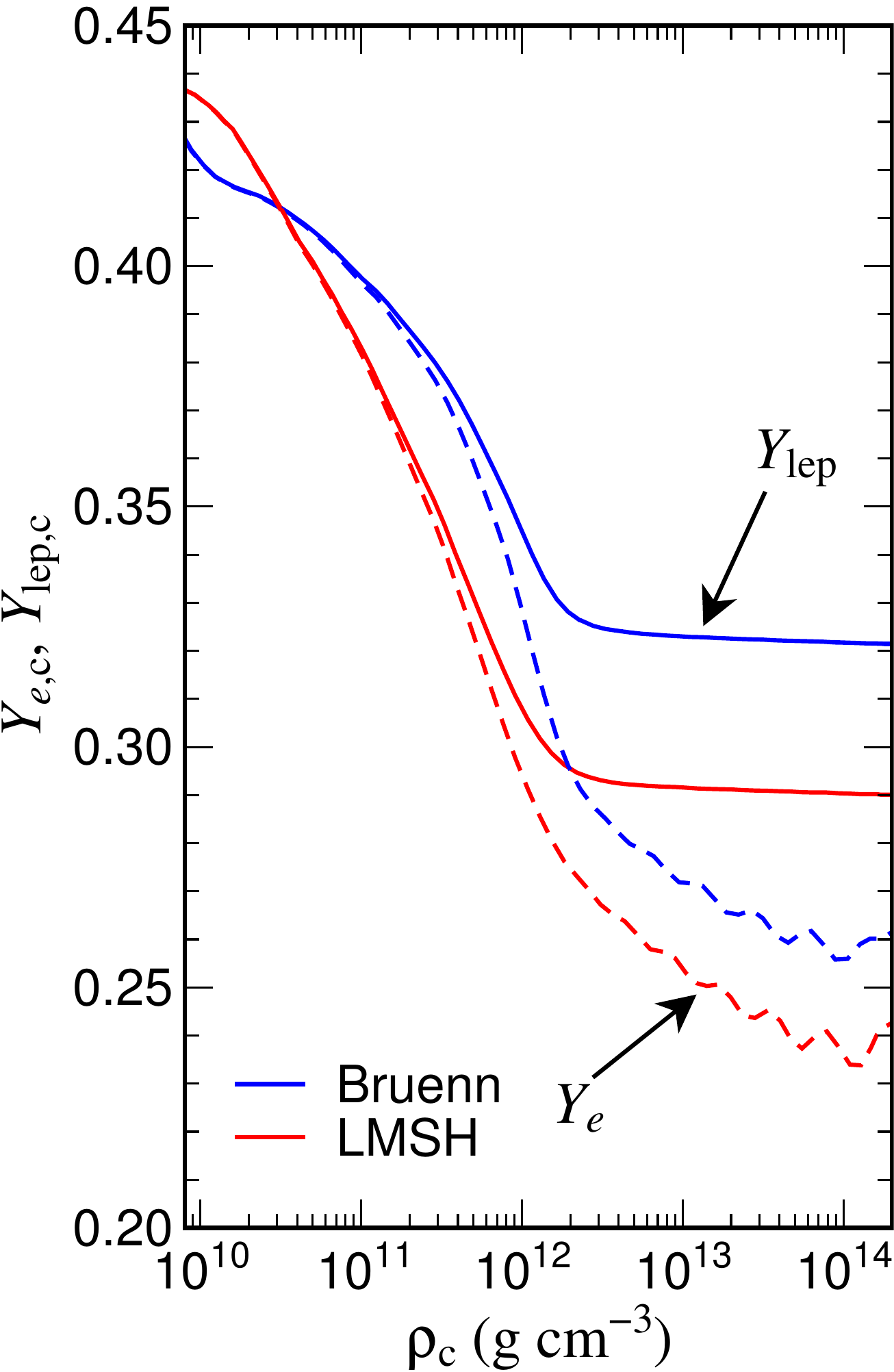}%
  \includegraphics[width=0.33\linewidth]{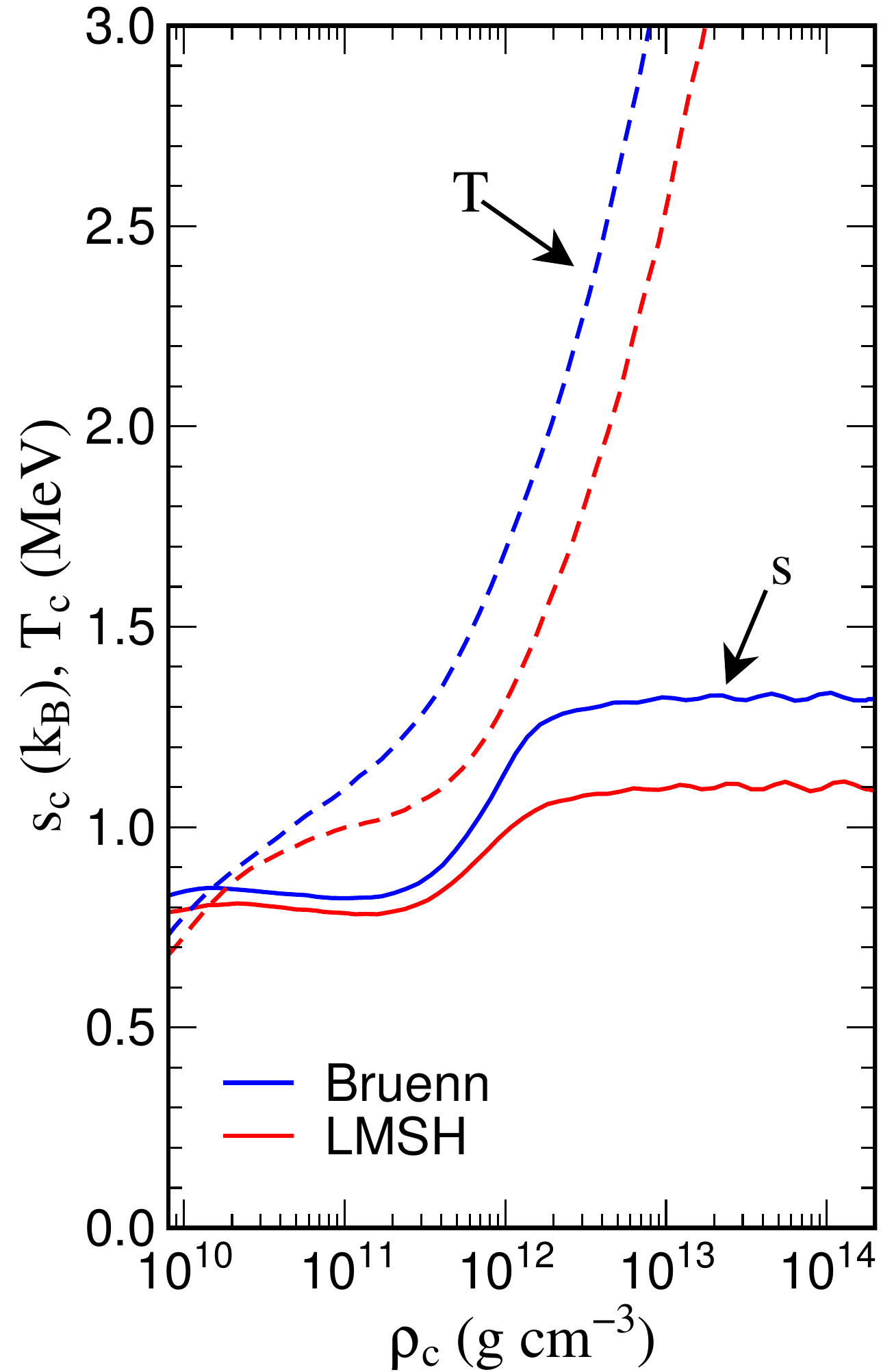}%
  \includegraphics[width=0.33\linewidth]{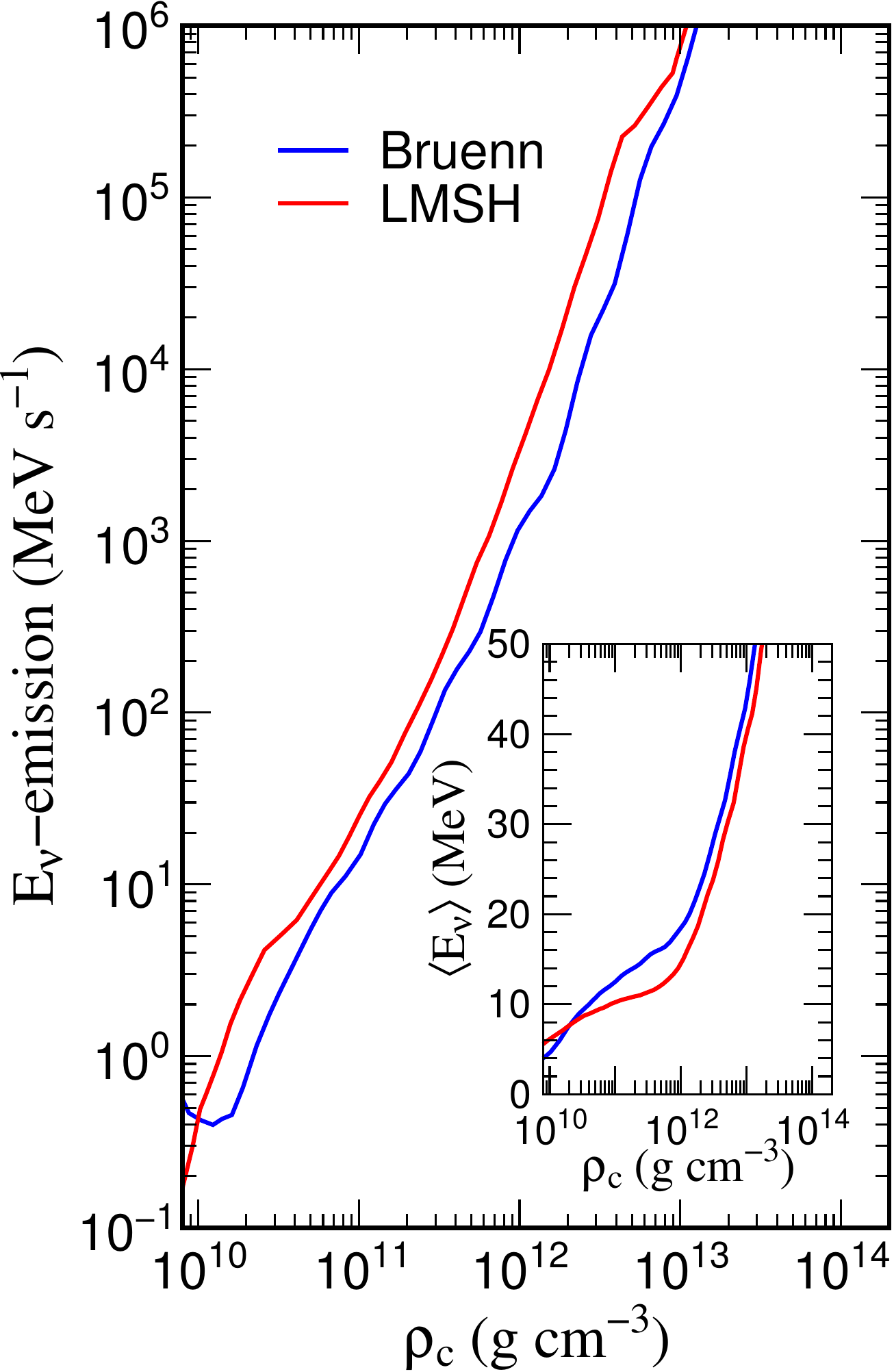}
  \caption{Comparison of the central values for the electron ($Y_e$)
    and lepton $Y_{lep}$ abundances (left panel), temperature and
    entropy (middle panel) and electron neutrino emission rate (right
    panel) obtained in a spherical simulation of the collapse of an
    18~M$_\odot$ star using the shell model electron capture rates
    (LMSH) or neglecting capture on heavy nuclei with $N>38$ (Bruenn
    rates \cite{Bruenn85}). The insert in the right panel compares the
    average neutrino energies. (courtesy Hans-Thomas Janka, adapted
    from~\cite{martinez-pinedo.liebendoerfer.frekers:2006})
    \label{fig:core-evolution}}  
\end{figure}

Neutrinos exchange energy with the core matter by inelastic
scattering, mainly on electrons.  As electrons are highly degenerate
at this stage of the collapse, they can only gain energy and hence
neutrinos are down-scattered in energy. By inelastic scattering on
nuclei neutrinos can both loose and gain energy (as is discussed in
Section 4). Simulations, however, show that this process is less
important during collapse than inelastic neutrino-electron scattering
which is found to be quite efficient in thermalizing the neutrinos
with the rest of the matter. This is effectively achieved at densities
in excess of $10^{12}$ g~cm$^{-3}$ as is demonstrated in
Fig. \ref{fig:core-evolution} which shows results from a (spherical)
collapse simulation of an 18~M$_\odot$ star. The figure shows that the
lepton abundance, i.e. the sum of abundances of electrons and electron
neutrinos, reach a constant value once neutrinos are trapped and a
Fermi sea of electron neutrinos builds up in the core. The figure also
clearly demonstrates the impact of the modern electron capture rates
(LMSH) on the collapse dynamics if compared to results based mainly on
the IPM approach (neglecting electron capture on nuclei with neutron
numbers $N>38$ \cite{Bruenn85}). As electron captures on nuclei
dominate over those on free protons the electron neutrino emission
rate is larger, while the central core values for temperature, entropy
and $Y_e$ are smaller. At densities $\rho < 3 \times 10^{10}$
g~cm$^{-3}$ nuclei from the mass region $A=45-65$ dominate. For these
nuclei the electron capture rates obtained by shell model
diagonalization \cite{Langanke00} are noticeably smaller than those
inspired by the IPM (and corrected by data whenever available)
\cite{FFN1,FFN2,FFN3,FFN4}. As neutron-rich nuclei have larger $Q$
values than protons, the average neutrino energies are smaller in
simulations incorporating the shell model capture rates.

After neutrino trapping, the collapse proceeds homologously
\cite{Goldreich}, until nuclear densities ($\rho_N = 2.5 \times
10^{14}$~g~cm$^{-3}$) are reached. Since nuclear matter has a much lower
compressibility, the homologous core decelerates and bounces in
response to the increased nuclear matter pressure. This drives a shock
wave into the outer core; i.e. the region of the iron core which lies
outside the homologous core and in the meantime has continued to fall
inwards at supersonic speed. The core bounce with the formation of a
shock wave is the mechanism that ultimatively triggers a supernova
explosion, but the exact mechanism of this physically appealing
scenario are still uncertain and controversial.  It appears as if the
energy available to the shock is not sufficient, and the shock uses up
its energy in the outer core mostly by the dissociation of nuclei into
nucleons~\cite{Bethe90,Janka07}. This change in composition results to
additional energy losses, because the electron capture rate on free
protons is significantly larger than on neutron-rich nuclei due to the
higher Q-values of the latter.  A large fraction of the neutrinos
produced by these capture behind the shock leave the star quickly in
what is called the neutrino burst at shock break out, carrying away
energy. This leads to further neutronization of the matter.  The shock
wave is weakened so much that it finally stalls and turns into an
accretion shock.

After the core bounce, a compact remnant begins to form in the center.
Depending on the stellar mass, this is either a neutron star (for
progenitor stars with masses roughly smaller than 25 $M_\odot$) or a
black hole.  The nascent proto-neutron star contains a large number of
degenerate electrons and neutrinos, the latter being trapped as their
mean free paths in the dense matter is significantly shorter than the
radius of the neutron star. As the trapped neutrinos diffuse out, they
convert most of their initially high degeneracy energy to thermal
energy of the stellar medium~\cite{Burrows90}.  The cooling of the
protoneutron star then proceeds by pair production of neutrinos of all
three generations which diffuse out. After several tens of seconds the
star becomes transparent to neutrinos and the neutrino luminosity
drops significantly \cite{Burrows88}.

In what is called the `delayed neutrino-heating mechanism' the stalled
shock wave can be revived by neutrinos~\cite{Bethe.Wilson:1985}. These
carry most of the energy set free in the gravitational collapse of the
core \cite{Burrows90} and deposit some of it in the layers between the
nascent neutron star and the stalled prompt shock, mainly by
absorption on nucleons.  This lasts for a few 100 ms, and requires
about $1\%$ of the neutrino energy to be converted into nuclear
kinetic energy. The energy deposition increases the pressure behind
the shock. However, it appears that energy transport by neutrinos is
not enough to explode models with spherical symmetry
\cite{Buras03,Rampp00,Mezzacappa01}.  However, multi-dimensional
hydrodynamical simulations in two dimensions
\cite{Brandt11,Herant94,Mueller10,Nordhaus10} and in 3 dimensions
\cite{Fryer02,Takiwaki12,Hanke12,Burrows12,Ott13} show that the region
of low density, but rather high temperature (named 'hot neutrino
bubble'), which forms between the neutron star surface and the shock
front, is convectively unstable developing large convective overturns
which enhance the neutrino energy deposit behind the
shock~\cite{Herant94,Burrows95,Fryer99,Fryer04}. This is additionally
aided by instabilities of the shock to non-radial deformation
(standing accretion shock instability SASI) which increases the time
matter stays in the heating
region~\cite{Blondin03,Scheck04,Scheck06,Burrows06,Scheck08,Foglizzo12}. 

The simulations sensitively depend on the nuclear Equation of State
(EoS) which is still not sufficiently well-known. It strongly
influences the formation, the strength and the evolution of the shock
wave, the latter indirectly via the compactness and the contraction of
the nascent neutron star.  For densities less than about $\rho_N/10$,
nucleonic matter exists of individual nuclei.  At larger densities
matter exists in a variety of complex shapes caused by a competition
of repulsive Coulomb forces and nuclear attraction until it reaches
uniform nuclear matter.  There are several
EoS~\cite{Lattimer91,Shen98,Wolff84,Hempel.Schaffner-Bielich:2010,%
Steiner.Hempel.Fischer:2013,Suwa.Takiwaki.ea:2013,Fischer.Hempel.ea:2014}
frequently used in supernova simulations. A novel ansatz, based on the
relativistic mean-field with density-dependent couplings and the
explicit consideration of light-particle bound states, has recently
been
presented~\cite{Typel12}. Refs.~\cite{Sagert09,Fischer.Sagert.ea:2011}
has studied the implications of an Equation of State with an explicit
QCD phase transition to quark matter on the postbounce supernova
evolution.

In the dense environment encountered in the late stage of the collapse
or in the nascent neutron star, neutrino processes are modified by
ion-ion correlations~\cite{Burrows.Reddy.Thompson:2006}. This includes
the coherent elastic scattering of neutrinos on nuclei
\cite{Horowitz97,Bruenn97,Marek05}.  Correlations are particularly
important for neutrino-pair bremsstrahlung, $NN \leftrightarrow NN\nu
{\bar \nu}$. Calculations performed on the basis of Chiral Effective
Field Theory have recently shown that noncentral contributions to the
strong interaction, i.e.  tensor and spin-orbit forces, reduce the
neutrino pair production rate significantly compared to one-pion
exchange approximations usually adopted in supernova simulations
\cite{Bacca09,Bacca12}.  Neutrino-pair bremsstrahlung, together with
neutrino-pair annihilation into pairs of other flavor (mainly $\nu_e
{\bar \nu_e} \rightarrow \nu_x {\bar \nu_x}$, $x=\mu,\tau$), is
crucial for transport of $\mu$ and $\tau$ neutrinos
\cite{Hannestad98,Buras03a} and its inclusion in supernova simulations
has resulted in noticeable changes of the spectrum of these two
neutrino types towards smaller average energies.

\begin{figure}
  \begin{center}
    \includegraphics[width=0.6\linewidth]{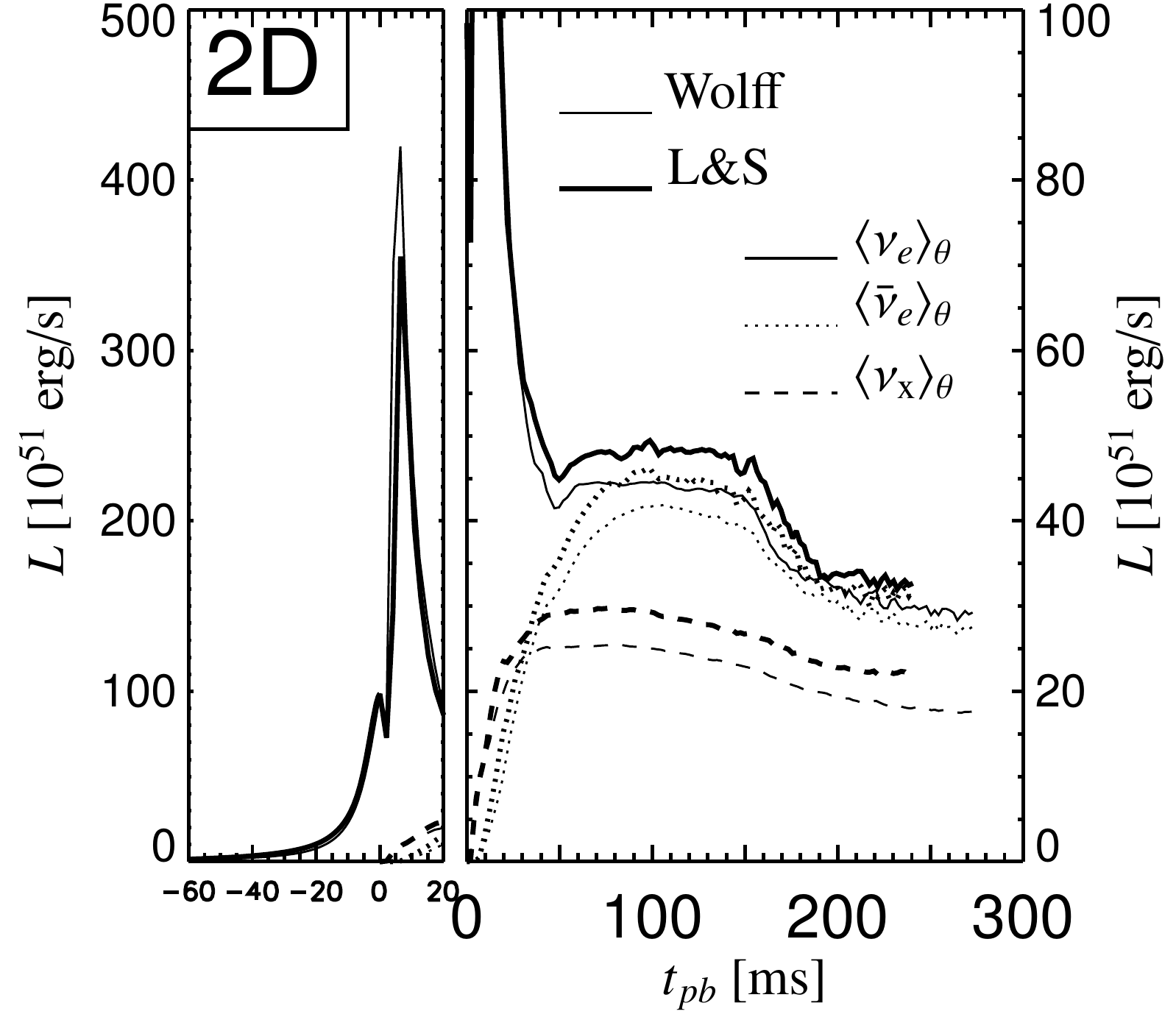}
    \caption{ Results from 2D core-collapse simulations with two
      different equations of states (``Wolff'' \cite{Wolff84} and
      ``L\&S'' \cite{Lattimer91}) for a 15~M$_\odot$ star, using the
      neutrino transport technique of Refs.~\cite{Rampp02,Buras06a}.
      The figure displays the neutrino luminosities with the prompt
      $\nu_e$ burst in the left window and the postbounce luminosities
      of $\nu_e$ (solid lines), $\bar\nu_e$ (dotted) and heavy-lepton
      $\nu$'s and $\bar\nu$'s (dashed) in the right window.  The
      results of the 2D runs were averaged over latitudinal angles
      $\theta$.  (from \cite{Janka07}) }
    \label{fig:neutrino-luminosity}
  \end{center}
\end{figure}

Fig. \ref{fig:neutrino-luminosity} shows the time evolution of the
luminosities in the different neutrino flavors obtained in a
two-dimensional simulation for two different Equation of
States. During the collapse phase the supernova overwhelmingly emits
electron neutrinos, produced by electron capture on nuclei (the
production rate of other neutrino types by nuclear deexcitation during
collapse is much less and is discussed in Section 4).  The sharp
reduction in the $\nu_e$ luminosity just before bounce (at $t_{pb}=0$)
is related to neutrino trapping in the last phase of the collapse. It
is followed by the sharp peak of the neutrino burst just after bounce,
caused by photodissociation of nuclei into free nucleons and the
associated fast electron captures on free protons. After about 50 ms
the neutrino luminosities are mainly due to neutrino pair production
caused by electron-positron annihilation in the dense and hot
environment. Consequently supernova shine now in all neutrino
flavors. As $\mu$ and $\tau$ neutrino pairs can couple to $e^+ e^-$
pairs only by neutral current, their production rate is smaller than
the one of $\nu_e {\bar \nu_e}$ pairs.

The radius from which neutrinos can be considered to be 'free streaming'
is called the neutrinosphere~\cite{Bethe90}. For electron neutrinos
the neutrinosphere approximately corresponds to the region with
trapping densities of a few $10^{11}$ g~cm$^{-3}$, which is at radii of
about 20 km in the later postbounce
phase~\cite{Fischer.Martinez-Pinedo.ea:2012}.  As neutrino cross
sections depend on the neutrino energy, the neutrinosphere is
'diffuse'; i.e. its radius decreases with neutrino energy.

On quite general grounds one expects a hierarchy in the average
energies of neutrinos of different flavors: As $\mu$ and $\tau$
neutrinos have not sufficient energies to interact with matter by
charged-current reactions, they decouple at smaller radii, and higher
temperatures, than electron neutrinos. Furthermore, as the matter is
neutron rich, anti-electron neutrinos, which by charged-current only
interact with protons, decouple at smaller radii and higher
temperature than electron neutrinos, which interact by charged-current
reactions with neutrons. The hierarchy in temperatures at which the
neutrino flavors decouple translates then into a hierarchy of average
energies: $\langle E_{\nu_x} \rangle \gtrsim \langle E_{\bar \nu_e} \rangle
> \langle E_{\nu_e} \rangle$.  For current estimates for the average
neutrino energies see refs.~\cite{janka12,gmp14}.

It is one of the major aims for future supernova neutrino detections
to confirm this hierarchy. However, these measurements will depend
strongly on neutrino flavor oscillations along the way to the
detector.  One possibility are adiabatic flavor
conversions~\cite{Dighe00} which implies noticeable mixing of the
${\bar \nu_e}$ and ${\bar \nu_x}$ spectra in case of the normal mass
hierarchy and a complete swap for the inverted mass hierarchy
\cite{Dighe00}.  For our nucleosynthesis discussion in Section 5 the
flavor conversion induced by neutrino-neutrino refraction
\cite{Duan10} is more important.  In fact, several studies
\cite{Dasgupta09,Duan11,Wu11,Wu.Qian.ea:2015} indicate that collective
neutrino flavor oscillations, which occur in the high-neutrino-density
environment surrounding the neutron star, swap the spectra of ${\bar
  \nu_e}$ and ${\bar \nu_x}$ neutrinos in certain energy intervals
bounded by sharp spectral splits. This can have interesting
consequences for the nucleosynthesis in the $\nu p$ process (see
Section 5),

\section{Theoretical background and experimental validation} 

In this Section we want to discuss the theoretical background to
describe neutrino-induced reactions on nuclei at neutrino energies
relevant for supernova dynamics and nucleosynthesis.  We will at first
focus on charged- and neutral-current reactions on nuclei in their
ground state.  These reactions are directly relevant for studies of
the neutrino-induced response of detectors for supernova
neutrinos. The finite temperature in the supernova environment implies
that nuclei exist in an ensemble with excited nuclear states thermally
populated. This requires special attention. As charged-current
$(\nu_e,e^-)$ and $(\bar{\nu}_e ,e^+)$ reactions are the inverse of
stellar electron and positron captures, the relevant reaction rates
can be obtained from the latter by applying the principle of detailed
balance. In fact, in this way $(\nu_e,e^-)$ and $(\bar{\nu}_e ,e^+)$
reactions on nuclei are included in supernova simulations for a long
time. Furthermore, charged-current reactions induced by muon and tau
neutrinos are neglected in supernova simulations as their energies are
not sufficient to initiate these processes at supernova conditions
where individual nuclei exist.  In the supernova neutral-current
reactions on nuclei might alter the neutrino spectrum and/or
contribute to the energy exchange between matter and neutrinos.  Hence
finite temperature effects are relevant and we will discuss below how
these can be estimated and which influence they have on cross
sections. The neutrino energies are often large enough to excite the
nucleus to states above particle thresholds.  In the hot supernova
environment the subsequent particle decay can be neglected as the
matter composition can be described in NSE. However, the decay is
important for nucleosynthesis studies at lower temperatures as it
alters the abundance distributions. The relevant decay probabilities
will be described within the framework of the statistical model.

This Section is divided in 3 subsections in which we will derive the
relevant cross sections, summarize briefly the hierarchy of nuclear
models adopted to describe the nuclear structure problem at the
various energy scales involved in the neutrino-nucleus reactions of
interest and finally validate this ansatz by comparison to relevant
experimental data.

\subsection{Cross sections}
For the semileptonic processes
\begin{equation}
  \nu_e +{} _{Z}A_{N} \rightarrow {} _{Z+1}A^{*}_{N-1} + e^{-}; \quad 
  \bar\nu_e + {} _{Z} A_{N} \rightarrow {} _{Z+1}A^{*}_{N-1} + e^{+} ;\quad
  \nu +{} _{Z}A_{N} \rightarrow {} _{Z}A^{*}_{N} + \nu'
\end{equation}
the excitation cross section from the ground states of a nucleus
(defined by proton and neutron numbers $Z$ and $N$, respectively) to a
discrete final state is given in the extreme relativistic limit (final
lepton energy $\epsilon_{f} >\!>$ lepton mass $m_{f}c^{2}$) by
\cite{WaWu,Co72}:

\begin{equation}
  \label{eq:Cr01a}
  \left ( \frac{d \sigma_{i \rightarrow f}}
    {d \Omega_{e}} \right )_{\nu,\bar\nu} =
  \frac{G_F^{2} \cdot \epsilon^{2}_{f}}{2 \pi^{2}} \cdot
  \frac{4 \pi \cos^{2} \frac{\Theta}{2}}{(2 J_{i} + 1)} \cdot F(Z\pm1,
  \epsilon_{f}) \cdot \left [ \sum_{J=0}^{\infty} \sigma_{CL}^J +
    \sum_{J=1}^{\infty} \sigma_{T}^J \right ]
\end{equation}
where
\begin{equation}
  \label{Cr01b}
  \sigma_{CL}^J =   | \langle J_{f} || \bm{M}_{J} (q)
  + \frac{\omega}{{q}} \bm{L}_{J}(q) || J_{i} \rangle|^{2}
\end{equation}
and
\begin{equation}
  \label{Cr01ic}
  \begin{array}{ll}
    \sigma_T^J =
    \mbox{} & \left ( -\frac{q^{2}_{\mu}}{2 {q}^{2}} + \tan^{2}
      \frac{\Theta}{2} \right ) \times \left [
      |\langle J_{f} ||
      \bm{J}_{J}^{mag}(q) || J_{i} \rangle|^{2} + |\langle J_{f}||
      \bm{J}^{el}_{J} 
      (q) || J_{i} \rangle |^{2} \right ] \\
    \mbox{} &  \mp 2\tan \frac{\Theta}{2} \sqrt{\frac{-q^{2}_{\mu}}{
        {q}^{2}}
      + \tan^{2} \frac{\Theta}{2}} \times
      \text{Re}\left[ \langle J_{f} || \bm{J}^{mag}_{J}(q) || J_{i}
      \rangle \langle J_{f} || \bm{J}
      ^{el}_{J}(q) || J_{i}\rangle^{*} \right ]  \, .
  \end{array}
\end{equation}
Here $G_F$ is the Fermi coupling constant, $\Theta$ the angle between
the incoming and outgoing lepton, and $q_{\mu} = (\omega, \vec{q})$
$(q = |\vec{q}|)$ the four-momentum transfer. The minus-(plus) sign in
Eq. (4) refers to the neutrino (antineutrino) cross section. The
quantities $\bm{M}_{J}, \bm{L}_{J}, \bm{J}^{el}_{J}$ and
$\bm{J}^{mag}_{J}$ denote the multipole operators for the charge,
and the longitudinal and the transverse electric and magnetic parts of
the four-current, respectively. Following Ref. \cite{WaWu} they can be
written in terms of one-body operators in the nuclear many-body
Hilbert space. The cross section involves the reduced matrix elements
of these operators between the initial state $J_{i}$ and the final
state $J_{f}$.

The Fermi function $F(Z, \epsilon_f)$ appears in the cross section for
the charged-current process and accounts for the Coulomb interaction
between the final charged lepton and the residual nucleus.  It can be
derived relativistically as the Coulomb correction obtained from a
numerical solution of the Dirac equation for an extended nuclear
charge \cite{Ha82,Behrens.Buehring:1982}:

\begin{equation}
  \label{Cr02}
  \begin{array}{ll}
    \mbox{} & F(Z,\epsilon_f) = F_{0}(Z,\epsilon_f) \cdot L_{0} \\
    {\rm {with}} & \displaystyle{F_{0}(Z,\epsilon_f) = 4(2p_{l}R)^{2(\gamma - 1)}
    \left | \frac{\Gamma (\gamma + i y)}{\Gamma(2 \gamma + 1)} \right |^{2}
    \cdot e^{\pi y}} ,
  \end{array}
\end{equation}
where $Z$ denotes the atomic number of the residual nucleus in the
final channel, $\epsilon_f$ the total lepton energy (in units of
$m_l c^{2}$) and $p_{l}$ the lepton momentum (in units of $m_l c$),
$R$ is the nuclear radius (in units of $\lambda = \hbar/(m_{l}c)$) and
$\gamma$ and $y$ are given by ($\alpha =$ 
fine structure constant):
\begin{equation}
  \gamma = \sqrt{1 - (\alpha Z)^{2}} \quad
  \text{and} \quad y = \alpha Z 
  \frac{\epsilon_f}{p_{l}} .
\end{equation}
The numerical factor $L_{0}$ in (5), which describes the finite charge
distribution and screening corrections, is nearly constant($\approx
1.0$), and can be well approximated by a weakly decreasing linear
function in $p_{l}$~\cite{Behrens.Buehring:1982}.

At higher energies, the Fermi function for $s$ waves becomes a poor
approximation when higher partial waves also contribute for $p_{l}R
\ge 1$. Guided by the distorted-wave approximation of quasielastic
electron scattering,  in that case the Coulomb effects can be treated in the
`Effective Momentum Approximation' in which the outgoing lepton
momentum $p_{l}$ is replaced by the effective momentum
\begin{equation}
  p_{\text{eff}} = \sqrt{E_{\text{eff}}^2 - m_{l}^2}, ~~E_{\text{eff}}
  = E_{l} - V_C(0) ~, 
\end{equation}
where $V_C(0) = 3 e^2 Z /(2 R)$ is the Coulomb potential at the
origin. In the formalism above~\cite{Engel93}, the Coulomb effect is
taken into account not only by using the effective momentum, but also
by replacing the phase space factor $p_{l} E_{l}$ by $p_{\text{eff}}
E_{\text{eff}}$ (see also \cite{Engel98} where this procedure is
called modified effective momentum approximation, and shown to work
quite well).  In practice, a smooth interpolation between these two
regimes of treatment of the Coulomb effects is used.

The total cross section is obtained from the differential cross
sections by summing over all possible final nuclear states and by
numerical integration over the solid angle $d \Omega$.

Supernova neutrino energies are sufficiently high to excite nuclei in
the final channel to energies above particle thresholds. The nuclear
state will then subsequently decay by particle emission. Under certain
conditions the final nucleus might even be excited to energies high
enough to allow for multi-particle emissions. Such a situation occurs
in neutral-current reactions induced by supernova $\nu_x$ neutrinos as
their expected average energies of about 18 MeV often suffices to
excite nuclei above the 2-particle emission threshold. Such a decay is
also relevant for charged-current reactions induced by supernova
$\nu_e$ neutrinos if the $Q$ value is large enough. Such a situation
will certainly occur for r-process nuclei if they are subject to
charged-current reactions with supernova $\nu_e$ neutrinos. As nuclei
on the r-process path also have very small neutron separation
energies, the state excited by the neutrino process in the daughter
nucleus will decay by the emission of several neutrons
\cite{Kolbe01,Kolbe02}. It has been pointed out that neutrino-induced
fission of r-process nuclei will also be accompanied by multiple
neutrons \cite{Qian02,Kolbe04}.

Neutrino-induced reactions with subsequent particle emission can be
described in a two-step process. In the first step a nuclear level $f$
at energy $E_f$ with angular momentum $J_f$ and parity $\pi_f$ is
excited. The respective reaction cross section is described by
Eq. \ref{eq:Cr01a}. The subsequent particle decay is calculated on the
basis of the statistical model. At relatively low excitation energies,
where one-particle decays dominate, these calculations are preferably
done on the basis of statistical model codes which explicitly consider
experimentally known nuclear spectra. There are several such codes
existing: SMOKER \cite{Cowan91}, MOD-SMOKER and SMARAGD
\cite{Rauscher11}, TALYS \cite{Koning12}.  There have been a few
applications where such codes have been extended to treat also the
cascade of two-particle emissions.  However, these treatments are
computationally quite demanding, thus multi-particle emissions are
usually studied within a Monte-Carlo treatment of the decay
cascades. Such a dynamical code is ABLA07 \cite{Kelic10} that
describes de-excitation of the compound nucleus through the
evaporation of light particles (protons, neutrons, $\alpha$ and
$\gamma$) and fission. The particle evaporation is considered in the
framework of the Weisskopf formalism \cite{Weisskopf37}, while in the
description of fission the ABLA code treats explicitly the relaxation
process in deformation space and the resulting time-dependent fission
width using an analytical approximation \cite{Jurado03} to the
solution of the Fokker-Planck equation. The wide range of excitation
energies and the large variety of nuclei of interest demands for a
consistent treatment of level densities as a function of excitation
energy and deformation. The nuclear level density is calculated
according to the Fermi-gas formula taking into account the shell
corrections, collective excitations and pairing correlations; details
on the implementation of these effects in the nuclear level density,
their damping with excitation energy and their deformation dependence
are described in details in
Refs. \cite{Junghans98,Schmidt82,Ignatyuk75}. The
angular-momentum-dependent fission barriers are taken from the
finite-range liquid-drop model \cite{Sierk86}. In order to
describe the fission-fragment mass and charge distributions, as well
as their kinetic energies, the ABLA code is coupled to the
semi-empirical Monte-Carlo fission model PROFI \cite{Benlliure98}.

\subsection{Hybrid model for allowed and forbidden contributions} 

\begin{figure}
  \begin{center}
    \includegraphics[width=0.5\linewidth]{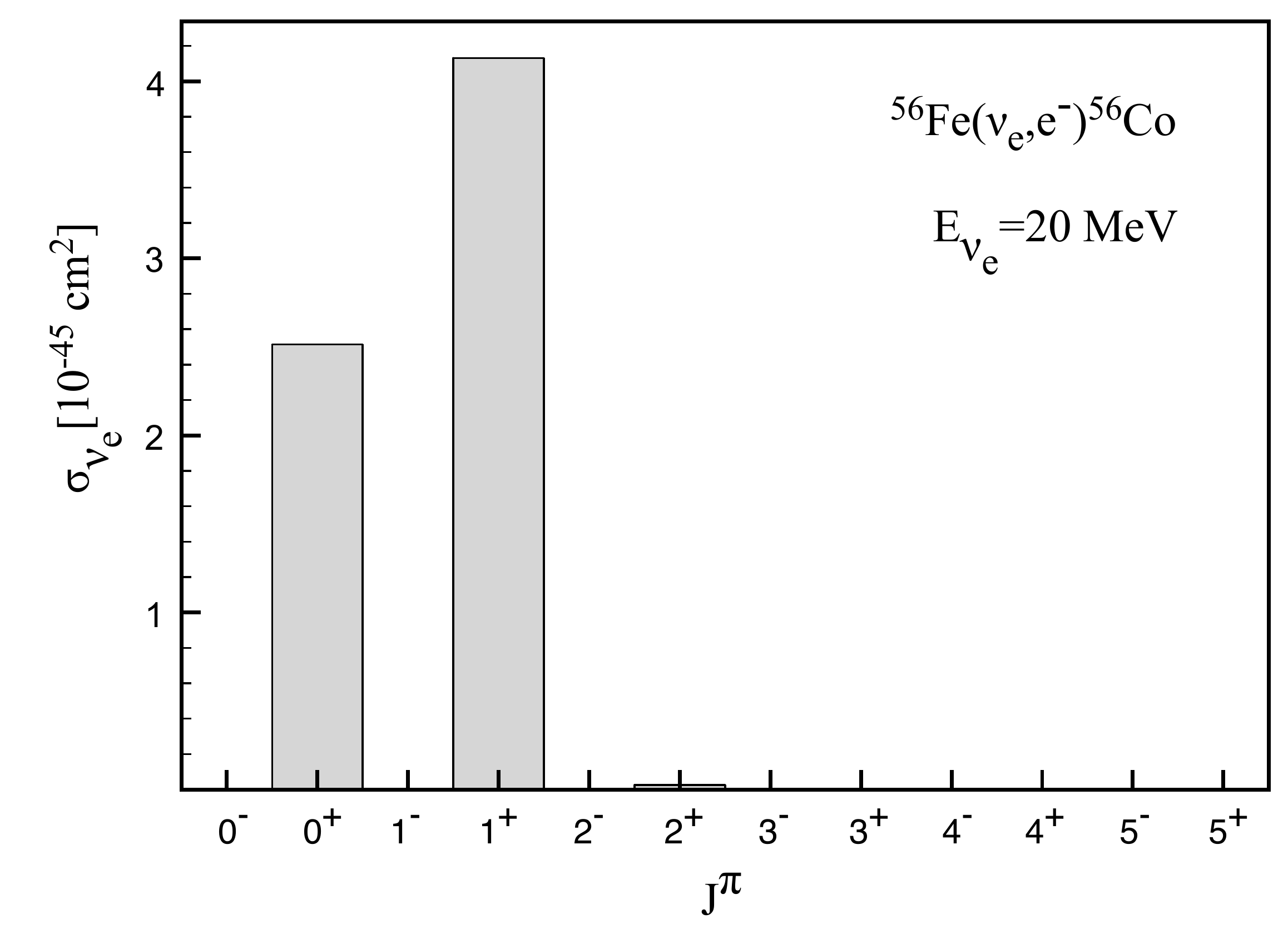}%
    \includegraphics[width=0.5\linewidth]{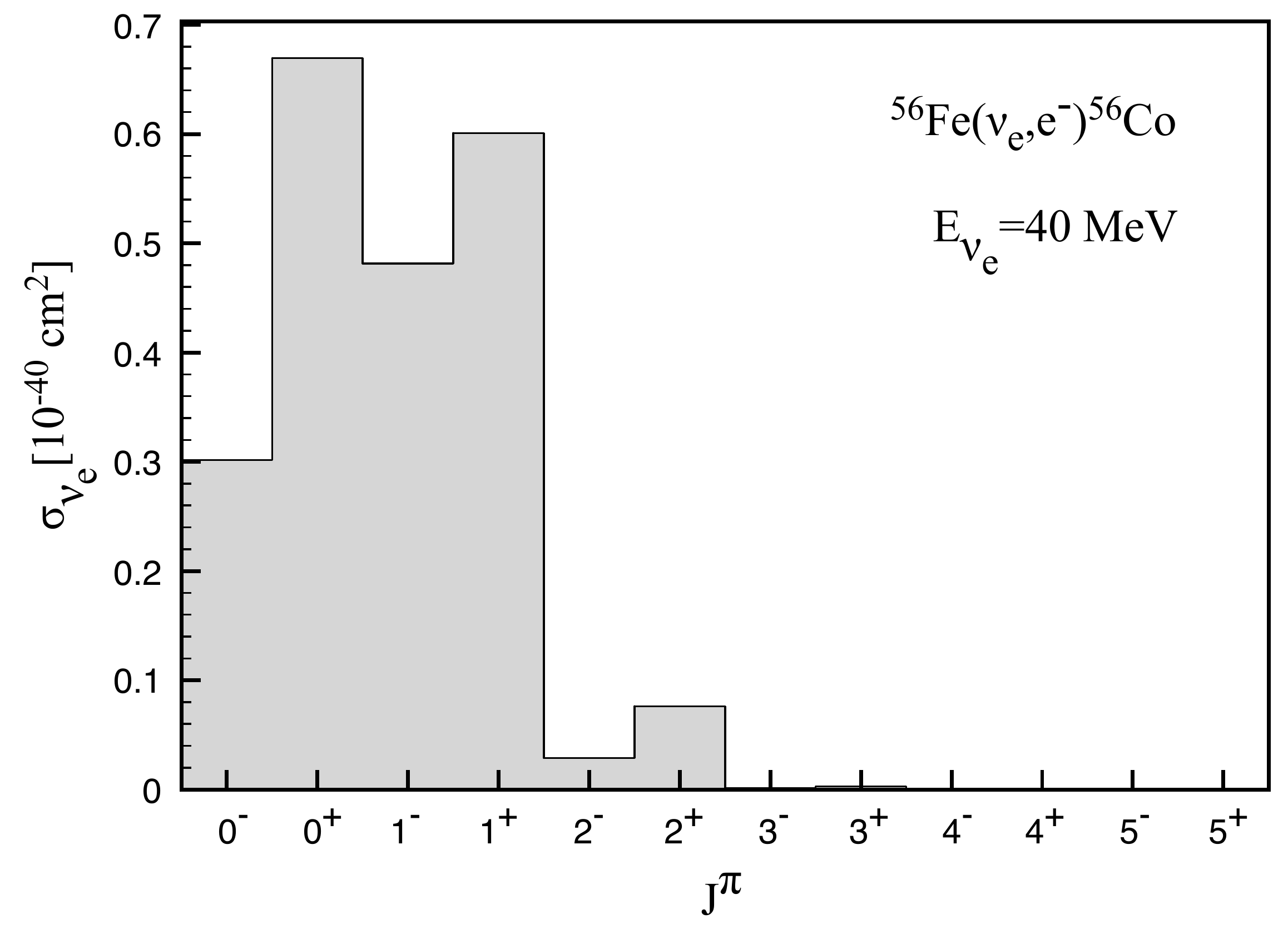}
    \caption{Multipole decomposition of $(\nu_e,e^-)$ cross sections
      on $^{56}$Fe for two different neutrino energies.  (from
      \cite{Paar08}) \label{fig:multipoles}} 
  \end{center}
\end{figure}

Reactions of low-energy neutrinos with $E_\nu \le 20$~MeV
off nuclei are dominated by allowed $J=0^+$ and $J=1^+$ transitions
(see Fig. \ref{fig:multipoles}). In the limit of vanishing momentum
transfers $q=0$, which is justified at the low neutrino energies
involved, these two transitions between initial ($i$) and final ($f$)
nuclear states reduce to the Fermi and Gamow-Teller transitions,
respectively:

\begin{equation}
  \label{eq:fermiT}
  B_{if}(F) = T(T+1)-T_{z_i} T_{z_f},
\end{equation}
where the transition is only possible between Isobaric Analog states
and where we have neglected the reduction in the overlap between
nuclear wave functions due to isospin mixing which is estimated to be
small ($\approx 0.5$\%~\cite{Towner95}), and
\begin{equation}
  \label{eq:bgt}
  B_{if}(GT) = \left(\frac{g_A}{g_V}\right)^2_{\text{eff}}
  \frac{\langle f||\sum_k \bm{\sigma}^k \bm{t}^k_\pm || i
    \rangle^2}{2 J_i +1},
\end{equation}
where the matrix element is reduced with respect to the spin operator
$\bm{\sigma}$ only (Racah convention~\cite{Edmonds60}) and the sum
runs over all nucleons.  For the isospin rising and lowering
operators, $\bm{t}_\pm = (\bm{\tau}_x \pm i \bm{\tau}_y)/2$, we use
the convention $\bm{t}_+ p = n$; thus, `$+$' refers to electron
capture and $\beta^+$ transitions and `$-$' to positron capture and
$\beta^-$ transitions.  Finally, $(g_A/g_V)_{\text{eff}}$ is the
effective ratio of axial and vector coupling constants that takes into
account the observed quenching of the GT
strength~\cite{Osterfeld92}. One often uses
~\cite{Wildenthal88,Langanke95,Martinez97}
\begin{equation}
  \label{eq:gaeff}
  \left(\frac{g_A}{g_V}\right)_{\rm{eff}} = 0.74
  \left(\frac{g_A}{g_V}\right)_{\rm{bare}},
\end{equation}
with $(g_A/g_V)_{\rm{bare}} = -1.2599(25)$~\cite{Towner95}. If the
parent nucleus (with isospin $T$) has a neutron excess, then the
GT$_-$ operator can connect to states with isospin $T-1$, $T$, $T+1$
in the daughter, while GT$_+$ can only reach states with $T+1$. This
isospin selection is one reason why the GT$_+$ strength is more
concentrated in the daughter nucleus (usually within a few MeV around
the centroid of the GT resonance), while the GT$_-$ is spread over
10-15~MeV in the daughter nucleus and is significantly more
structured.

The GT response is in turn sensitive to nuclear structure effects and
hence its contribution to the cross section has to be derived from a
model which is capable to describe the relevant nuclear structure and
correlation effects. This model is the diagonalization shell model
\cite{Caurier03}.  Higher multipoles will, however, contribute to the
cross section for larger neutrino energies. For each of these
multipoles the response of the operator will be fragmented over many
nuclear states. However, most of the strength resides in a collective
excitation, the giant resonance, whose centroid energy grows with
increasing rank $\lambda$ roughly like $\lambda \hbar \omega \approx 41 \lambda
/ A^{1/3}$ MeV. As the phase space prefers larger final neutrino
energies, the average nuclear excitation energy grows noticeably
slower than the initial neutrino energy. As a consequence, initial
neutrino energies are noticeably larger than the energy of a giant
resonance, when the later will contribute to the neutrino-nucleus
cross section.  Fortunately, the neutrino-nucleus cross section
depends then mainly on the total strength of the multipole excitation
and its centroid energy, and not on the detailed energy distribution
of the strength (as this is the case for the Gamow-Teller response at
low neutrino energies).  Thus the higher multipole
contributions to the neutrino-nucleus cross section can be derived within the RPA
making use of the fact that the RPA describes the energy centroid and
total strength of multipoles other than the Gamow-Teller quite well.

In such a hybrid model, the total reaction cross section consists then
of two parts
\begin{equation}
  \sigma_\nu^{\text{tot}}(E_\nu) =
  \sigma_\nu^{\text{sm}}(E_\nu)
  + \sigma_\nu^{\text{rpa}}(E_\nu),
  \label{eq-sigma-tot}
\end{equation}
where the shell model part, $\sigma^{sm}$, describes the allowed
contributions to the cross section for $(\nu_e,e^-)$ (GT$_-$) and
$(\bar\nu_e,e^+)$ (GT$_+$) reactions and is given by:
\begin{equation}
  \sigma (E_\nu) = \frac{G_F^2 {\rm cos}^2 \theta_C}{\pi}
  \sum_{f} 
  k_e^{f} E_e^{f}
  F(\pm Z\pm1,E_e^{f}) \left( B_f (F) + B_{f} (GT_\pm) \right)
\end{equation}
where $G_F$ is the Fermi constant, $\theta_C$ the Cabibbo angle, and
$k_e$ and $E_e$ the momentum and energy of the outgoing electron or
positron.  The sum is over the final ($f$) nuclear states with
energies $E_f$. There is no Fermi contribution to $(\bar\nu_e,e^+)$
reactions on nuclei with neutron excess.  For inelastic neutrino
scattering to an excited nuclear state at energy $E_f$ the allowed
contribution to the cross section is given by:
\begin{equation}
  \sigma_{\nu}(E_\nu)=\frac{G_F^2\cos^2\theta_C}{\pi}\sum_f
  E_{\nu^\prime,f}^2B_{f}({\rm GT}_0)
  \label{eq1}
\end{equation}
where energy conservation yields $E_{\nu^\prime,f} =E_\nu-E_f$.  These
allowed contributions replace the $\lambda = 0^+$ and $\lambda =1^+$
contributions in the cross section formula above. All other multipole
contributions can be derived adopting the RPA as the nuclear model to
describe the initial and final nuclear states. We mention that, for
light nuclei like $^{12}$C, shell model calculations are now also
possible for the first-forbidden transitions ($\lambda=0^-,1^-,2^-$)
requiring model spaces spanned by $3
\hbar\omega$~\cite{Hayes00,Volpe00,Suzuki13}.

\subsection{Inelastic neutrino-nucleus reactions at finite
  temperature}

So far, our formalism assumes that the target nucleus is in the ground
state corresponding to an environment with temperature $T=0$. This is,
of course, appropriate for neutrino-induced reactions in earthbound
detectors and it is a reasonable approximation for the supernova
environment in which neutrino-nucleus reactions occur in
nucleosynthesis processes ($T < 100$ keV). However, the $T=0$
approximation is not valid if we want to study inelastic
neutrino-nucleus reactions in the hot supernova environment (with
temperatures $T \gtrsim 1$ MeV). (We recall that charged-current
reactions under these conditions can be derived from electron and
positron captures by detailed balance.) Here nuclei exist as a thermal
ensemble with excited nuclear states at energy $E_i$ with spin $J_i$
populated with a Boltzmann weight of $G_i=(2J_i+1)\exp(-E_i/kT)$.
There are two interesting effects which occur at finite $T$: i)
neutrinos can be up-scattered in energy; i.e. the neutrino picks up
energy from the nucleus which by the scattering reaction deexcites;
ii) at zero temperature for even-even nuclei with their $J=0^+$ ground
states GT transitions can only be induced by neutrinos with minimum
energies to bridge the gap to the first excited $J=1^+$ state in the
nucleus.  Transitions to excited states at lower excitation energies
can be induced by higher multipoles, but the respective cross sections
are usually quite small. As a consequence inelastic neutrino-nucleus
cross sections for even-even nuclei drop quite fast with decreasing
small neutrino energies. This is not expected to be the case for
inelastic neutrino-nucleus reactions at finite temperatures.  Due to
the exponential increase in the nuclear level density states at modest
excitation energies are likely to have states in their energy vicinity
with which they can be connected by GT transitions. Furthermore, it is
even likely that GT transitions (or with less strength by other
multipoles) can lead to levels at lower energies, a transition which
is in principle possible for all neutrino energies. In summary, we
expect on general grounds that the cross section is in particular at
low neutrino energies modified at finite temperature.

Unfortunately an evaluation of the inelastic cross section at finite
temperature by explicitly summing over the thermally populated states
is computationally unfeasible due to the many states involved at
temperatures of order MeV.  There are two suggestions how to overcome
this problem and treat inelastic neutrino scattering at finite
temperature. We briefly discuss both in turn.
 
Refs.  \cite{Sampaio,Juodagalvis05} apply Brink's hypothesis which
states that for a given excited nuclear level $i$ the strength
distribution built on this state, $S_i(E)$, is the same as for the
ground state, $S_0(E)$, but shifted by the excitation energy $E_i$:
$S_i(E)=S_0(E-E_i)$.  The validation of Brink's hypothesis has been in
detailed studied in ref. \cite{Misch14}, however, for electron capture
on sd-shell nuclei (see also \cite{Langanke00}). Upon applying the
Brink hypothesis, the down-scattering part becomes independent of
temperature and can be solely derived from the ground state GT
distribution.  With this approximation, the Gamow-Teller (shell model)
contribution to the cross section becomes:
\begin{equation}
  \sigma_\nu^{\text{sm}}(E_\nu)= \frac{G_F^2}{\pi}
  \left[
    \sum_{f} E_{\nu',0{f}}^{2} B_{0f} (GT_0) +
    \sum_{f} E_{\nu',if}^{2} B_{if}
    (GT_0) \frac{G_{i}}{G}
  \right],
  \label{eq-sm-cross-section}
\end{equation}
where $G_F$ is the Fermi constant, and $E_{\nu', if}$ is the
energy of the scattered neutrino, $E_{\nu',if}=E_\nu+(E_i-E_f)$,
with $E_i,$ $E_f$ denoting the initial and final nuclear energies.

The first term in Eq. (\ref{eq-sm-cross-section}) arises from Brink's
hypothesis.  By construction, this Brink term does not allow for
neutrino up-scattering. These contributions to the cross section are
comprised in the second term, where the sum runs over both initial ($i$)
and final states ($f$). The former have a thermal weight of
$G_i=(2J_i+1)\exp(-E_i/kT)$, where $J_i$ is the angular momentum,
$E_i$ is energy of the initial state, and $T$ is temperature;
$G=\sum_i G_i$ is the nuclear partition function.  The up-scattering
contributions are the more important i) the lower the nuclear
excitation energy (Boltzmann weight in the thermal ensemble), ii) the
larger the GT transition strength $B_{if}$, and iii) the larger the
final neutrino energy. Guided by these general considerations
Ref. \cite{Juodagalvis05} has approximated the second term in
Eq. (\ref{eq-sm-cross-section}) by explicitly considering GT
transitions between nuclear states with $E_i > E_f$, where the sum
over the final nuclear states is restricted to the lowest nuclear
levels.  As is usual in astrophysical applications, experimental data
(in this case mainly excitation energies) have been used whenever
available.

As stressed above, the derivation of the first (down-scattering) term
in the cross section explicitly assumes that the distributions on
excited states is the same as for the ground state. This is likely not
the case as the vanishing of pairing correlations with increasing
temperature should effect the relative excitation energy of the
strength centroid and move it slightly down in energy.  In fact Shell Model
Monte Carlo studies, performed at finite temperature, confirm this
conjecture \cite{Radha97}. To account for this effect, Dzhioev {\it et
  al.}  developed a model which extends the QRPA approach to finite
temperature (Thermal Quasiparticle Random-Phase Approximation, TQRPA
\cite{Dzhioev11,Dzhioev13}). In the applications the authors consider
multipole contributions up to $\lambda=3$. However, it is illustrative
to assume that the scattering process is given solely by the
contribution from the allowed $J=1^+$ multipole. Then the cross
section induced by neutrinos with energy $E_\nu$ at temperature $T$
can be written as
\begin{equation}
  \sigma(E_\nu ,T) = \frac{G_F^2}{\pi} \sum_i^\prime (E_\nu - \omega_i)^2 \Phi_i
  +  \frac{G_F^2}{\pi} \sum_i (E_\nu + \omega)^2
  \exp(-\frac{\omega_i}{kT}) \Phi_i .
  \label{eq:Dzhioev}
\end{equation}
Analog to Eq. (\ref{eq-sm-cross-section}) the first term corresponds
to the down-scattering, and the second to the up-scattering
contributions. The sum in the down-scattering term runs over all $1^+$
thermal phonon states with positive energy $\omega_i < E_\nu$.  The
Boltzmann factor suppresses the contribution from highly thermally
populated states in the up-scattering term. These transitions are,
however, somewhat favored by the phase factor $(E_\nu + \omega)^2$
\cite{Dzhioev13}. In Eq.~(\ref{eq:Dzhioev}) $\Phi_i$ denotes the
transition strength of the GT operator. We note that the TQRPA
approach obeys detailed balance which, however, is slightly violated
in the approach based on Eq. (\ref{eq-sm-cross-section}).

Using iron isotopes as example we will discuss results from both
approaches in Section 4.

\subsection{Nuclear models and their validation by data}

Measurements of neutrino-nucleus cross sections are very rare.  For
charged-current reactions there exist the set of measurements for
solar electron neutrinos on the nuclei comprising the various solar
neutrino detectors. These data - despite being very valuable for solar
physics and the proof of neutrino oscillations - are only of
restricted use to check the applicability of nuclear models to
describe neutrino-nucleus reactions for supernova neutrinos.  Solar
neutrinos have an energy spectrum, which is well understood from the
reactions mediated by the weak interaction in solar hydrogen burning
and distorted by matter-enhanced oscillations in the Sun. However,
solar neutrino energies are too low to excite giant nuclear resonances
which in turn are expected to dominate the cross sections for
supernova neutrinos. This means that transitions induced by solar
neutrinos are strongly state-dependent and are quite a challenge to
model. There have been a few theoretical studies performed for solar
neutrinos within the framework of effective field
theory~\cite{Gudkov03}, shell model and RPA.

For the deuteron, which has been the target nucleus in the SNO
detector, neutrino-scattering calculations have been performed in
effective field theory (EFT).  This approach allows to split the
neutrino-induced cross section into two pieces $\sigma (E) = a(E) +
L_{1A} b(E)$ where $a(E)$ and $b(E)$ are known and all unknown effects
can be lumped together in one unknown parameter $L_{1A}$ (isovector
two-body axial current) which must be determined by experiment
\cite{Marcucci07,Vogel-Her}.  Although $a(E)$ contributes dominantly
to the cross section, the current uncertainty in $L_{1A}$ (about $50
\%$) is still large
\cite{Schiavilla98,Butler02,Chen03,Brown08}. Similar results have been
obtained based on the phenomenological Lagrangian
approach~\cite{Nakamura01}

For the two nuclei $^{12}$C and $^{56}$Fe there exist measurements of
cross sections induced by $\nu_e$ neutrinos
\cite{Drexlin91,Zeitnitz94,Albert95,Athanassopoulos97,Imlay98,Maschuw98}.
The neutrinos are produced by muons decaying at rest and have
therefore spectra which are slightly shifted to larger neutrino
energies than supernova neutrinos. Nevertheless the respective
neutrino-nucleus cross sections are expected to be dominated by
contributions from allowed (Fermi and GT) and first-forbidden
transitions and are therefore valuable test cases for nuclear
models. In turn measurements of these transition strengths serve as
strong indirect constraints of the ability of nuclear models to
describe neutrino-nucleus scattering at supernova neutrino energies.

In Ref. \cite{Kolbe03} it is argued on the basis of general
considerations that, depending on the neutrino energies, the nuclear
structure problem has to be treated with different demands and
details. As we will show below these requirements are fulfilled by a
hierarchy of models which is actually the basis of the hybrid model
defined above: the nuclear shell model for low neutrino energies and
the Random Phase Approximation (or its variants) for higher energies.

i) Low energy neutrino scattering requires the detailed reproduction
of the nuclear response (mainly allowed transitions) at low nuclear
excitation energy. The method of choice here is the shell model which,
for a fixed model space of valence nucleons, accounts for
nucleon-nucleon correlations among these valence particles via an
effective interaction. Shell model wave functions have good quantum
numbers for angular momentum, parity and isospin.  Modern shell model
codes allow now diagonalization in model spaces with up to a few
billion configurations. Based on a Monte Carlo strategy of importance
sampling shell model calculations for low energy spectra have been
extended to even larger model spaces. Hence shell model
diagonalization is now possible for medium-mass nuclei in the
iron-nickel region in the full $pf$ shell as is desirable for
calculations of GT transitions..  For lighter nuclei, like $^{12}$C
and $^{16}$O, shell model calculations can be performed in
multi-$\hbar \omega$ model spaces allowing also for fully correlated
studies of first-forbidden transitions. Such transition strengths are
calculated employing the Lanczos technique for a finite number (of
order 100) iterations. As a consequence the lowest transitions
correspond to well-defined (shell model) states, while at higher
excitation energies the calculated strength corresponds to the summed
strength per energy interval (which depends on the number of
iterations considered).  For a review on the shell model and its
abilities and applications the reader is refered to
Ref. \cite{Caurier03}.

ii) The RPA has been developed to describe the collective excitations
of a nucleus by considering the one-particle one-hole excitations of
the correlated ground state. Compared to the shell model, the RPA
calculations usually use model spaces with a (significantly) larger
set of single particle, but they consider, however, only a strongly
truncated part of the possible configurations. This has the
consequence that the strength is less fragmented in RPA than in shell
model calculations.

It is important that in the shell model and in the RPA the total
strength is fixed by sum rules. Of particular relevance for the
description of neutrino-nucleus reactions at supernova neutrino
energies are the Ikeda sum rule for the Gamow-Teller strength
\begin{equation}
  \sum_i GT(Z \rightarrow Z+1)_i - \sum_i GT(Z \rightarrow Z-1)_i
  = 3 (N- Z),
\end{equation}
and the Thomas-Reiche-Kuhn sum rule for the dipole strength
\begin{equation}
  \sum_i(E_i - E_0)B(E1;0 \rightarrow i) = \frac{9}{4 \pi}
  \frac{\hbar^2} {2 M_n} \frac{N Z}{A} e^2 ~.
\end{equation}
where $M_n$ is the nucleon mass. For nuclei with large neutron
excess, the Ikeda sum rule basically fixes the total GT$_-$ strength
relevant for $(\nu_e,e^-)$ reactions.

It is known from studies of $\beta$ decays and of GT distributions
\cite{Wildenthal88,Martinez97} that both shell model and RPA
calculations systematically overestimate the GT transition
strengths. For the shell model calculations this shortcoming can be
overcome by introducing a constant quenching factor via the
introduction of an effective axial-vector coupling constant which for
$pf$ shell nuclei has empirically been determined as $g_A^{\rm eff} =
0.74 g_A$ \cite{Martinez97}.  GT strengths, obtained in RPA
calculations, are often quenched by the same factor, although this
procedure in RPA studies is hampered by the limited consideration of
configurations within the model space.

Replacing the full $\lambda = 1^+$ operator by its approximation in
the $q=0$ limit (the GT operator) is only justified for small incident
neutrino energies. At modest neutrino energies the consideration of
the finite momentum transfer reduces the cross section.  In
Ref. \cite{Kolbe99a} it is suggested to correct cross sections
calculated from the shell model GT distribution by the ratio of RPA
calculations performed for finite momentum transfer and for $q=0$,
respectively.  For neutrinos produced by muon decay at rest the
correction is of order $20\%$.  For supernova neutrino energies the
correction is smaller.

Ref. \cite{Kolbe03} gives a detailed comparison of data obtained for
charged-current reactions on $^{12}$C induced by muon-decay-at-rest
(DAR) $\nu_e$ neutrinos with various model calculations. The exclusive
reaction to the $J=1^+$ ground state of $^{12}$N is given by a GT
transition.  In general models agree quite well with experiment (see
Table I in Ref. \cite{Kolbe03}).  However, it must be noted that the
calculations are usually constrained by other experimental data which
fix the GT matrix element (like the $\beta$ decays of the $^{12}$N and
$^{12}$B ground states or the muon capture to the $^{12}$B ground
state). Hence these calculations should be understood as proving that
the exclusive DAR from Karmen \cite{Drexlin91,Zeitnitz94} and from
LSND \cite{Albert95,Athanassopoulos97,Imlay98} are consistent with
other measurements probing the nuclear matrix element.

\begin{figure}[htb]
  \begin{center}
    \includegraphics[width=0.6\linewidth]{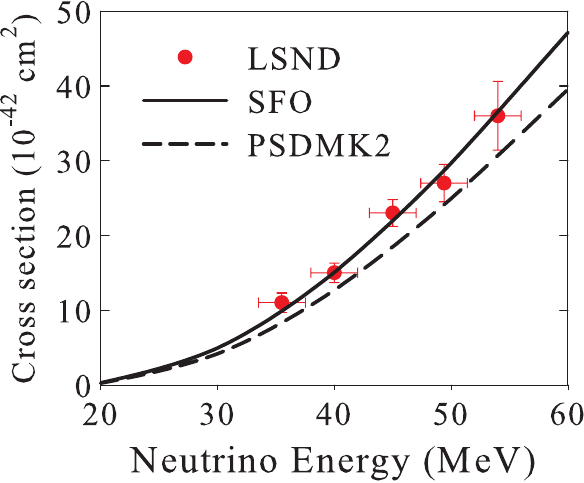}
    \caption{ Comparison of the exclusive
      $^{12}$C$(\nu_e,e^-)^{12}$N$_{gs}$ cross section data of the
      LSND collaboration \cite{Athanassopoulos97} with shell model
      calculations performed with two different residual interactions
      (from \cite{Suzuki13}). \label{fig:c12-exclusive}}
  \end{center}
\end{figure}

In fact the description of the GT transition between the $^{12}$C and
$^{12}$N ground states (or to the other members of the $T=1$
multiplett in the $A=12$ nuclei including the $^{12}$B ground state
and the $J=1^+$ state in $^{12}$C at 15.11 MeV) is a very demanding
nuclear structure problem involving the subtle breaking of the SU(4)
symmetry (in the limit of good SU(4) symmetry the GT transition
vanishes). Clearly RPA correlations are insufficient and overestimate
the transition strength by about a factor of 4
\cite{Kolbe94,Kolbe99,Hayes00}. Hayes {\it et al.} performed shell
model calculations in successively larger model spaces up to $6 \hbar
\omega$ and gained two important insights into the problem
\cite{Hayes03}: i) enlarging the model space increasingly breaks the
SU(4) symmetry. Relatedly the GT matrix elements increases moving
towards the experimental value. However, this convergence process is
quite slow. ii) The inclusion of a three-body interaction noticeable
improves the agreement with data.  The findings of Ref. \cite{Hayes03}
have recently been confirmed in no-core shell model (NCSM)
calculations using chiral NN+3N interactions \cite{Navratil07,
  Maris14}. In fact, the NSCM calculation of Maris {\it et al.}
\cite{Maris14}, performed in a model space with $8 \hbar \omega$ now
slightly overestimates the transition by about $15 \%$.  A good
reproduction of the GT transition strength, and relatedly of the
exclusive $^{12}$C$(\nu_e,e^-)^{12}$N$_{\text{gs}}$ has also been obtained
in a diagonalization shell model calculation performed in a $3\hbar
\omega$ model space~\cite{Suzuki13} using a novel residual
interaction~\cite{Otsuka10}. In fact, using a slightly quenched value
for the axial-vector coupling constant Otsuka {\it et al.}
\cite{Otsuka10} find very good agreement with the LSND exclusive cross
sections (see Fig. \ref{fig:c12-exclusive}).

The inclusive $^{12}$C$(\nu_e,e^-)^{12}$N cross section induced by DAR
neutrinos leads to many excited states in $^{12}$N. The typical
momentum transfer in this process, $q \simeq 50$~MeV, is noticeably
larger than the average excitation energy, $\omega \simeq$ 20
MeV. Under these conditions the cross section is dominated by
forbidden transitions, but it is not too sensitive to details of the
strength distribution. As discussed above, RPA studies should be
adequate to reproduce the cross sections which is indeed the case as
it is in details outlined in Ref. \cite{Kolbe03}.  Further proof that
$^{12}$C cross sections, which are dominated by contributions from
collective excitations, are well described by RPA calculations are
given by studies of the muon capture on $^{12}$C \cite{Kolbe94} and by
inclusive electron scattering. The successful reproduction of muon
capture rates by RPA calculations is not restricted to $^{12}$C, but
has been extended to the data set across the full nuclear chart
\cite{Zinner06,Marketin09}.

\begin{table}[htb]
  \caption{Comparison of $^{56}$Fe($\nu_e,e^-$)$^{56}$Co cross sections
    for DAR neutrinos calculated with different variants of the Random
    Phase Approximation, with the hybrid model and with data (exp)
    \cite{Maschuw98}.} 
  \begin{center}
    \begin{tabular}{lc}
      \hline\hline 
      \multicolumn{1}{c}{model} & cross section (in $10^{-40}$ cm$^2$) \\ \hline
      QRPA \cite{Samana12} &   2.646 \\
      PQRPA \cite{Samana12} &  1.973 \\
      RPA \cite{Athar06} &     2.77 \\
      QRPA \cite{Lazauskas07} &3.52 \\
      RQRPA \cite{Paar08}     &1.4 \\
      Hybrid \cite{Kolbe99a}  & 2.38   \\
      exp \cite{Maschuw98}  & ($2.56 \pm 1.08 \pm 0.43$) \\
      \hline\hline
    \end{tabular}
  \end{center}
\end{table}

The KARMEN collaboration has reported a measurement of the
$^{56}$Fe($\nu_e,e^-$)$^{56}$Co cross section for DAR neutrinos,
$\sigma = (2.56 \pm1.08\pm0.43 \cdot 10^{-40}$) cm$^2$
\cite{Maschuw98}.  Table 1 compares this cross section with results
from a hybrid model study and from different variants of the RPA. All
calculations are consistent with the data within their relatively
large uncertainty. We note, however, that the RPA cross sections show
a spread of about a factor of 2.5.  Looking at the results of the
hybrid model in more details one finds the largest contributions from
the allowed transitions.  The $\lambda=1^+$ part ($1.12 \times
10^{-40}$ cm$^2$) has been calculated from the shell model GT strength
distribution with the standard quenching factor and corrected for
effects of finite momentum transfer \cite{Kolbe99a}.  The
$\lambda=0^+$ contribution to the Isobaric Analog State at excitation
energy of 3.5 MeV in $^{56}$Co is readily calculated exploiting the
appropriate Fermi sum rule yielding a cross section of $0.53 \times
10^{-40}$ cm$^2$.  Higher multipoles, which have been treated within
an RPA approach, contribute a cross section of $0.73 \times 10^{-40}$
cm$^2$. The shell model $\lambda=1^+$ cross section is about 32$\%$
smaller than the one obtained within the RPA approach showing some
sensitivity to the detailed GT distribution.

\begin{figure}
  \begin{center}
    \includegraphics[width=0.60\linewidth]{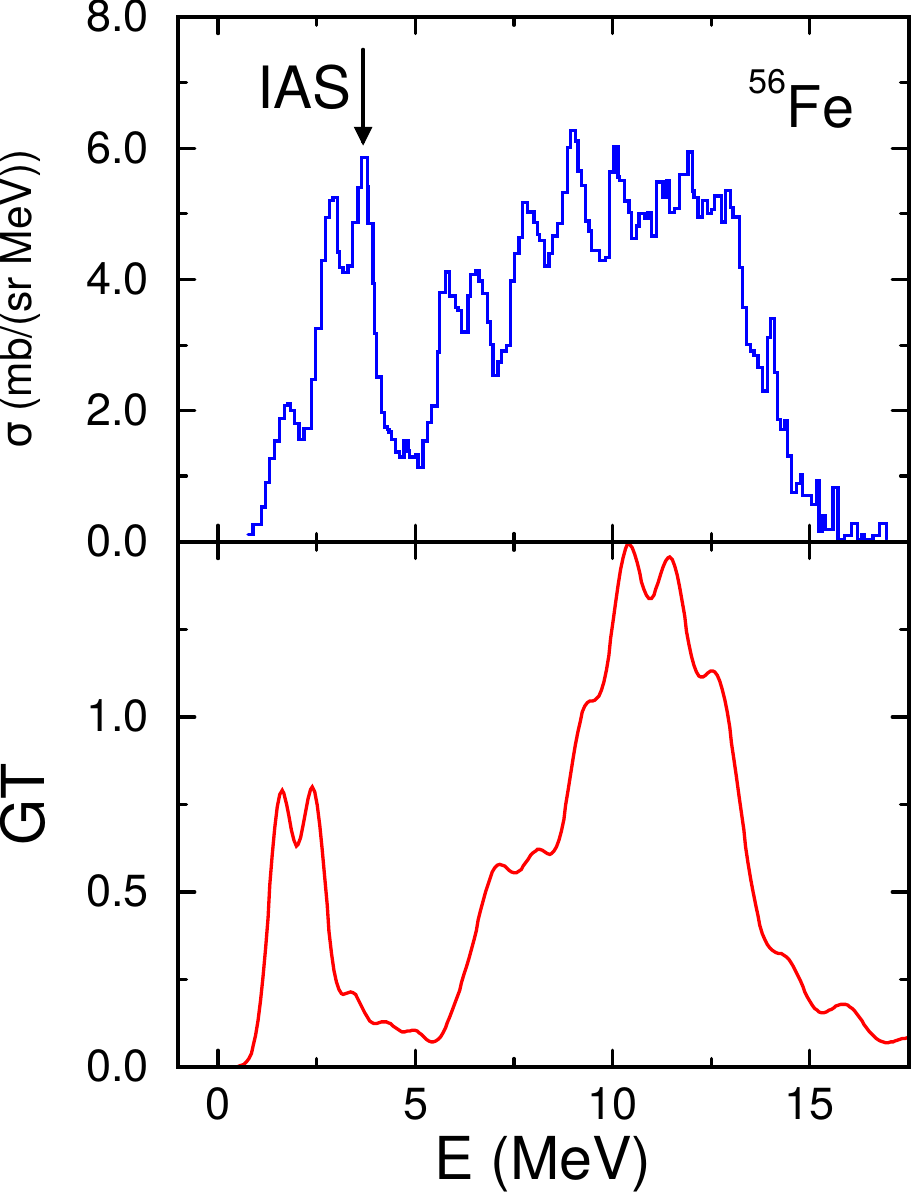}
    \caption{Comparison of the GT strength distributions for $^{56}$Fe
      obtained by $(p,n)$ charge-exchange experiment~\cite{Rapaport83}
      and a shell model calculation (folded with the experimental
      resolution). The data also include the Fermi contribution from
      the Isobaric Analog State (IAS) (from
      \cite{Caurier99}).\label{fig:FE56-GT-}}
  \end{center}
\end{figure}

Such GT distributions - as other relevant strength distributions for
dipole transitions or M1 transitions - serve as indirect constraints
to validate the nuclear input into neutrino-nucleus cross section
calculations.  For $^{56}$Fe Ref. \cite{Rapaport83} have measured the
forward-angle cross section in a charge-exchange $(p,n)$
reaction. These data are proportional to the GT$_-$ distribution,
which is of importance for the KARMEN measurement of the $(\nu_e,e^-)$
cross section.  In Fig. \ref{fig:FE56-GT-} the measured forward-angle
cross section is compared to the results obtained from a shell model
calculation \cite{Caurier99}, where the latter has been folded with
the experimental resolution. Overall a reasonable agreement is
observed. The total GT$_-$ strength has been calculated as 9.3 units,
while Ref. \cite{Rapaport83} quotes $9.9 \pm 2.4$ units. In agreement
with experiment, the calculation places the main strength at energies
between 6 and 15 MeV, however, it misses some of the detailed
fragmentation in the data. For low excitation energies the
forward-angle cross sections have been recently determined using the
$(^3\text{He},t)$ charge-exchange reaction with a noticeably improved
resolution compared to the pioneering $(p,n)$ experiments
\cite{Fujita06,Rubio11}, again showing a rather good agreement with
the shell model calculation.

\begin{figure}
  \begin{center}
    \includegraphics[width=0.47\linewidth]{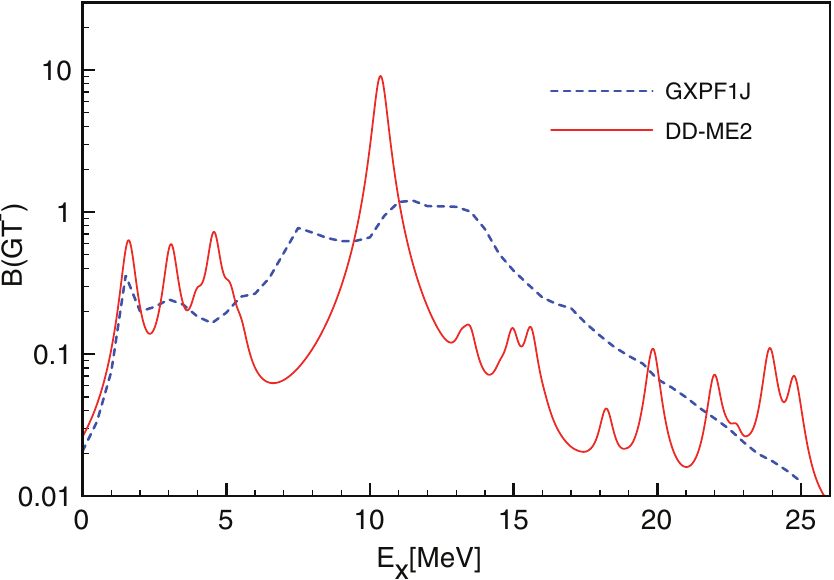}%
    \hspace{0.02\linewidth}%
    \includegraphics[width=0.49\linewidth]{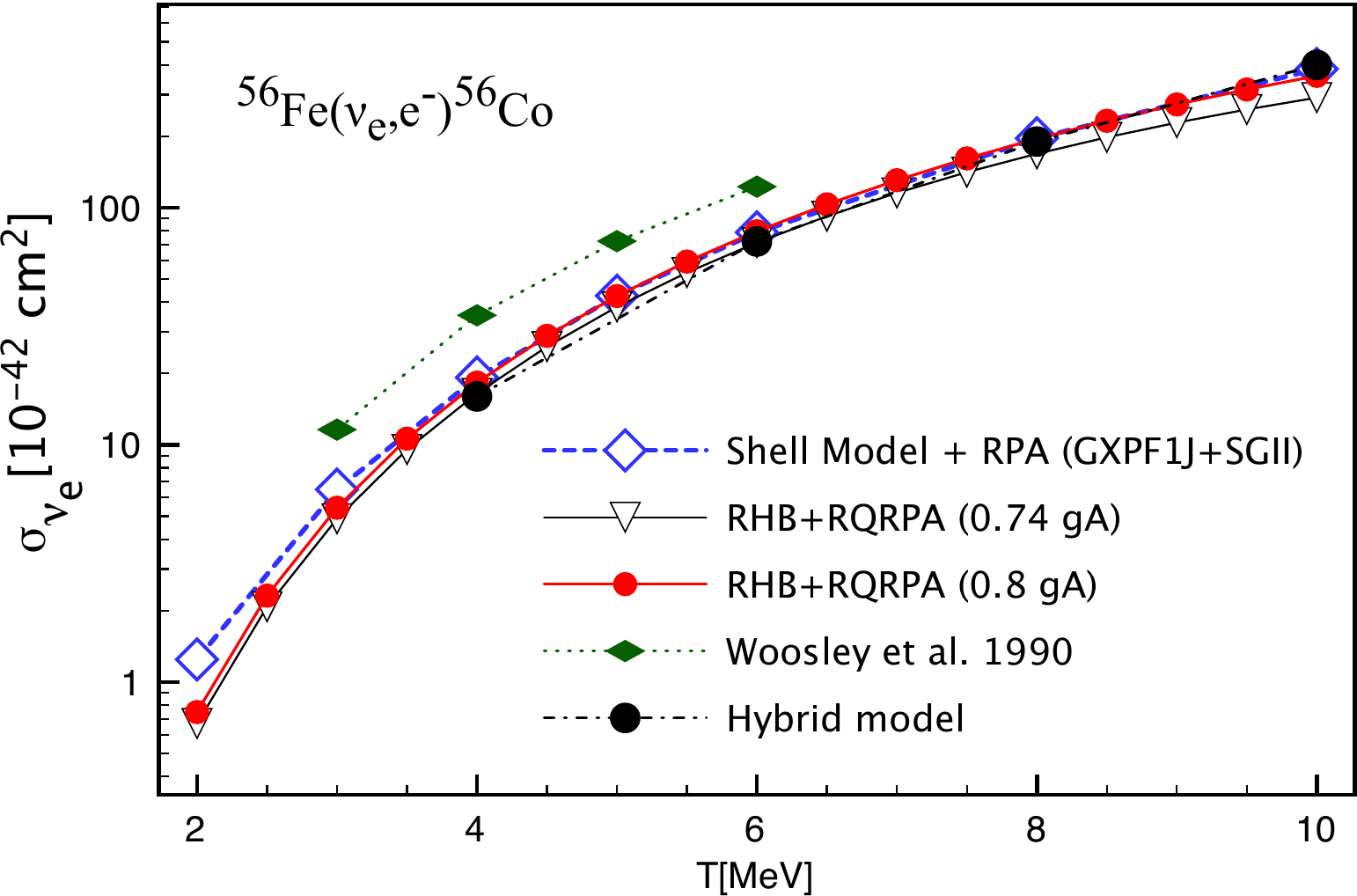}
    \caption{(left panel) GT$_-$ strength distribution for $^{56}$Fe,
      calculated within a relativistic QRPA approach (RQRPA with
      interaction DD-ME2) and a shell model study with the GXPF1J
      residual interaction. Both calculations have been folded with a
      Lorentzian of width 0.5 MeV.  (right panel) Comparison of
      $(\nu_e,e^-$) cross sections on $^{56}$Fe calculated within
      relativistic QRPA approaches with different quenchings of the
      axial-vector coupling constant (RHB+RQRPA) and within hybrid
      models where the GT$_-$ strength has been calculated within the
      shell model and the other multipole contributions within the
      RPA. The cross sections have been folded with a Fermi-Dirac
      distribution with temperature $T$
(corrected version from
      \cite{Paar11}). \label{fig:Fe56-neutrino}}
  \end{center}
\end{figure}

The left panel of Fig. \ref{fig:Fe56-neutrino} compares the GT$_-$
strength calculated within a relativistic QRPA approach with a shell
model calculation \cite{Paar11}. The shell model calculation uses a
different residual interaction than adopted in \cite{Caurier99}, but
yields very similar results. The total GT$_-$ strength with the GXPF1J
interaction is 9.5 units, but it is also well reproduced by the QRPA
study 9.7 units \cite{Paar11}. (Note that also the Ikeda sum rule is
quenched.)  The QRPA strength distribution has some strength at
excitation energies below 5 MeV, similar to the shell model, and it
places the largest portion of the strength in a resonance centred
around 10 MeV.  As expected, the shell model and experimental
distributions show a significantly stronger fragmentation of this
resonant strength which is not recovered in the QRPA study due to lack
of correlations.

As we have stressed above the detailed distribution of the GT$_-$
strength matters in the description of neutrino-nucleus reactions if
the incident neutrino energy is of the order of the nuclear excitation
energy (considering the $Q$ value of the reaction).  This is nicely
demonstrated in Fig. \ref{fig:Fe56-neutrino}.  For the distributions
dominated by low energy neutrinos (at $T=2$ and 3 MeV) the QRPA
predicts slightly smaller cross sections than the hybrid model.  But for
the neutrino distributions with temperatures $T> 4$ MeV the cross
sections are very similar indicating that the (average) neutrino
energies are large enough that the cross section is only sensitive to
the total strength and to the energy centroid of the distribution
(which are quite similar in the QRPA and shell model), but not to the
details of the distribution. Also the hybrid model calculation of
Ref. \cite{Kolbe01a}, using a different shell model residual
interaction and another RPA variant, yields cross sections which are
very close to the one obtained in Ref. \cite{Paar11}.  We note that
the relative importance of correlations for the description of the
total GT$_-$ strength decreases with growing neutron excess as for
nuclei with modest and large neutron excess this strength is dominated
by the (quenched) $3(N-Z)$ part stemming from the Ikeda sum rule.

We expect that the results found for $^{56}$Fe applies quite
generally: i.e. for charged-current $(\nu_e,e^-)$ reactions, RPA and hybrid
model calculations will give reliable estimates for supernova
neutrinos provided they give a reasonable account of the total GT$_-$
strength and of the energy centroid.

\begin{figure}
  \begin{center}
    \includegraphics[width=0.5\linewidth]{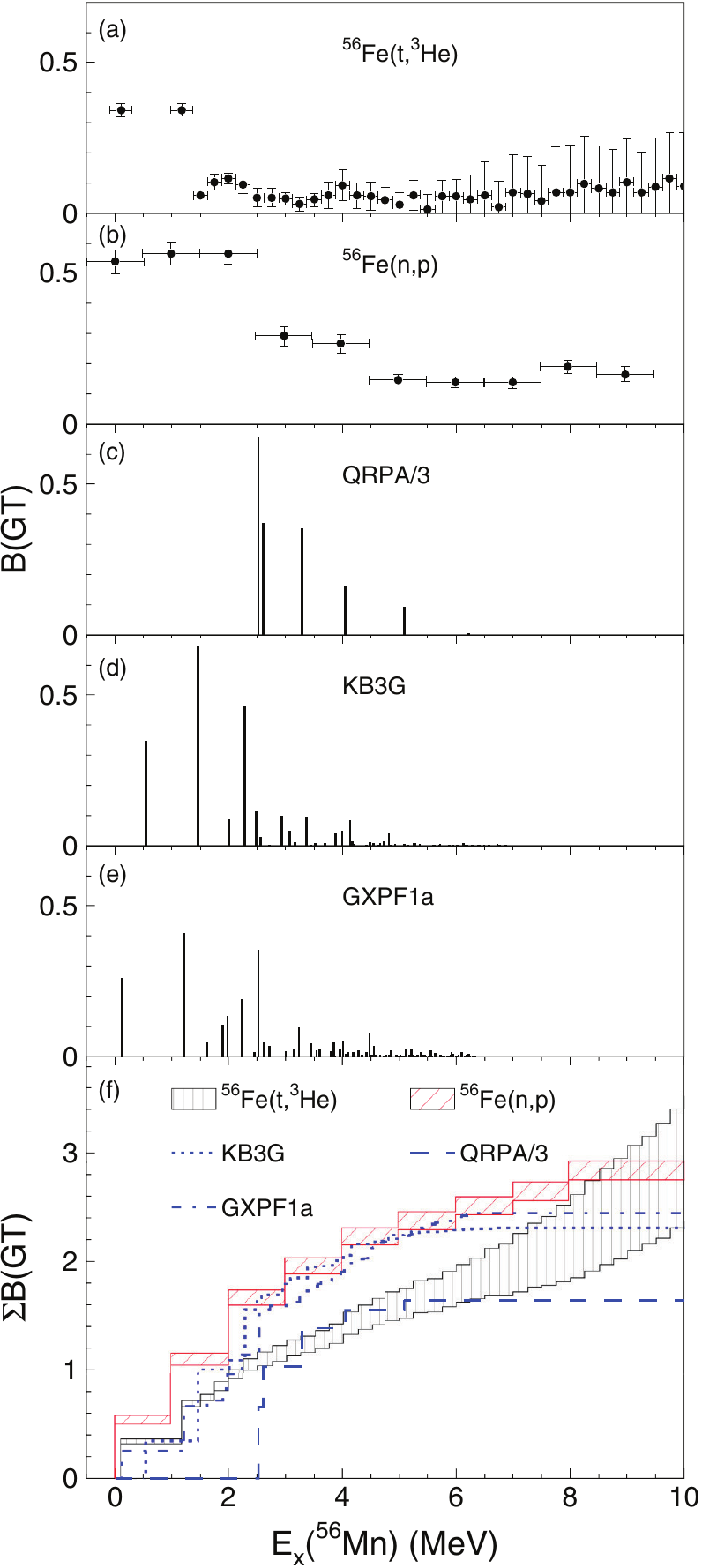}
    \caption{ Comparison of GT$_+$ strength for $^{56}$Fe deduced from
      $(n,p)$~\cite{Roennquist93,El-Kateb94} and
      $(t,^3\text{He})$~\cite{Scott14} charge-exchange experiments
      with distributions calculated within the QRPA approach (c) and
      by the shell model (d and e) with different residual
      interactions (KB3G and GXPF1a).  Panel f) shows the running sums
      of the strengths. The QRPA results have been divided by a factor
      of 3 (from \cite{Scott14}).\label{fig:Fe56-GT+}}
  \end{center}
\end{figure}

As the GT$_+$ strength for nuclei with neutron excess is unconstrained
by the Ikeda sum rule, its reproduction is a larger challenge to
nuclear models than the GT$_-$ strength. Importantly the shell model
could be established as a reliable tool for such calculations in
recent years \cite{Caurier99,Caurier03}.  Using again $^{56}$Fe as an
example, Fig. \ref{fig:Fe56-GT+} compares the experimental GT$_+$
strength with various calculations performed within the shell model
and the QRPA. At first we notice the significantly improved resolution
obtained in the $(t,^3\text{He})$ charge-exchange experiment
\cite{Scott14} (a similar resolution is achieved in the
$(d,^2\text{He})$ experiments) compared to the pioneering work using
the $(n,p)$ reaction with an energy resolution of about 1 MeV
\cite{Roennquist93,El-Kateb94}. Secondly, the shell model gives a fair
account of the total GT$_+$ strength as well as of the distribution,
although both residual interactions place slightly too much strength
into the energy interval between 2 and 4 MeV. The shell model
calculations, which agree quite well with each other, however, differ
from the QRPA predictions. The latter yields noticeably more total
strength than the data (about 5.5 units and $2.6 \pm 0.8$ units,
respectively).  Furthermore the lack of correlations in the QRPA has
the consequence that not sufficient strength is pulled to excitation
energies below 2 MeV, in contrast to data and the shell model. Such
shortcomings can lead to significant differences in the
$(\bar\nu_e,e^+)$ cross sections for low anti-neutrino energies.

\begin{figure}
  \begin{center}
    \includegraphics[width=0.9\linewidth]{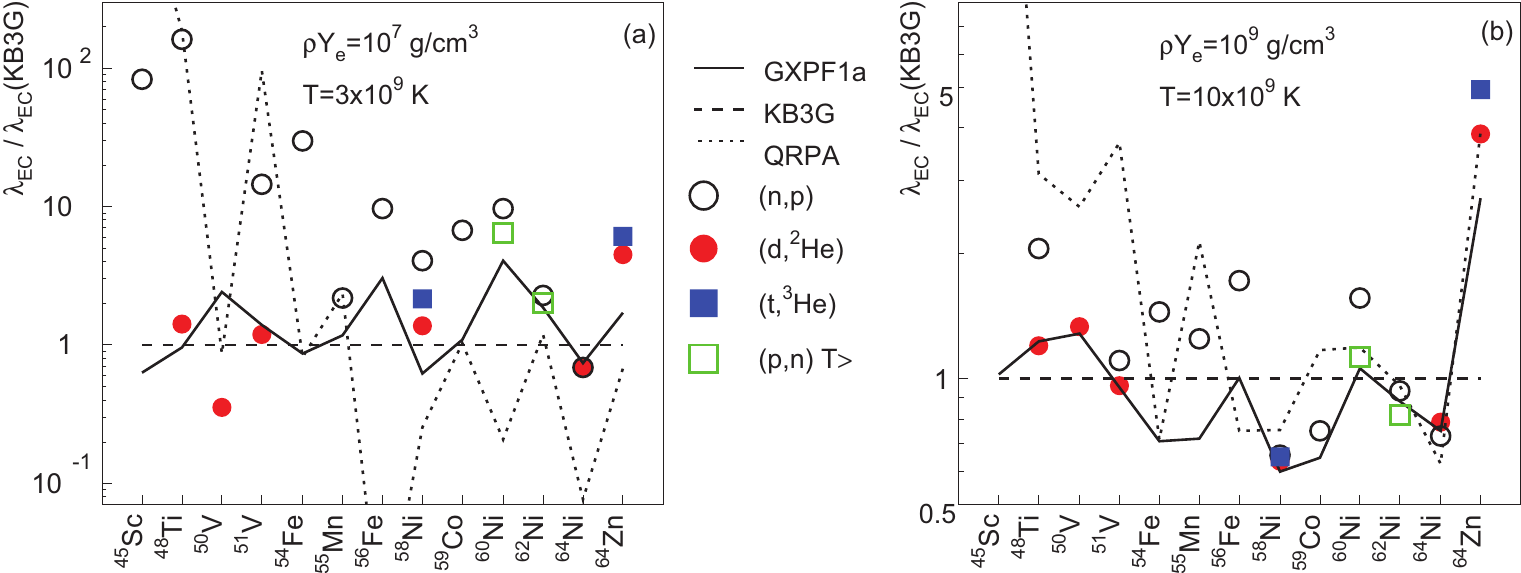}
    \caption{Comparison of electron-capture rates for 13 $pf$-shell
      nuclei calculated on the basis of theoretical (connected by
      lines) and experimental (indicated by markers) GT strength
      distributions. All rates are plotted relative to those
      calculated by the shell-model with the KB3G interaction. The
      rates are shown for two combinations of stellar density and
      temperature: (a) $\rho Y_e = 10^7$~g~cm$^{-3}$ and $T = 3
      \times 10^9$~K and (b) $\rho Y_e = 10^9$~g~cm$^{-3}$ and $T = 10
      \times 10^9$~K (from \cite{Cole12}).\label{fig:cole-comparison}}
  \end{center}
\end{figure}

Satisfyingly the shell model has been proven to describe all measured
GT$_+$ distributions for $pf$ shell nuclei quite well
\cite{Cole12}. This ability makes it the method of choice to describe
astrophysically important weak-interaction processes like electron
capture or $(\bar\nu_e,e^+)$ reactions if the astrophysical conditions
make the respective cross sections sensitive to a detailed description
of the GT$_+$ distribution.  As an example,
Fig. \ref{fig:cole-comparison} compares electron captures rates
calculated from the measured GT$_+$ distributions for all $pf$ shell
nuclei, for which these data exist, and from QRPA and shell model
strength distributions (for two different residual interactions). The
astrophysical conditions in the right panel correspond to the early
onset of collapse where the electron Fermi energy is of the same order
of magnitude as the reaction $Q$ value making the capture rate
sensitive to details of the GT$_+$ distribution. Fig.
\ref{fig:cole-comparison} shows a very satisfying agreement between
the rates obtained from data and shell model distributions. The
capture rates obtained from the QRPA calculations agrees usually also
within a factor of 2.

The sensitivity of the rates to the details of GT$_+$ distributions
decreases with increasing Fermi energy (or density).  Fig.
\ref{fig:cole-comparison} confirms this fact by showing a noticeably
larger variation between the different calculations and the data for
supernova conditions at smaller density. In particular two points are
noteworthy: i) the better energy resolution achieved in the
$(d,^2\text{He})$ and $(t,^3\text{He})$ reactions usually leads to
smaller rates than those obtained from the $(n,p)$ data, in good
agreement with the shell model; ii) the missing fragmentation of the
QRPA strengths can lead to large differences in either directions. We
stress, however, that the $pf$ shell nuclei shown in the figure have
small abundances at the relatively low densities and hence do not
contribute much to the overall electron capture rate.

\begin{figure}
  \begin{center}
    \includegraphics[width=0.48\linewidth]{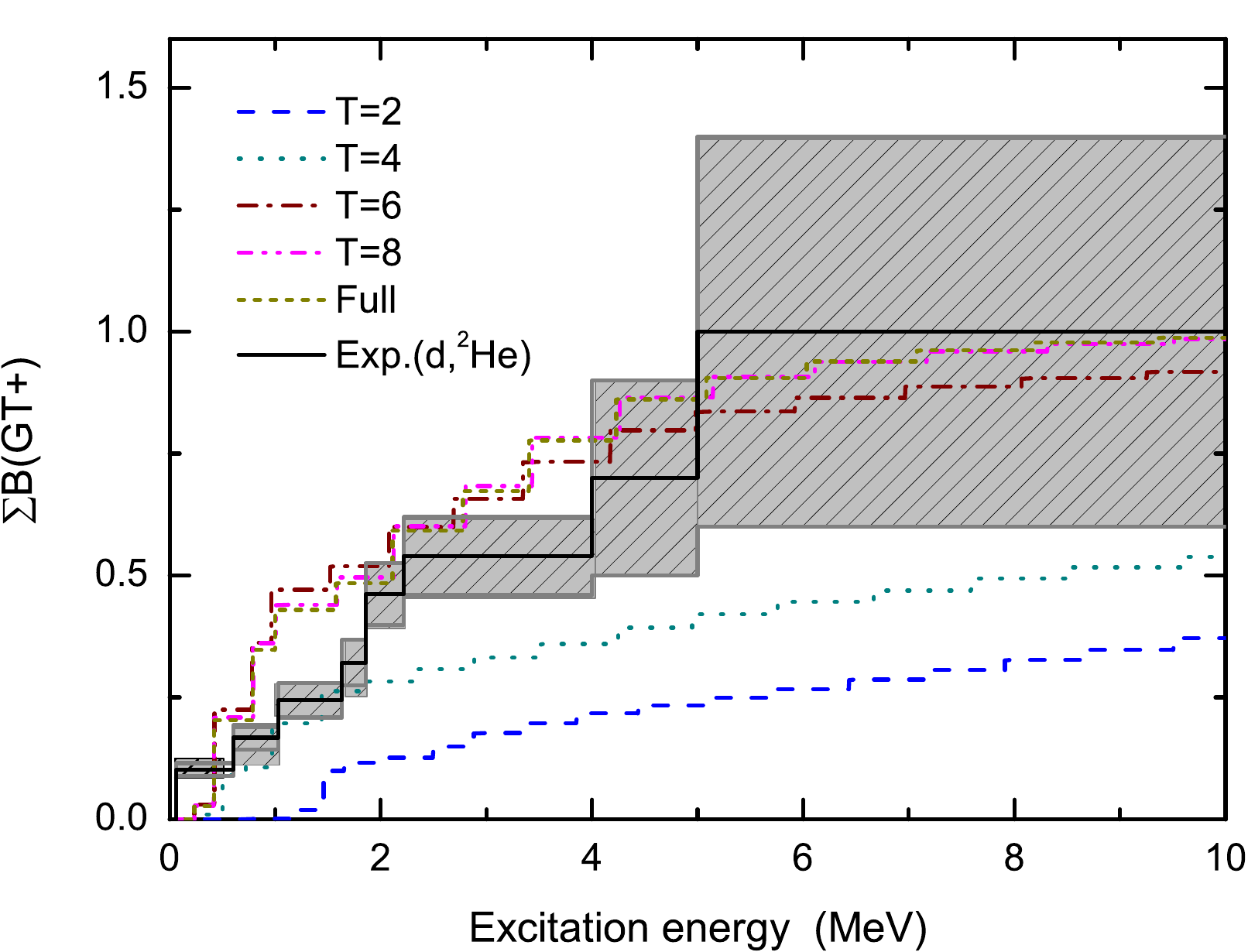}%
    \includegraphics[width=0.48\linewidth]{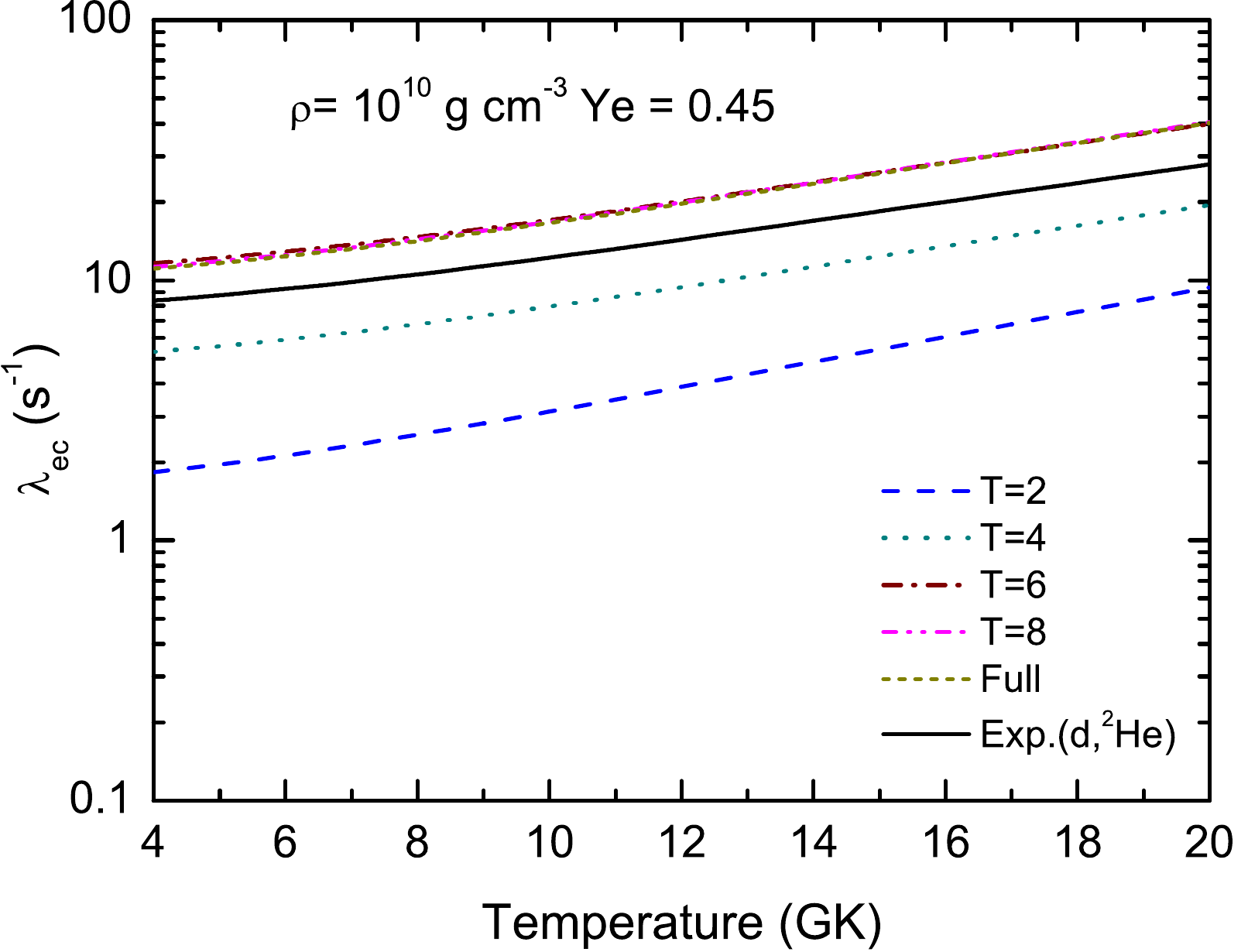}
    \caption{(left panel) Running GT$_+$ sum for different values of
      the particle-hole excitations across the $N=40$ shell
      gap. (right panel) Electron capture rates calculated with
      normalized GT strength distributions at different truncation
      levels  (from \cite{Zhi11}).   \label{fig:se76-trunc}} 
  \end{center}
\end{figure}

A similar study as performed for electron capture in
Ref. \cite{Cole12} would be also desirable for $(\bar\nu_e,e^+$) cross
sections for supernova neutrinos.  As energies of supernova
$\bar\nu_e$ neutrinos are larger than the electron Fermi energies for
both collapse situations depicted in Fig.  \ref{fig:cole-comparison}.
we expect from the general considerations above that the respective
$(\bar\nu_e,e^+$) cross sections are less sensitive to the details of
the GT$_+$ distribution than shown at the conditions of Fig.
\ref{fig:cole-comparison} and that the shell model, and also the QRPA,
should give a fair account of the GT contribution to the cross
sections, supplemented by higher multipole contributions obtained from
RPA calculations. This statement is likely correct for the chosen $pf$
shell nuclei, but it can unfortunately not be generalized. For these
$pf$ shell nuclei the neutron number is smaller than 40 and hence
GT$_+$ transitions, in which a proton is changed into a neutron, is
even possible within the IPM. However, for nuclei with proton numbers
below a major oscillator shell closure (say $Z<20$ or $Z<40$) and
neutron numbers above ($N>20$ or $N>40$, respectively) GT$_+$
transitions are Pauli blocked in the IPM. However, they can open up by
correlations which move nucleons across the shell closure. Thus such
correlations can enable GT$_+$ transitions for protons in the upper
oscillator shell or into neutron holes in the lower oscillator shell.
Detailed calculations have shown that describing cross-shell
correlations is a slowly converging process requiring many-particle
many-hole configurations usually beyond those considered in QRPA
studies~\cite{Caurier01,Lisetzki}.  To illustrate this point we use
the electron capture rate on $^{76}$Se with $N=42$ and $Z=34$ as an
example (see Section 2).  the left panel of Fig.~\ref{fig:se76-trunc}
shows the shell model results for the $^{76}$Se GT$_+$ strength. Note
that the shell model space is adequate to describe the correlations of
valence nucleons across the shell gap, it lacks, however, some
spin-orbit partners making the use of the universal quenching factor
inappropriate. In Ref. \cite{Zhi11} the quenching has been simply
chosen to match the total experimental GT$_+$ strength
\cite{Grewe08}. This choice does not affect our argument here which is
the evolution of the strength with truncation level; i.e. with the
number of nucleons allowed to be promoted from the $pf$ shell across
the shell gap by correlations. In particular, truncation level $t=2$
includes up to two-particle two-hole correlations like in QRPA
calculations.  We observe from Fig.~\ref{fig:se76-trunc} that, with
increasing correlations, i) the total GT$_+$ strength grows, mainly
due to an increase of neutron holes in the $pf$ shell, and ii) more
strength is shifted to lower excitation energies.  This has an
important effect on the capture rates. This is illustrated in the
right panel of Fig.~\ref{fig:se76-trunc} at conditions with electron
Fermi energies of nearly 10 MeV, where the rates differ by up to an
order of magnitude between the calculation with truncation level $t=2$
and the converged calculation. We mention that in the IPM the GT$_+$
strength and the rates vanish. As the average energies of supernova
$\bar\nu_e$ neutrinos are slightly larger than the electron Fermi
energy in our example, we expect that the differences in
$(\bar\nu_e,e^+$) cross sections for supernova neutrinos will be still
appreciable, but somewhat smaller.

The only neutral-current cross section measured for nuclei beyond the
deuteron is the transition from the ground state to the $T=1$ state at
15.11 MeV in $^{12}$C \cite{Zeitnitz94,Auerbach01}. This is the
Isobaric Analog State of the $^{12}$B and $^{12}$N ground states. Thus
from a nuclear structure point of view this measurement yields the
same informations as the exclusive charged-current reactions on
$^{12}$C discussed above. Hence we have to use indirect experimental
information to validate cross section calculationsi for inelastic
neutrino-nucleus scattering. Due to the relatively low energies of
supernova neutrinos allowed transitions dominate inelastic
$(\nu,\nu'$) cross sections and the relevant nuclear quantity is the
GT$_0$ strength distribution.  This strength can be experimentally
studied for example in $(p,p')$ experiments.  However it is also
intimitely related to M1 excitations of nuclei which have been studied
with great precision by inelastic electron
scattering~\cite{Richter:2000}. We will now use such electron
scattering data to show that the shell model is capable to describe
the allowed response needed to determine inelastic cross sections for
supernova neutrinos.

$M1$ transitions are mediated by the operator
\begin{equation}
  \label{eq:bm1}
  \bm{O}(M1) = \sqrt{\frac{3}{4\pi}} \sum_k [ g_l(k) \bm{l}(k) + g_s(k)
  \bm{s}(k)] \mu_N
\end{equation}
where the sum runs over all nucleons.  The orbital and spin
gyromagnetic factors are given by $g_l=1$, $g_s= 5.585694675(57)$ for
protons and $g_l=0$, $g_s=-3.82608545(90)$ for neutrons \cite{Mohr00};
$\mu_N$ is the nuclear magneton. Using isospin quantum numbers
$\bm{t}_z=\bm{\tau}_3/2=\pm 1/2$ for protons and neutrons,
respectively, Eq.~(\ref{eq:bm1}) can be rewritten in isovector and
isoscalar parts. Due to a strong cancellation of the $g$-factors in
the isoscalar part, the $M1$ transitions are of dominant isovector
nature.  The respective isovector $M1$ operator is given by
\begin{equation}
  \bm{O}(M1)_{\mathrm{iv}} = \sqrt{\frac{3}{4\pi}} \sum_k [
  \bm{l}(k)\bm{t}_z(k) + (g_s^p-g_s^n)\bm{s}(k)\bm{t}_z(k) ] \mu_N
\end{equation}
We note that the spin part of the isovector $M1$ operator is the zero
component of the $GT$ operator,
\begin{equation}
  \bm{O}(GT_0) = \sum_k 2 \bm{s}(k)\bm{t}_z(k) = \sum_k \bm{\sigma}
  (k)\bm{t}_z(k),
\end{equation}
however, enhanced by the factor $\sqrt{3/4\pi}(g_s^p -
g_s^n)/2=2.2993$.  The allowed contribution to the inelastic
neutrino-nucleus scattering at low energies, where finite momentum
transfer corrections can be neglected, from an initial nuclear state
($i$) to a final state ($f$) is then given by
\begin{equation}
  \sigma (E_\nu, i\rightarrow f) = \frac{G_F^2 g_A^2}{\pi (2J_i+1)}
  (E_\nu-\omega)^2 
  | \langle f ||\sum_k \bm{\sigma} (k) \bm{t}_z (k) ||i \rangle |^2
\end{equation}
where $E_\nu$ is the energy of the scattered neutrino and $\omega$ is
the difference between final and initial nuclear energies. The nuclear
dependence is contained in the $B(GT_0) = g_A^2 | \langle f ||\sum_k
\bm{\sigma} (k) \bm{t}_z (k) ||i \rangle |^2/(2 J_i+1)$ reduced transition
probability between the initial and final nuclear states. In shell
model calculations of the M1 strength the spin part is quenched by the
same factor used for studies of the GT
strengths~\cite{VonNeumann.Poves.ea:1998}.

Thus, experimental $M1$ data yield the desired $GT_0$ information,
required to determine inelastic neutrino scattering on nuclei at
supernova energies, to the extent that the isoscalar and orbital
pieces present in the $M1$ operator can be neglected.  On general
grounds one expects, as stated above, that the isovector component
dominates over the isoscalar piece. Furthermore, it is wellknown that
the major strength of the orbital and spin $M1$ responses are
energetically well separated in the nucleus. In $pf$ shell nuclei,
which are of interest for supernova neutrino-nucleus scattering, the
orbital strength is located at excitation energies $E^\star \simeq
2-4$ MeV \cite{Guhr90}, while the spin $M1$ strength is concentrated
between 7 and 11 MeV.  A separation of spin and orbital pieces is
further facilitated by the fact that the orbital part is strongly
related to nuclear deformation \cite{Enders99}. For example, the
famous scissors mode \cite{Bohle84}, which is the collective orbital
$M1$ excitation, has been detected in well-deformed nuclei like
$^{56}$Fe~\cite{Fearick03}. Thus one can expect that in spherical
nuclei the orbital $M1$ response is not only energetically well
separated from the spin part, but also strongly suppressed.

\begin{figure}
  \begin{center}
    \includegraphics[width=0.5\linewidth]{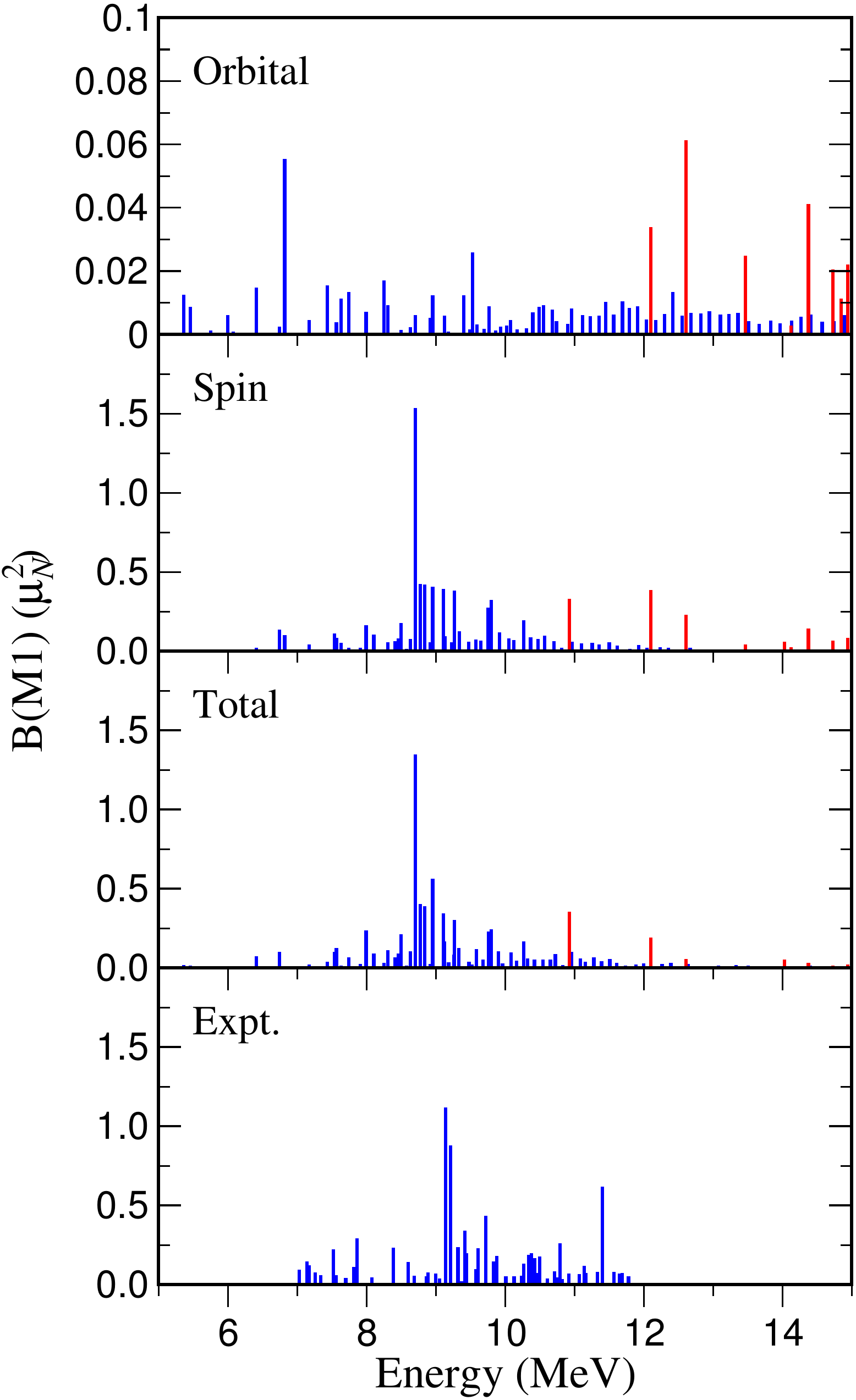}
    \caption{Comparison of experimental $^{52}$Cr $M1$ strength
      distribution (bottom panel) with shell-model result (upper
      panels) showing the orbital, spin components and the total. Note
      the different scales of the ordinate for the spin and orbital
      pieces, respectively. The shell-model calculations show isospin
      $T=2$ states as blue spikes and $T=3$ as red spikes (adapted
      from \cite{Langanke04}). \label{fig:cr52-m1}}
  \end{center}
\end{figure}

\begin{figure}
  \begin{center}
    \includegraphics[width=0.6\linewidth]{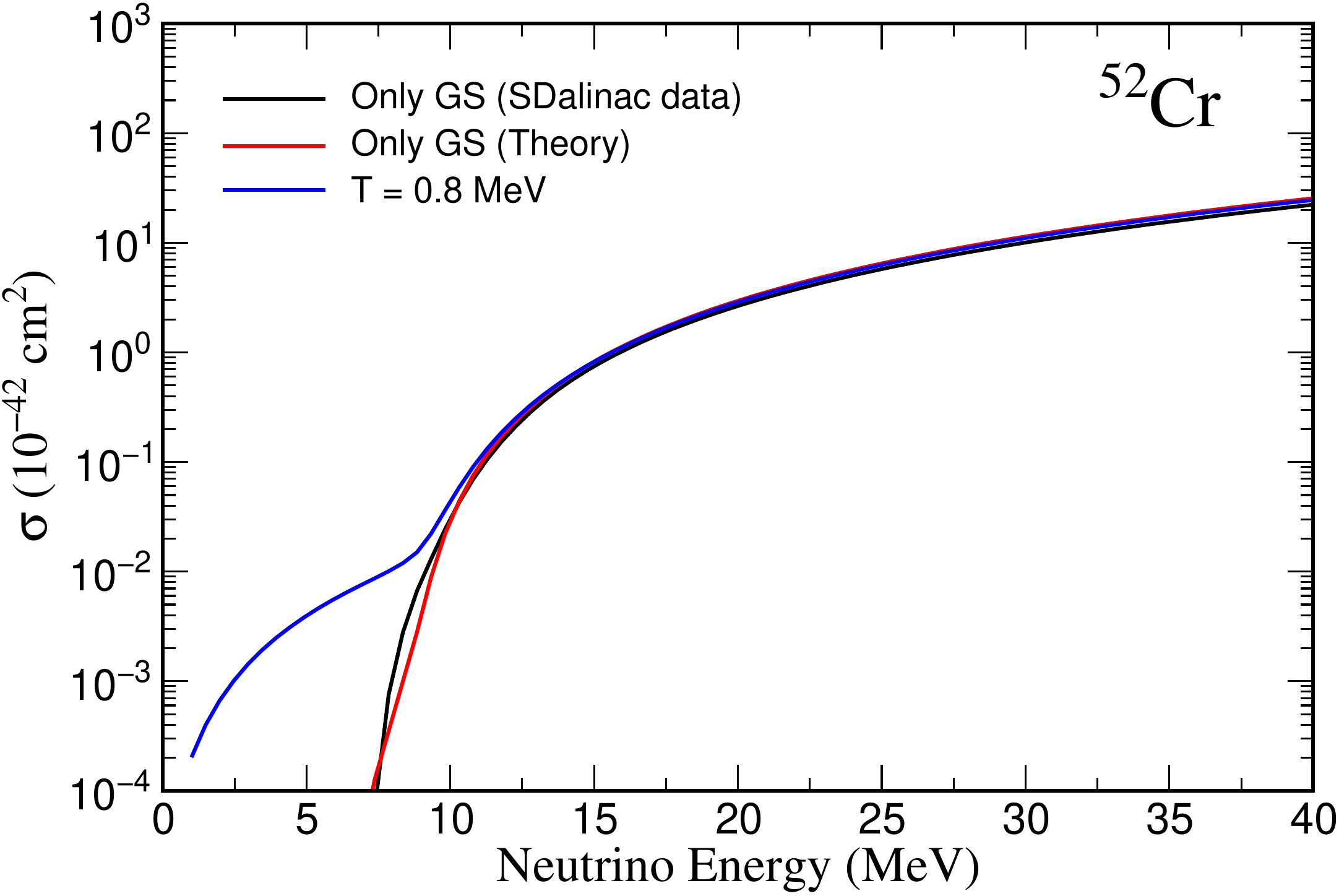}
    \caption{Neutrino-nucleus cross sections on $^{52}$Cr calculated
      from the $M1$ data (black line) and the shell-model GT$_0$
      distribution (red line).  The blue line shows the cross section
      for a temperature of 0.8~MeV  (adapted from \cite{Langanke04}).
    \label{fig:cr52-cross}}
  \end{center}
\end{figure}

Examples of spherical $pf$ shell nuclei are $^{50}$Ti, $^{52}$Cr and
$^{54}$Fe. In Fig. \ref{fig:cr52-m1} we show the total M1 strength and
its spin and orbital contributions from a shell model calculation for
$^{52}$Cr and compare it to the experimental data
\cite{Langanke04}. We observe a very good agreement between data and
calculation for the M1 strength distribution, with the strength mainly
located at excitation energies between 7-10 MeV. We also find that the
calculated M1 strength is dominated by its spin contribution which
exceeds the orbital part by more than an order of magnitude. Finally,
the calculation yields no indication for a scissors mode or any
orbital strength at low excitation energies, as expected for spherical
nuclei. Thus, the calculation strongly supports the argument of
Ref. \cite{Langanke04} that, except for an overall normalization, the
M1 strength for spherical nuclei reflects in a good approximation the
GT$_0$ distribution. Based on this assumption, Ref. \cite{Langanke04}
has calculated the allowed contributions to the inelastic neutrino
scattering on $^{52}$Cr from both the M1 data (properly normalized to
represent the GT$_0$ strength) and the GT$_0$ strength calculated in
the shell model. Both cross sections are compared in
Fig. \ref{fig:cr52-cross} and show excellent agreement for scattering
from the ground state. For the cross section at finite temperature
contributions of thermally populated states have to be included. These
have been calculated from appropriate shell model GT$_0$ strength
distributions and added to the ground state cross section as described
above, eq.~(\ref{eq-sm-cross-section}).  Fig. \ref{fig:cr52-cross}
indicates that the contributions from excited states have noticeable
effects at low neutrino energies. This is particularly enhanced for
spherical nuclei like $^{52}$Cr where there is no significant GT$_0$
strength at low nuclear excitation energies. For neutrino energies
which are sufficient to excite the GT$_0$ resonance, related to the
spin part in the M1 strength, from the ground state the finite
temperature corrections become negligible.  For applications to
supernova neutrinos contributions from forbidden transitions have to
be added to the cross section. These can be calculated within RPA
approaches and become increasingly relevant with growing neutrino
energies.

The importance of a detailed reproduction can be deduced from a study
presented in Ref. \cite{Dzhioev13}. These authors performed a
comparison of inelastic neutrino-nucleus cross sections for $^{56}$Fe
calculated within the hybrid model (shell model for allowed
contributions and RPA for other multipoles, \cite{Kolbe01a} and with
studies in which all multipole contributions are determined within
different variants of the QRPA \cite{Chasioti07,Dapo12,Dzhioev13}. The
cross sections are summarized in Table \ref{tab:neutral} for different
incident neutrino energies.  We note some differences in the cross
section predictions between the QRPA calculations.  As mentioned in
Ref. \cite{Dapo12} these are related to the choices of single particle
energies and the residual interaction.  and should become less
important with increasing neutrino energy. However, the calculations
also treated the quenching of the various multipole contributions
differently: only quenching the GT$_0$ contribution (like in the
hybrid model) \cite{Dzhioev13} or quenching all multipole
contributions \cite{Dapo12}. As Ref. \cite{Dzhioev13} took special
care in reproducing the centroid and strength of the GT$_0$
distribution and of the dipole giant resonance by slight adjustment of
the residual interaction, this study allows a detailed comparison to
the hybrid model calculation \cite{Kolbe01a}.  We find indeed a very
good agreement between the cross sections except at the lowest
neutrino energy $E_\nu=10$ MeV. Here the cross section is sensitive to
a detailed description of the GT$_0$ strength: the shell model
describes the fragmentation of the strength while the QRPA
calculations predict it to be quite strongly concentrated around
excitation energies of $E_x=10$ MeV (see
Figs. \ref{fig:Fe-isotopes-GT0} and \ref{fig:Dzi-GT} below).

\begin{table}[h]
\caption
{Comparison of cross sections (in units of $10^{-42}$ cm$^2$)
for inelastic neutrino scattering on the ground state of $^{56}$Fe
calculated within the hybrid model \cite{Kolbe01a} and different
variants of the QRPA. The incident neutrino energy is $E_\nu$. Exponents
are given in parenthesis (from \cite{Dzhioev13}).
\label{tab:neutral}}
\begin{center}
  \begin{tabular}{cccccc}
    \hline\hline 
    $E_\nu$ [MeV] & hybrid model \cite{Kolbe01a} & QRPA \cite{Dzhioev13} &
    QRPA \cite{Dapo12} & QRPA \cite{Chasioti07} \\ \hline
    10 & 1.91(-1) & 1.69(-2) & 1.87(-1) & 1.01(+0) \\
    20 & 6.90(+0) & 5.64(+0) & 9.78(+0) & 5.79(+0) \\
    30 & 2.85(+1) & 2.41(+1) & 4.08(+1) & 1.87(+1) \\
    40 & 7.86(+1) & 6.65(+1) & 1.05(+2) & 5.51(+1) \\
    50 & 1.72(+2) & 1.49(+2) & 2.16(+2) & 1.43(+2) \\
    60 & 3.20(+2) & 2.87(+2) & 3.89(+2) & 3.09(+2) \\
    70 & 5.25(+2) & 4.83(+2) & 6.33(+2) & 5.63(+2) \\
    80 & 7.89(+2) & 7.36(+2) & 9.59(+2) & 8.82(+2) \\
    90 & 1.11(+3) & 1.03(+3) & 1.38(+3) & 1.22(+3) \\
    100& 1.49(+3) & 1.36(+3) & 1.92(+3) & 1.52(+3) \\ 
    \hline\hline
  \end{tabular}
\end{center}
\end{table}

\section{Applications to supernova dynamics}

As we have stressed before neutrino-induced processes play a crucial
role in the dynamics of core-collapse supernovae. In accordance with
the main theme of this review we will focus here on two types of
neutrino-nucleus reactions which both have recently for the first time
been incorporated in supernova simulations and their impact on the
dynamics has been explored. The first subsection deals with inelastic
neutrino-nucleus scattering. We have outlined in the last section how
progress in nuclear modelling, constrained by experimental data from
inelastic electron scattering and charge-exchange experiments, has
made possible to reliably estimate respective cross sections for
scattering on ground states, despite the absence of direct data, and
has shown ways to extend the calculations to finite temperature as is
appropriate for supernova simulations.  In the first subsection we
will discuss the results of the nuclear model calculations, focussing
on the effects introduced by finite temperature, and then turn to the
impact elastic neutrino-nucleus scattering has in supernova
simulations. In the second subsection we deal with nuclear
deexcitation by neutrino pairs; i.e. the transition of a nucleus from
a state at higher energy to a state at lower energy by neutrino pair
production.  Obviously this is only possible at finite temperature and
can therefore not be directly studied in the laboratory. However, from
a nuclear structure point of view this neutral current process is
dominated by the same allowed Gamow-Teller and forbidden response as
inelastic neutrino scattering, but in different kinematics.  We will
discuss two quite distinct approaches to derive the respective cross
sections. Both yield, however, similar results. Supernova simulations,
which include nuclear deexcitation by neutrino pair emission, indicate
that this process does not impact the dynamics of the collapse. The
process is, however, a source of $\nu_x$ and $\bar\nu_e$ neutrinos
during the collapse phase.

\subsection{Inelastic neutrino-nucleus scattering and its impact in
  supernova simulations}

Inelastic neutrino-nucleus scattering, $A_i + \nu \rightarrow A_f +
\nu'$, changes the nucleus $A$ from an initial state $A_i$ with energy
$E_i$ to a final state $A_f$ with energy $E_f$. In this process the
neutrino either gains energy if $E_f < E_i$ or it looses energy if
$E_f > E_i$.  The energy exchange between the nucleus and the neutrino
can have potentially interesting consequences for the supernova
dynamics.  Inelastic neutrino-nucleus scattering might i) transfer
energy to nuclei outside the shock front after bounce before arrival
of the shock \cite{Haxton88} enabling an easier revival of the shock,
ii) add to inelastic neutrino-electron scattering speeding up the
thermalization of neutrinos with matter after neutrino trapping during
the collapse phase, and iii) change the neutrino opacity which will in
turn modify the spectra of neutrinos released in the supernova and
subsequently observed by earhbound detectors.

Investigations of inelastic neutrino-nucleus scattering in SN
simulations has been pioneered in an exploratory study
\cite{Bruenn91}, approximating the matter composition by a
representative nucleus, $^{56}$Fe.  The reaction cross sections were
based on a nuclear model appropriate for temperature $T=0$, combining
a truncated shell model evaluation of the allowed Gamow-Teller
response to the cross section with estimates of forbidden components
derived from the Goldhaber-Teller model.  The study concluded that
inelastic neutrino-nucleus scattering rates can compete with those of
neutrino-electron scattering at moderate and high neutrino energies
($E_\nu > 25\,$MeV), while they are significantly smaller for low
$E_\nu$.  No significant effects of neutrino-nucleus interactions on
the stalled shock by preheating the accreted matter were found in
\cite{Bruenn91}.

Approximating the composition of supernova matter by the ground state
of the even-even nucleus $^{56}$Fe is too simple an assumption for the
calculation of inelastic neutrino-nucleus interaction rates for two
reasons. i) Gamow-Teller transitions, which as we recall dominate the
cross sections at the neutrino energies of interest, can only connect
the $0^+$ ground state of $^{56}$Fe to final states with spin and
parity $J=1^+$. As the lowest $1^+$ state in $^{56}$Fe is at an
excitation energy of $E_x = 3\,$MeV, there exists a threshold for
inelastic neutrino scattering on $^{56}$Fe (similarly on other
even-even nuclei) and the cross sections are rather small for low
neutrino energies.  Supernova matter consists of a mixture of many
nuclei with even and odd proton and neutron numbers.  Since odd-$A$
and odd-odd (i.e., with odd proton and neutron numbers) nuclei miss
the strong pairing gap that lowers the ground state in even-even
nuclei relative to excited states, and have usually $J_i \neq 0$,
Gamow-Teller transitions from the ground state to levels at rather low
excitation energies are possible, reducing the threshold for inelastic
neutrino scattering on the ground state and generally increasing the
cross sections at low $E_\nu$. ii) More importantly, however,
supernova matter has a non-zero temperature of order 1$\,$MeV or
higher, requiring the description of nuclei as a thermal
ensemble. This completely removes the energy threshold for inelastic
neutrino scattering, as we have shown in Section 3.

\begin{figure}
  \begin{center}
    \includegraphics[width=0.585\linewidth]{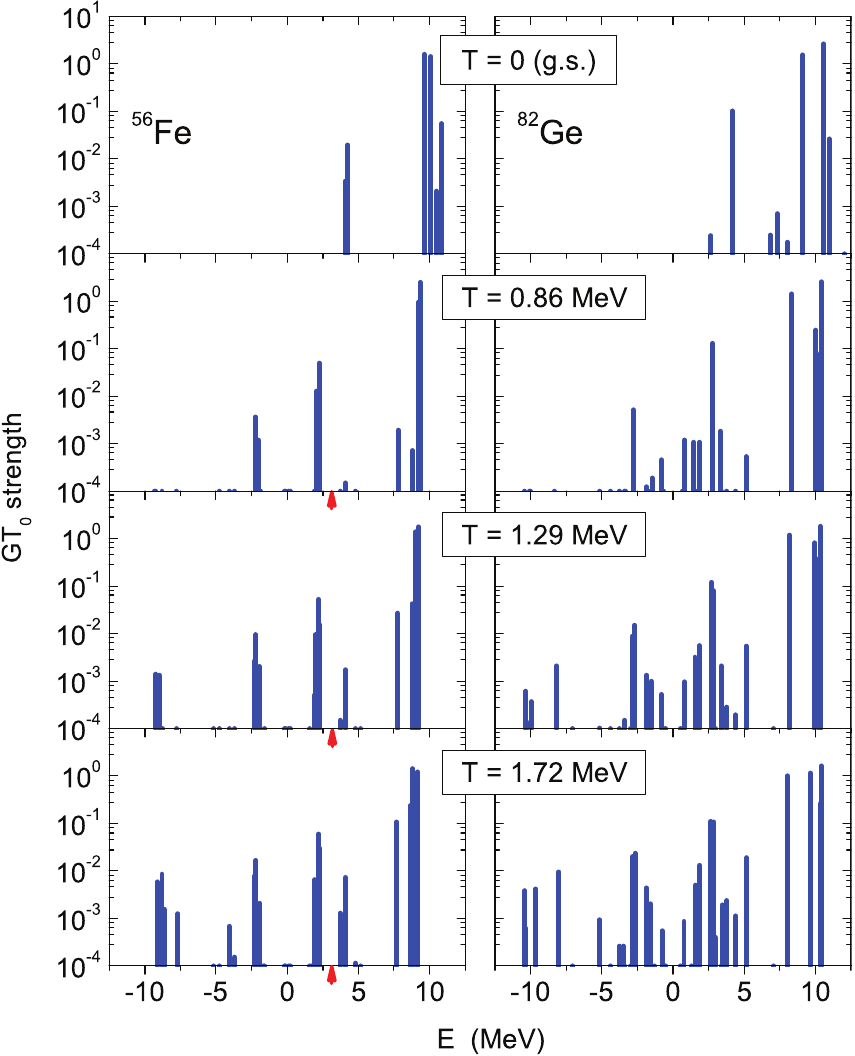}
    \caption{Temperature evolution of the GT$_0$ strength
      distributions for $^{56}$Fe (left panels) and $^{82}$Ge (right
      panels) calculated in the TQRPA approach. The $x$ axis denotes
      the neutrino energy transfer to the nucleus.  The red arrows
      indicate the threshold for inelastic neutrino scattering on
      $^{56}$Fe at temperature $T=0$ which is given by the excitation
      energy of the lowest $J=1^+$ state. Here the experimental value
      3.12 MeV is chosen, which is slightly lower than the calculated
      value  (from \cite{Dzhioev13}). \label{fig:Dzi-GT}}
  \end{center}
\end{figure}

The finite temperature effect on the inelastic neutrino cross section
off the two even-even nuclei $^{56}$Fe and $^{82}$Ge has been in
detailed explored recently by Dzhioev and collaborators
\cite{Dzhioev13} using the TQRPA approach.  As their general findings
for the two nuclei are similar, we concentrate on $^{56}$Fe here. The
temperature evolution of the GT$_0$ strength is exhibited in
Fig. \ref{fig:Dzi-GT}.  The ground state ($T=0$) distribution shows
that the strength is overwhelmingly concentrated at excitation
energies around 10~MeV, corresponding to a spin excitation.  The
sttrength at low energies around 4 MeV is significantly lower. We note
that these states are more pronouncedly seen in M1 data as they
predominantly reflect orbital excitations, including the scissors mode
at 3.45 MeV. At finite temperature the GT$_0$ distributions show
several effects \cite{Dzhioev13} which are relevant for the subsequent
discussion of neutrino-nucleus scattering. For positive neutrino
energy transfer ($E>0$ in Fig. \ref{fig:Dzi-GT}) the main effect is a
shift of the strength towards smaller energies, by about 1.5 MeV. This
shift is attributed to the pairing phase transition \cite{Dzhioev13}:
at temperatures above the critical one no extra energy has to be
invested into breaking Cooper pairs and as a consequence the GT$_0$
strength moves to lower energies. This behavior will not be present in
calculations which are based on Brink's hypothesis.  Furthermore, at
finite temperature strength appears now also at negative neutrino
energy transfers ($E<0$). This relates to thermally populated nuclear
states which decay to states at lower energies. Obviously the downward
transitions increase noticeably with temperature. The strength
observed at $E=-9$ MeV can be attributed to the deexcitation of the
GT$_0$ resonance. Dzhioev {\it et al.} point out that the assumption
of Brink's hypothesis also affects the downward strength
distribution. In their approach the shift of the upward strengths to
lower energies due to the melting of pairing makes it easier to
thermally populate these states which, in turn, leads to an enhanced
downward strength as well. Furthermore we remark that the smearing of
the Fermi surface at finite temperature allows for a larger
fragmentation of the strength, which affects more the low lying
transitions (and $^{82}$Ge).

\begin{figure}
  \begin{center}
    \includegraphics[width=0.7\linewidth]{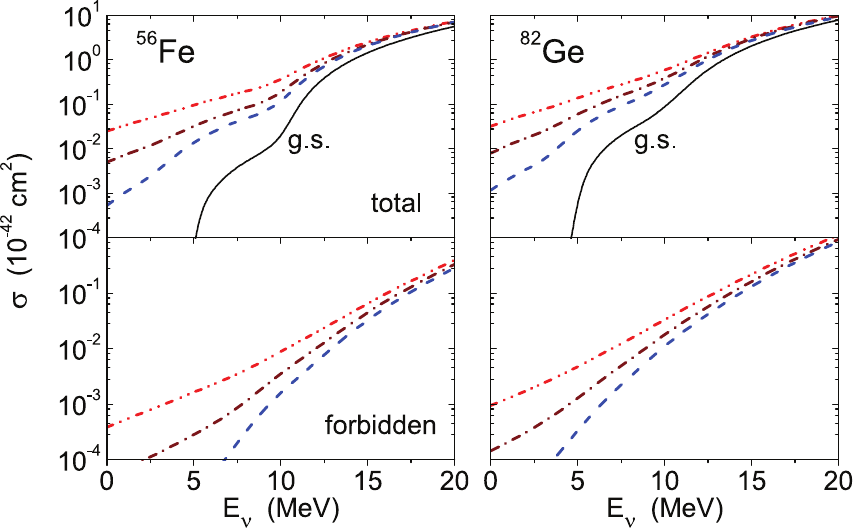}
    \caption{(upper panels): Cross sections for inelastic
      neutrino-nucleus scattering on $^{56}$Fe (left) and $^{82}$Ge
      (right) at different temperatures: $T=0$ (full line), $T=0.86$
      MeV (blue dashed line), $T=1.29$ MeV (brown dot-dashed line),
      $T=1.72$ MeV (red double-dot-dashed line).  (lower panels):
      Contribution of the forbidden multipoles to the cross sections
      at selected temperatures  (from
      \cite{Dzhioev13}). \label{fig:Dzi-cross}} 
  \end{center}
\end{figure}

The Gamow-Teller transitions and contributions from forbidden
multipoles up to $\lambda=3$ have been used to calculate inelastic
neutrino-nucleus cross sections for $^{56}$Fe and $^{82}$Ge at
different temperatures \cite{Dzhioev13}. The results are shown in
Fig. \ref{fig:Dzi-cross}. There are several interesting
observations. i) At $T=0$, i.e.  using the ground state transitions,
the cross section drops at low neutrino energies. This is related to
the gap in the GT distribution requiring neutrinos to have a certain
minimum energy (about 4 MeV in the TQRPA $^{56}$Fe calculation) to
excite the lowest $J=1^+$ state.  ii) There is no gap in the cross
sections at the finite temperatures mainly caused by the fact that
deexciation of thermally populated nuclear states contributes to the
cross section at all neutrino energies. Obviously the contributions of
the thermally populated states increase with temperature, so does the
scattering cross section at low energies.  iii) With increasing
neutrino energies the cross sections at different temperatures
converge. At neutrino energies $E<15$ MeV, excitation of the GT$_0$
resonance dominates the cross section. At the energies shown in
Fig. \ref{fig:Dzi-cross} the shift of the centroid energy due to the
melting of pairing leads to a small increase in cross section at the
finite temperatures compared to the $T=0$ case. Due to the energy
dependence of the phase space factor the relative enhancement of the
cross section caused by this energy shift of the centroid decreases
with increasing neutrino energy.  iv) At the neutrino energies shown,
which are typical for supernova neutrinos, the cross sections is
dominated by the GT contribution at all temperatures. Forbidden
transitions contribute less than $10\%$.

\begin{figure}
  \begin{center}
    \includegraphics[angle=270,width=0.8\linewidth]{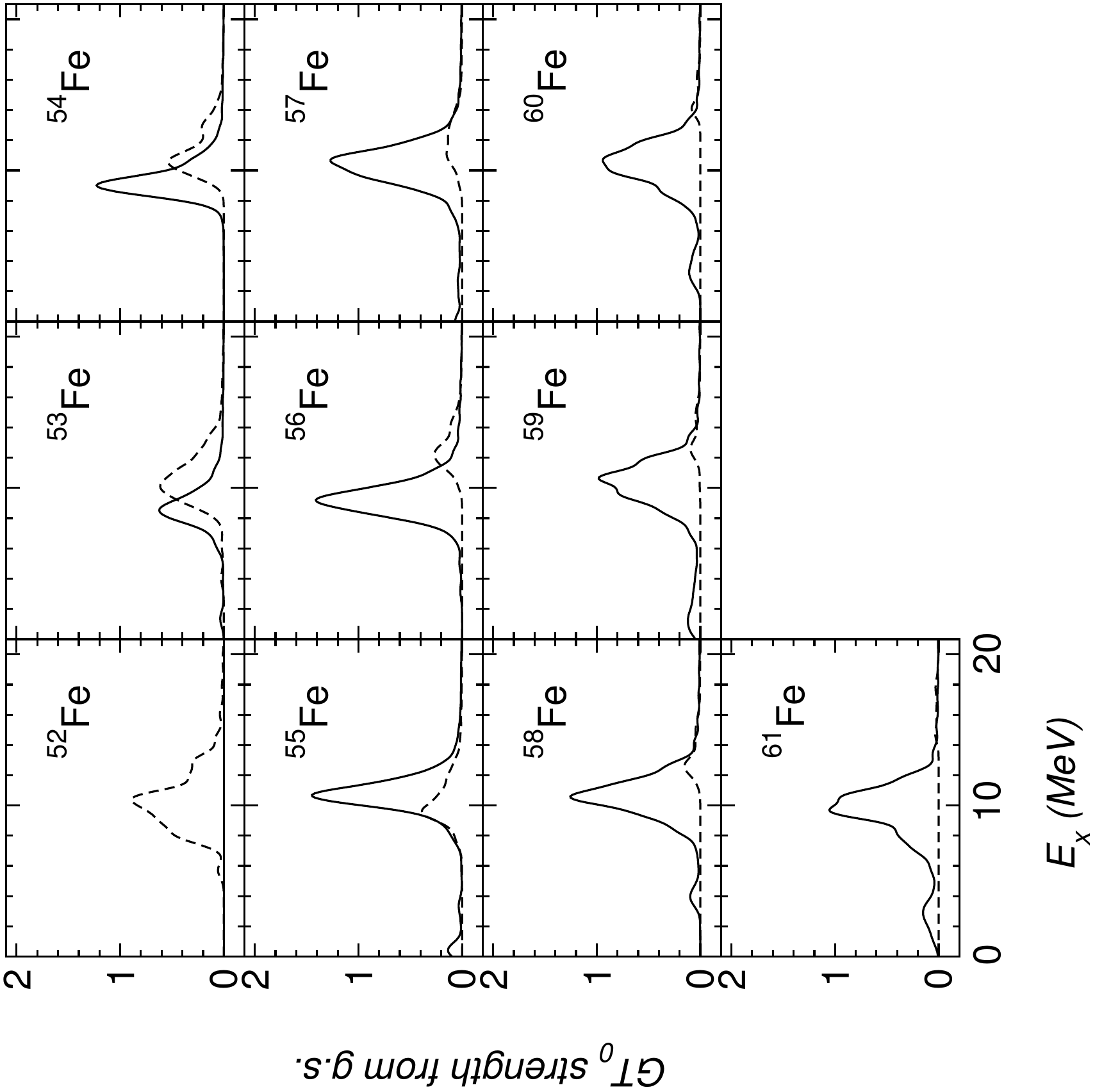}
    \caption{ Gaussian-smoothed neutral Gamow-Teller (GT$_0$)
      distributions from the ground state in Fe isotopes. Solid lines
      indicate $\Delta T=0$ strength, and dashed lines $\Delta T=1$.
      $E_x$ is the nuclear excitation energy (from
      \cite{Juodagalvis05}). \label{fig:Fe-isotopes-GT0}} 
  \end{center}
\end{figure}

We mentioned above briefly that the GT distributions and the inelastic
neutrino-nucleus cross sections at supernova neutrino energies are
quite similar for $^{56}$Fe and $^{82}$Ge.  This is actually more
globally true. Neutral-current $(\nu,\nu')$ cross sections, folded
over appropriate supernova neutrino energy distributions and
calculated for $T=0$, show relatively small variations from nucleus to
nucleus \cite{Woosley90,Kolbe01,Kolbe02}.  The reason is related to
the fact that the cross section at the supernova energies is dominated
by GT transitions and that the GT strength is strongly concentrated in
a resonance at excitation energies around 9-10 MeV for medium-mass and
heavy nuclei.  To support this point we show the GT$_0$ distributions
from shell model calculations \cite{Juodagalvis05} for the iron
isotopes $^{52-61}$Fe in Fig. \ref{fig:Fe-isotopes-GT0} (see also
\cite {Toivanen01}).  In fact for all isotopes the dominant part of
the strength resides in a resonance around 10 MeV. Also the total
resonant strength does not vary much. Very similar strength functions
are found for the mangan, cobalt and nickel isotopic chains
\cite{Juodagalvis05}.  It is worth mentioning that the GT$_0$
distributions for this set of Mn, Fe, Co and Ni isotopes does not
depend on the pairing structure of the ground state (even-even,
odd-$A$, odd-odd).

We recall that the hybrid model (introduced in Section 3) explicitly incorporates Brink's
hypothesis for the downscattering part of the cross section. Hence it
does not account for the energy shift in the GT$_0$ strength due to
the melting of pairing, in contrast to the TQRPA approach.  As a
consequence the TQRPA cross sections for $^{56}$Fe (the only nucleus
which has been studied in both approaches) are somewhat larger (by
about a factor of 2-3) at $E_\nu < 10$ MeV than those calculated in
the hybrid model \cite{Dzhioev13}.  For higher energies the cross
sections of the two models become nearly identical.

\begin{figure}
  \begin{center}
    \includegraphics[angle=270,width=0.8\linewidth]{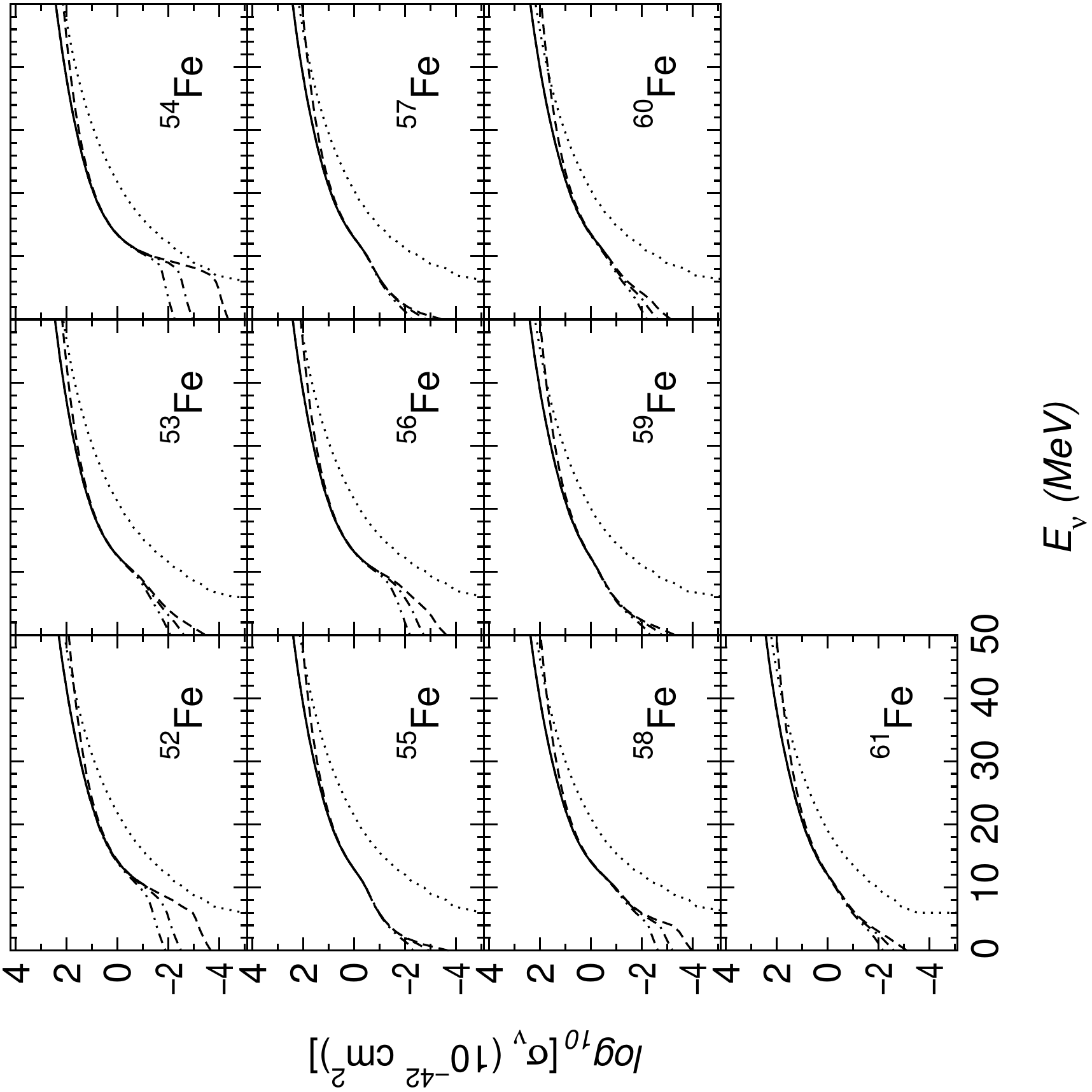}
    \caption{ Neutral current neutrino-nucleus inelastic
      cross-sections for Fe isotopes. The dashed lines show shell
      model contributions at three different temperatures ($T=0.86$
      MeV dashed line, 1.29 MeV dash-dotted line, and 1.72 MeV
      dash double-dotted line); the dotted line is the RPA
      contribution; the solid line corresponds to the total cross
      sections. At low neutrino energies the total cross section
      coincides with the SM contributions  (from
      \cite{Juodagalvis05}). \label{fig:Fe-isotopes-cross}} 
  \end{center}
\end{figure}

Juoadagalvis {\it et al.} have calculated inelastic neutrino
scattering cross sections on Mn, Fe, Co and Ni isotopes at finite
temperatures \cite{Juodagalvis05}. The study has been performed within the hybrid model
(taking the GT$_0$ distributions from the shell model and other
multipole contributions from RPA calculations).  The finite
temperature effects have been derived by inverting the shell model
GT$_0$ distributions for the lowest excited states. The cross sections
for the iron isotopes are shown in
Fig. \ref{fig:Fe-isotopes-cross}. We again observe the same trend in
the cross sections as discussed above for TQRPA calculations. At low
neutrino energies the cross sections is given by the contributions
arizing from deexcitation of thermally populated states and there are
small variations in the cross section between the different
isotopes. Once the neutrino energy is large enough to excite the
GT$_0$ resonance the cross sections for all isotopes become very
similar (which is due to the similarity of the GT$_0$ distributions)
and finite-temperature effects become rather small. At energies $E_\nu
<20$ MeV forbidden contributions are at least an order of magnitude
smaller than the GT contribution, become, however, increasingly more
important with energy.




\begin{figure}
  \begin{center}
    \includegraphics[width=0.48\linewidth]{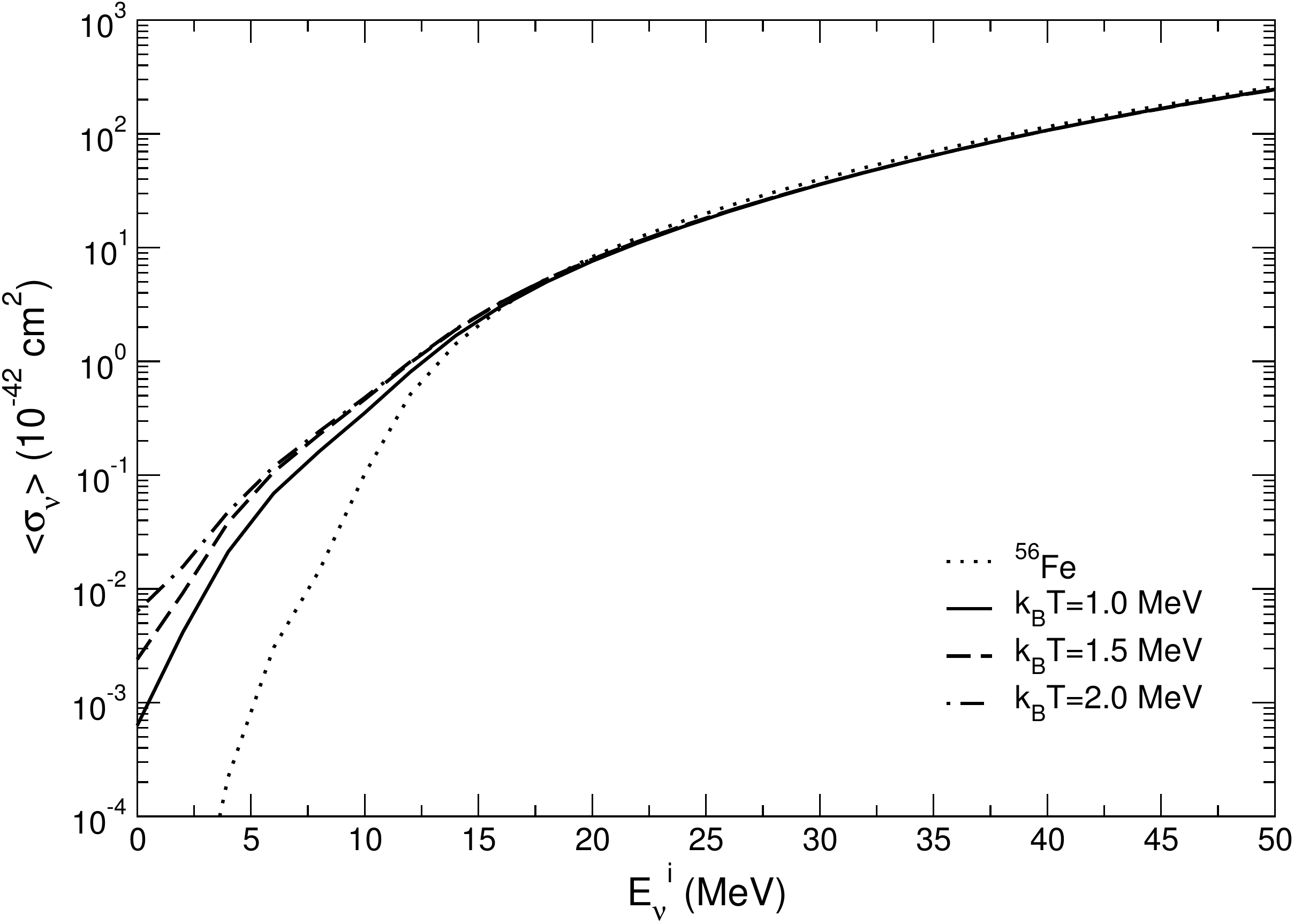}%
    \hspace{0.02\linewidth}%
    \includegraphics[width=0.48\linewidth]{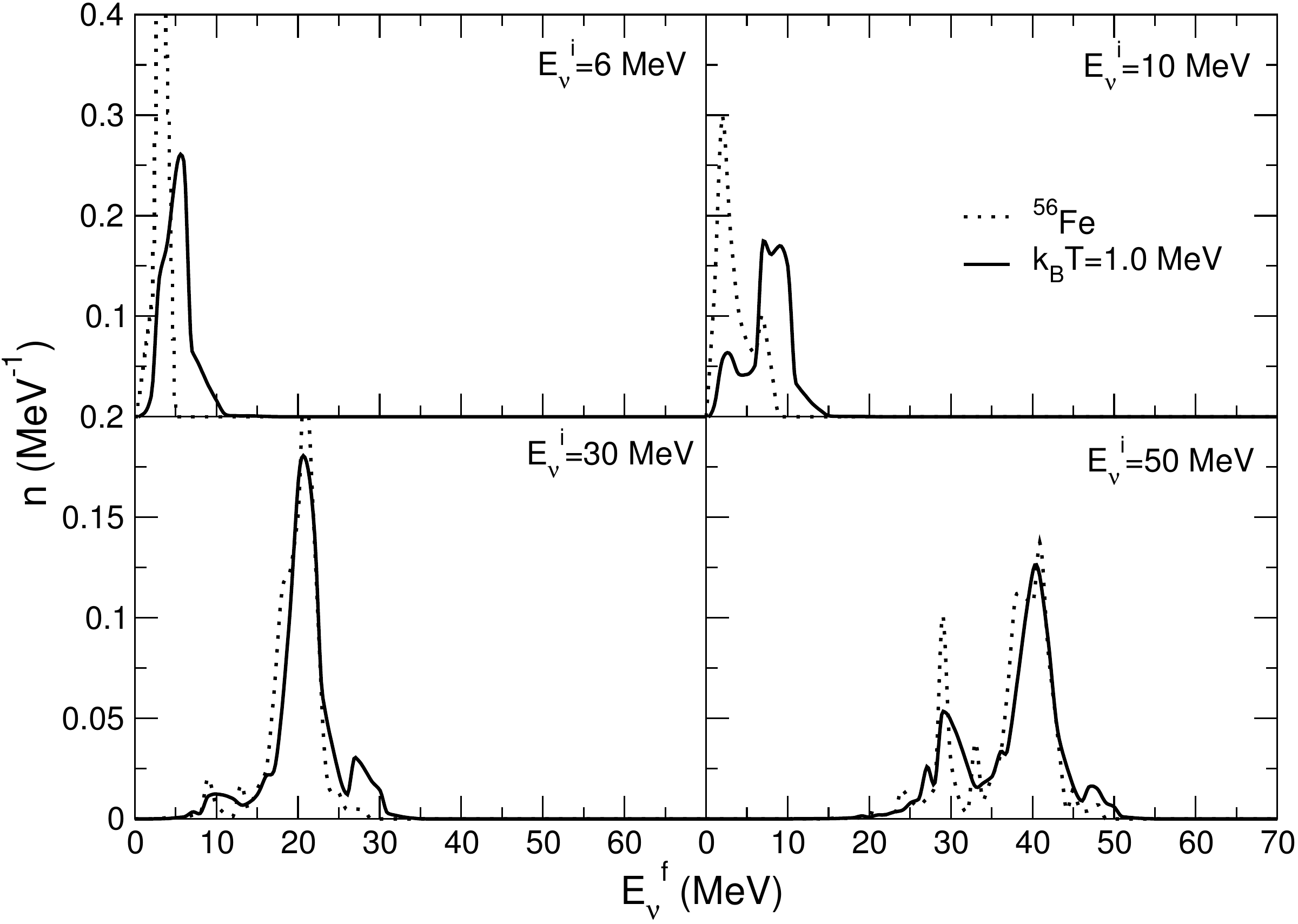}
    \caption{ (left panel) Inelastic neutrino scattering cross
      sections for supernova conditions (see text) at 3 different
      temperatures.  (right panels) Normalized neutrino spectra
      derived from inelastic neutrino scattering on a supernova matter
      composition for 4 different initial neutrino energies and at
      temperature $T=1$ MeV.  For a comparison the dotted lines in the
      various panels give the respective quantities for the individual
      nucleus $^{56}$Fe at $T=0$  (based on the simulations presented
      in \cite{Langanke08}). }
    \label{fig:sn-nunucleus}
  \end{center}
\end{figure}

The effect of inelastic neutrino-nucleus scattering on the supernova
dynamics has been in details explored in Ref.  \cite{Langanke08} by
simulating the collapse of a 15$\,M_\odot$ progenitor star (model
s15a28 of \cite{Woosley02}).  The simulation has been performed in
spherical symmetry with the neutrino-hydrodynamics code VERTEX of the
Garching group \cite{Rampp02,Buras06a} using a state-of-the-art
description of the interactions of neutrinos and antineutrinos of all
flavors as defined in \cite{Buras06a,Langanke03,Marek05}.

The inelastic neutrino-nucleus scattering cross sections were derived
on the basis of the hybrid model for a pool of about 50 nuclei
covering the $Z=24-28$ chains \cite{Juodagalvis05,Langanke08}.
Importantly, for temperatures $T > 1\,$MeV, these calculations do not
show large variations between the individual cross sections, for the
reasons explained above. This justifies to approximate the relevant
neutrino-nucleus cross sections by replacing the supernova matter
composition by the results drawn from the restricted pool of nuclei.
Hence, Ref. \cite{Langanke08} defined an average neutrino-nucleus
cross section as $\langle \sigma_{\nu n}\rangle = \sum_i Y_i \sigma_i/\sum_i
Y_i$, where the sum is restricted to the pool of nuclei for which
individual cross sections exist. The abundances of the individual
nuclei $Y_i$ were determined from an appropriate NSE matter
distribution \cite{Hix96}. A rate table of the cross section has been
derived as a function of temperature, density and electron-to-baryon
ratio $Y_e$ allowing for cross section determination at all
encountered astrophysical conditions during the simulation.  As an
example the pool-averaged total cross section is shown in Fig.
\ref{fig:sn-nunucleus} at 3 different temperatures. (The densities and
$Y_e$ values are different at the different temperatures. They
influence the average cross section by change of the NSE distribution
with very minor impact.)  The average cross section reflects the
general trends discussed at length above.

Keeping track of neutrino energies is a crucial ingredient in
supernova simulations. Inelastic neutrino scattering on nuclei changes
the energy of the incident neutrino. This change depends on the
incident energy and on temperature.  The right panel in Fig.
\ref{fig:sn-nunucleus} shows the normalized outgoing neutrino spectra
at different initial neutrino energies and for temperature $T=1$ MeV.
For $E_\nu =6$ and 10 MeV down-scattering contributes significantly to
the cross section; i.e., the de-excitation of thermally populated
nuclear levels produces neutrinos with $E_\nu^\prime >
10\,$MeV. Down-scattering becomes essentially irrelevant at higher
neutrino energies. For $E_\nu = 30\,$MeV the cross section is
dominated by the excitation of the GT$_0$ centroid, giving rise to a
cross section peak around $E_\nu^\prime = 20\,$MeV.  For even higher
neutrino energies forbidden transitions contribute noticeably to the
cross section. The peaks in the differential cross sections for $E_\nu
= 50\,$MeV correspond to excitations of the centroids of the GT$_0$
and dipole transition strengths.  With increasing population of
excited nuclear states the down-scattering contributions get
relatively enhanced with increasing temperature
\cite{Juodagalvis05,Dzhioev13}.

\begin{figure}
  \begin{center}
    \includegraphics[width=0.7\linewidth]{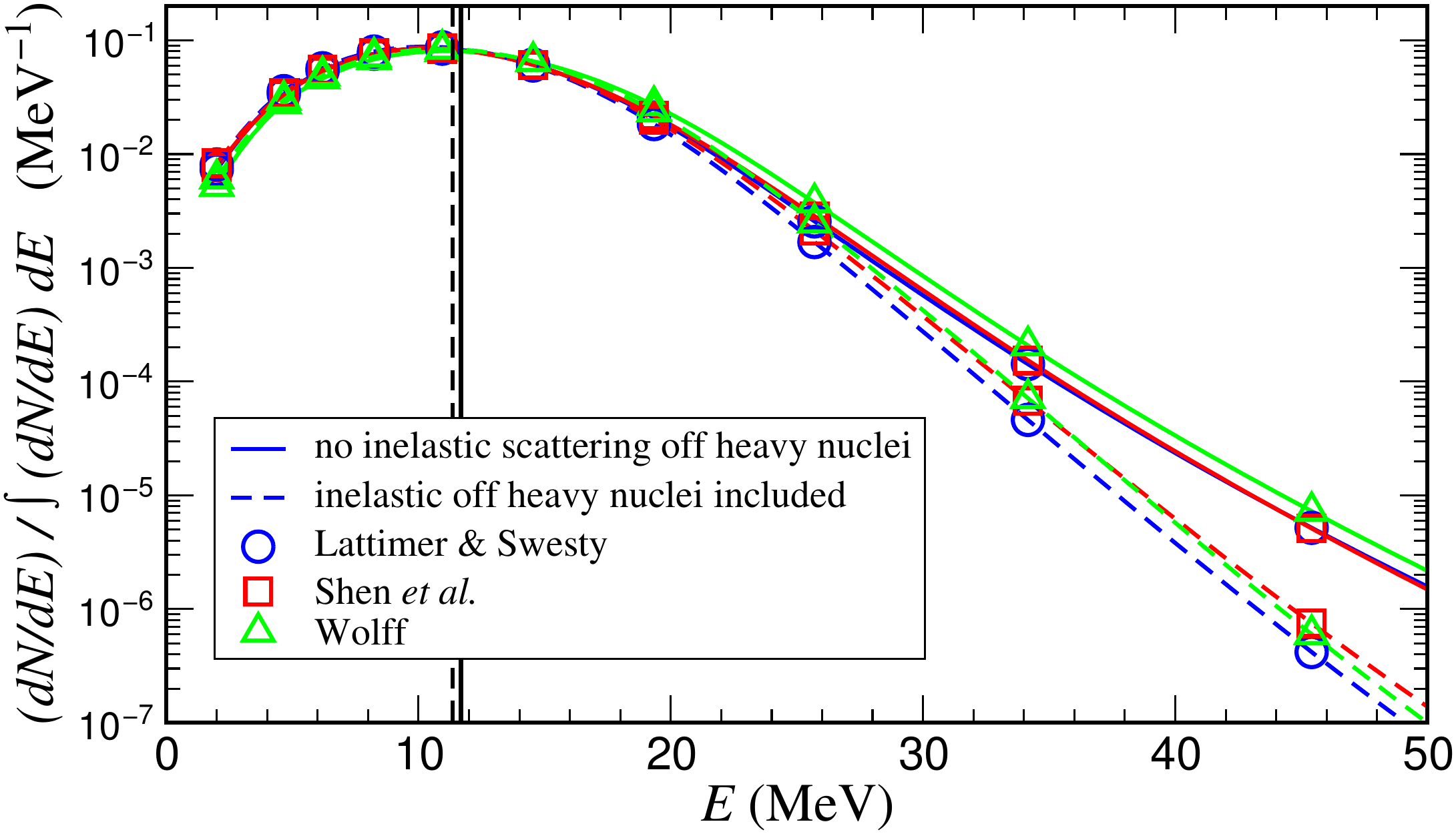}
    \caption{Normalized $\nu_{\mathrm{e}}$ number spectra radiated
      during the shock breakout burst as seen by a distant observer at
      rest. Results are shown for simulations with three different
      nuclear EoSs.  For better comparison of
      the strongly time-dependent spectra during this evolution phase,
      integration in a window of 8$\,$ms around the peak luminosity
      was performed.  Inelastic neutrino scattering off nuclei (dashed
      lines) leads mostly to energy losses of high-energy neutrinos
      and thus reduces the high-energy tails of the spectra. The
      vertical line marks the mean spectral energy (from
      \cite{Langanke08}). \label{fig:nuespectra}}
  \end{center}
\end{figure}

Despite causing a higher neutrino opacity by increasing the
neutrino-matter interaction, inelastic neutrino-nucleus scattering has
virtually no impact on the supernova evolution \cite{Langanke08}.  As
suggested by Haxton \cite{Haxton88} the scattering increases the
neutrino-matter coupling and thus the total energy transfer rate
mainly to matter (nuclei and electrons) ahead of the shock by a factor
2-3 during the first 50 ms after shock bounce. However, the time
period during which this additional heating acts on the supersonically
infalling matter is too short to have consequences on the supernova
simulation. The simulation of \cite{Langanke08} thus confirms the
expectation of Bruenn and Haxton \cite{Bruenn91} that neutrino
preheating of matter ahead of the shock does not add significantly to
the shock revival.

Inelastic neutrino-nucleus scattering has, however, a significant
influence on the spectra of neutrinos emitted after bounce, in
particularly during the period of the so-called $\nu_e$ neutrinoburst
(which, as we recall, is caused by fast electron captures on free
protons after photodissociation of nuclei by the shock, see Section
2).  During this period matter ahead of the shock still consists of
heavy nuclei in NSE. Furthermore the density is still high with a
relatively large optical depth for neutrinos.  Under these conditions
neutrinos interact frequently with nuclei by inelastic scattering. Due
to the energy dependence of the cross section the scattering occurs
more often for high-energy neutrinos, which will spend part of their
energy by mainly exciting the GT$_0$ or giant dipole resonances (see
Fig. \ref{fig:sn-nunucleus}).  In fact, inelastic neutrino-nucleus
scattering shifts the mean spectral energy of the $\nu_e$ burst
spectrum only slightly by about 0.5 MeV, but it strongly suppresses
the high-energy tail of the spectrum.  This is shown in
Fig.~\ref{fig:nuespectra} which compares the $\nu_e$ burst spectrum in
supernova simulations with and without consideration of inelastic
neutrino-nucleus scattering. A similar suppression of the high-energy
tail, caused by inelastic neutrino-nucleus scattering, is also
observed for $\bar{\nu_e}$ and $\nu_x$ neutrinos during this early
phase of the explosion. At later times inelastic neutrino scattering
on nuclei becomes negligible for the neutrino emission spectra as the
abundance of heavy nuclei has dropped drastically due to dissociation
of nuclei

\begin{table}
  \caption{Electron neutrino cross sections for scattering off
    electrons and for charged-current interactions with nuclei of
    different detector materials, averaged over the supernova neutrino spectrum
    of Fig.~\ref{fig:nuespectra}, using
    the results obtained with the Lattimer-Swesty EoS.
    The different cross sections were taken from:
    e~\cite{Hooft71},
    d~\cite{Nakamura02},
    $^{12}$C~\cite{Kolbe99},
    $^{16}$O~\cite{Kolbe02a},
    $^{40}$Ar~\cite{Kolbe03},
    $^{56}$Fe and $^{208}$Pb~\cite{Kolbe01} (from \cite{Langanke08}).}
  \label{tab:crosssections}
  \renewcommand{\arraystretch}{1.1}
  \begin{center}
    \begin{tabular}{lccc}
      \hline\hline
      Material & \multicolumn{2}{c}{$\langle \sigma \rangle$ (10$^{-42}$
        cm$^2$)} & Reduction \\ \cline{2-3}
      & With IS & Without IS &  \\ \hline
      e    & 0.114  & 0.123 & 7\% \\
      d    & 5.92  & 7.08 & 16\% \\
      $^{12}$C & 0.098 & 0.17 & 43\% \\
      $^{12}$C (N$_{\textrm{gs}}$) &     0.089 &   0.15 & 41\% \\
      $^{16}$O & 0.013 & 0.031 & 58\% \\
      $^{40}$Ar & 17.1 & 21.5  & 20\% \\
      $^{56}$Fe &  8.81 & 12.0  & 27\% \\
      $^{208}$Pb & 147.2 & 201.2 & 27\% \\ \hline\hline
    \end{tabular}
  \end{center}
\end{table}

The observation of the $\nu_e$ burst neutrinos is one of the aims
of supernova neutrino detectors \cite{Scholberg12,Thompson03,Kachelriess05}.
The change of the $\nu_e$ burst spectrum due to inelastic neutrino-nucleus
scattering will have consequences on the expected event rates for
such detectors. We note again that, due to the energy dependence of  
neutrino-nucleus cross sections, the weight of high-energy neutrinos
gets enhanced in the event rates.
In Table \ref{tab:crosssections} we compare the relevant detection
cross sections for several target nuclei 
calculated for the $\nu_e$ burst spectra obtained
with and without consideration of inelastic neutrino-nucleus scattering.
The largest reduction occurs for $^{12}$C and $^{16}$O which are detector
materials in Borexino, MiniBooNe, KamLAND, SNO+, and Super-Kamiokande.
For these nuclei only neutrinos with relatively high energies
($E>17$ MeV for $^{12}$C and $E>15$ MeV for $^{16}$O) can trigger
charged-current reactions. As the neutrino scattering cross section 
on electrons increases linearly with energy, the reduction of the event rate
for electrons, is less. Ref \cite{Langanke08} estimates the improved 
event rates (including the effect of inelastic neutrino-nucleus scattering)
arizing from the $\nu_e$ burst of a supernova going off at 10~kpc distance. 
 
In summary, supernova simulations imply that inelastic
neutrino-nucleus scattering has no impact on the supernova evolution,
but noticeably changes the $\nu_e$ burst spectrum. The latter effect
is mainly caused by scattering on nuclei induced by high energy
neutrinos, for which the respective cross sections are likely under
control and can be calculated with sufficient accuracy, also at finite
temperature.  Finite temperature corrections significantly affect the
cross sections for low-energy neutrinos. Here recent developments
achieved within the TQRPA approach \cite{Dzhioev13} are very promising
and show that the cross section estimates based on the
finite-temperature treatment in the hybrid model \cite{Juodagalvis05}
are probably somewhat too small.  This uncertainty in the low-energy
neutrino-nucleus cross section has, however, probably no impact on the
supernova evolution or the emitted neutrino spectra.

\subsection{Deexcitation of hot nuclei by neutrino-pair emission in supernova
environment}

During the final phase of the collapse the temperature in the inner
core can reach values above 1.5 MeV. Under such conditions the average
excitation energy of a nucleus with mass number $A \sim 120$, (a
representative value for the matter composition of heavy nuclei) is
about $\langle E_x \rangle = 35$ MeV. It has been argued that the
deexcitation of such highly excited states can be a significant
neutrino-pair producing process \cite{Fuller91} which can compete with
electron-positron annihilation or nucleon-nucleon bremsstrahlung. As
the presence of electron neutrinos does not block the production of
muon and tau neutrino-antineutrino pairs, the nuclear deexcitation
process might further reduce the entropy of the collapsing core if the
neutrinos produced by the process can leave the core during the
dynamical timescale of the collapse \cite{Fuller91}. Recently nuclear
deexcitation by neutrino-pair emission has, for the first time, been
incorporated into a supernova simulation \cite{Fischer14}. We will
here briefly summarize the results of this study.

The formalism how to derive the rate for nuclear dexxecitation,
$A_i \rightarrow A_f + \nu+ \bar{\nu}$, and its inverse reaction
neutrino-pair absorption, $A_i + \nu+\bar{\nu} \rightarrow A_f$,
and how to incorporate it into supernova simulations
is presented in Ref. \cite{Fischer14}. For our discussion here it
is sufficient to recall that both processes are mediated by
neutral current. In fact for neutrino-pair absorption the nuclear
input is given by the same allowed and forbidden strength functions
which determine inelastic neutrino-nucleus scattering.
Our discussions above have shown that the hybrid model and the TQRPA
calculations both indicate that these allowed and forbidden strengths
are both concentrated around the GT$_0$ and giant dipole resonances.
Furthermore the calculations (and the available M1 data) show that the position
and total strengths do only mildly vary between nuclei. 
The authors of Ref. \cite{Fischer14} used these findings to
approximate the strength for the absorption process by:
\begin{equation}
S^{\text{abs}}(\Delta) = S_A g_A(\Delta) + S_F g_F(\Delta)
\label{eq:deexcite-strength}
\end{equation}
where $\Delta = E_\nu + E_{\bar \nu}$ is the energy sum of the
neutrino-antineutrino pair. The normalized strength functions $g_A$
for allowed transitions and $g_F$ for forbidden transitions have been
chosen as Gaussians. The centroids of these Gaussians $\mu_A, \mu_F$
have been chosen guided by the shell model and TQRPA calculations for
allowed transitions ($\mu_A =9$~MeV,
\cite{Dzhioev13,Juodagalvis05,Toivanen01}) and by RPA calculations for
forbidden transitions ($\mu_F=22$ MeV, \cite{Kolbe01a,Dzhioev13}).
Assuming that the strengths for excited states is somewhat more
fragmented and wider spread in energy the standard deviations
$\sigma_A, \sigma_F$ have been adopted slightly larger than for the
ground state; i.e $\sigma_A = 5$ MeV, $\sigma_F = 7$ MeV. Finally the
total allowed strength, $S_A=5$ and the total forbidden strength,
$S_F=7$, has been adopted from shell model
\cite{Juodagalvis05,Toivanen01} and RPA calculations
\cite{Kolbe01a}. Finally we mention that the approximate ansatz for
the absorption strength function, eq.~(\ref{eq:deexcite-strength}),
explicitly assumes the Brink hypothesis which makes the strength
independent of the initial nuclear excitation energy and consequently
the absorption strength becomes independent of temperature.

The process of relevance for supernova simulations is
nuclear deexcitation by neutrino pair emission. As this is the inverse
of the neutrino-pair absorption process, the relevant strength
function  $S^{emi}$ can be obtained by detailed balance
\begin{equation}
S^{emi} (\Delta,T) = S^{abs} (\Delta) \times \exp \left( - \frac{\Delta}{kT}
\right).
\label{eq:deexcite-balance}
\end{equation}
While the absorption strength is independent of temperature by
construction, the emission strength function depends on temperature by
the Boltzmann factor $\exp{( - \Delta/kT}) = \exp
( - (E_\nu+ E_{\bar\nu})/kT)$.

The ansatz Eq. (\ref{eq:deexcite-strength})
is a rather simple approximation. For example, it assumes that it is
the same for all nuclei and  it neglects the dependence
of the widths on temperature. An alternative approach to
describe the nuclear deexcitation rate has been suggested by
Fuller and Meyer \cite{Fuller91}. These authors derived analytical
expressions for the allowed and forbidden contributions to the
emission and absorption strength functions
at finite temperature based on the Fermi gas model. The parameters 
in their expressions were adjusted to reproduce results obtained
in an independent single-particle shell model. We note that the 
strength functions proposed in Ref. \cite{Fuller91} depend on temperature
as well as on charge and mass number of the nucleus.

\begin{figure}
  \begin{center}
    \includegraphics[width=\linewidth]{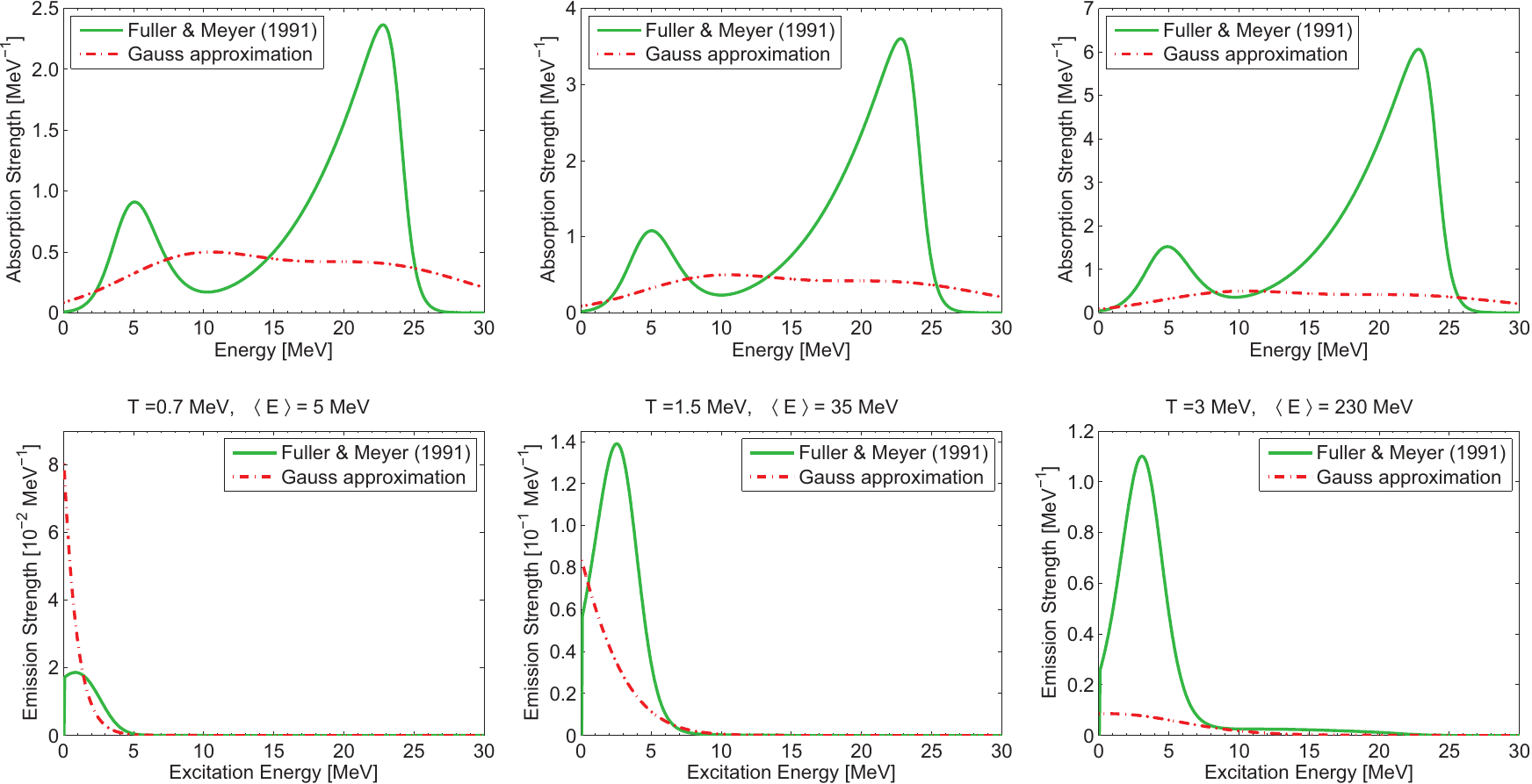}
    \caption{Strength for absorption (top panels) and emission (bottom
      panels) of neutrino pairs based on the Fermi-gas model of
      ref.~\cite{Fuller91} (green lines) and the Gaussian ansatz,
      eq. (\ref{eq:deexcite-strength}), (red lines) at temperatures of
      $T=0.7$~MeV (left panel), $T=1.5$~MeV (middle panel), and
      $T=3.0$~MeV (right panel). The average thermal energy at the
      different temperatures is given above the lower panels (from
      \cite{Fischer14}). \label{fig:deexcite-spectrum}}
  \end{center}
\end{figure}

The absorption and emission strengths of the two models are plotted in Fig.
\ref{fig:deexcite-spectrum}. 
The absorption strength functions
of Ref. \cite{Fuller91} show two distinct peaks for allowed
and forbidden transitions, while the Gaussian ansatz 
(Eq. \ref{eq:deexcite-strength}) distributes the strength over a wider
energy range. While the Gaussian ansatz is independent of charge
and mass number of the nuclei, the absorption strength of Ref. \cite{Fuller91}
is proportinal to the number of nucleons and increases with temperature
as the NSE composition gets shifted to heavier nuclei with growing temperature.
For example, the average mass number is $\langle A \rangle =124$
at $T=1.5$ MeV using the Lattimer-Swesty Equation of State \cite{Lattimer91}
and $\langle A \rangle =205$ at $T=3$ MeV.
In summary, the two absorption strengths  are quite distinct 
and can be understood as extreme cases allowing to explore how sensitive
supernova simulations are to an accurate description of the strength.
Nevertheless it might be desirable that the 
absorption strength function will be calculated within the TQRPA approach
for several nuclei and at different temperatures. 
 
The emission strength function is obtained by multiplying the absorption
strength with the appropriate Boltzmann factor, see Eq. 
\ref{eq:deexcite-balance}. The differences between the two absorption
strength functions lead to strong deviations in the emission strength
functions. We note that there is also emission strength at energies
above the thermal average energy due to the thermal population of states
at higher energies. In the emission decay rate the high-energy
tails of the emission strength function get enhanced by an $E^5$ factor
in the phase space factor, while this factor relatively decreases the contributions
from lower energies. 
 
Recently the downwards strength function has been calculated for fixed
exciation energies for several nuclei
within the diagonalization shell model \cite{Misch13}. To translate the
downward strength into a thermal emission strength, the excitation
energy has been adequated with the average thermal energy of a Fermi
gas model. Hence by construction this approach does not allow 
for the production of neutrino pairs with a sum energy larger than
the thermal average energy. Due to the phase space factor such
high-energy pairs do noticeably contribute to the deexcitation rate.

\begin{figure}
\begin{center}
  \includegraphics[width=0.7\linewidth]{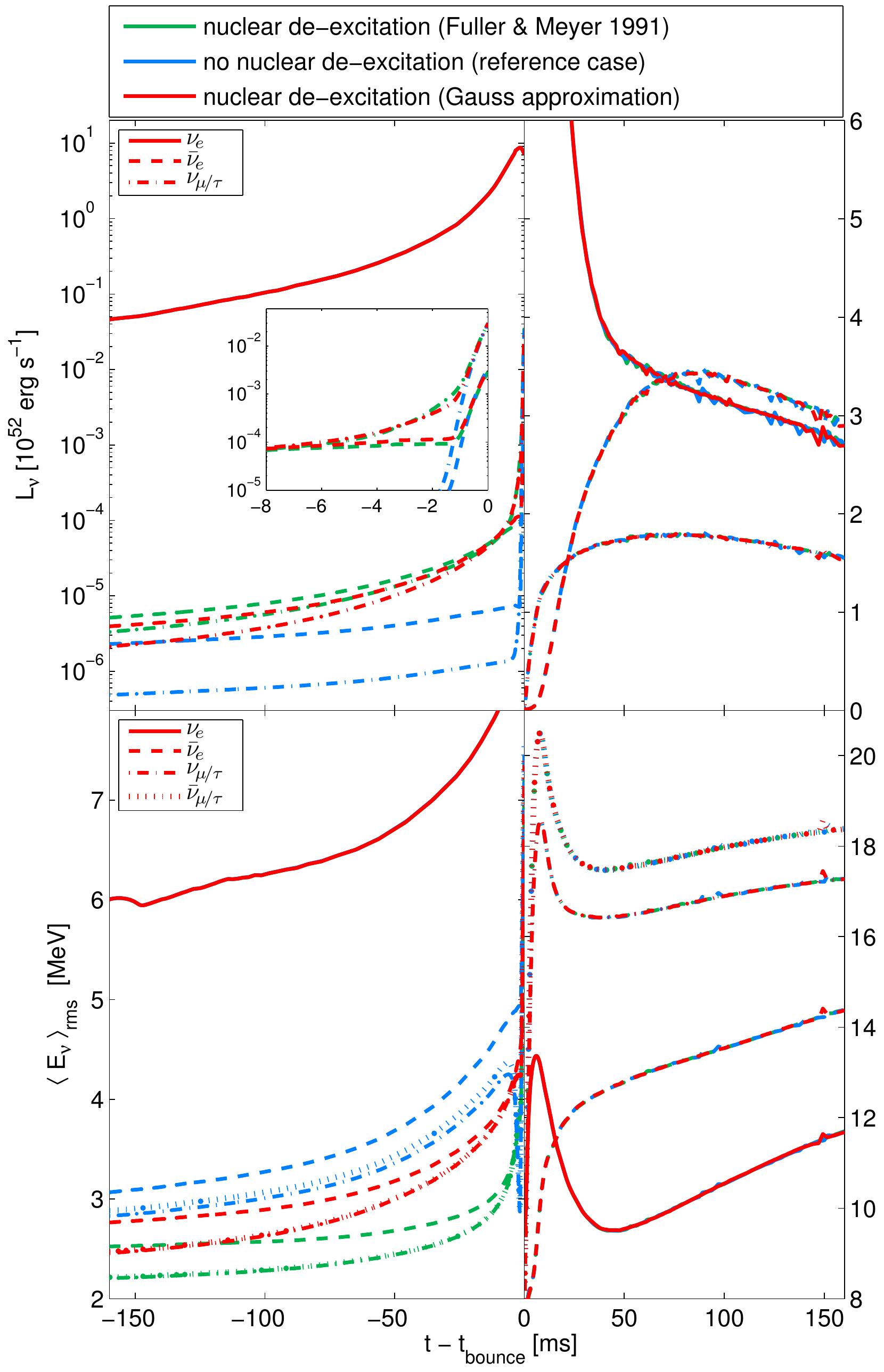}
  \caption{Evolution of neutrino luminosities and average energies
    from a core-collapse supernova simulation of an 11.2~M$_\odot$
    progenitor star including the production of neutrino pairs from
    heavy-nuclei de-excitations, based on Ref.~\cite{Fuller91} (green
    lines) and the Gauss ansatz (see text, red lines), in comparison
    to a simulation that uses identical input physics, but neglects
    the nuclear deexcitation process (blue lines)  (from
    \cite{Fischer14}). \label{fig:deexcite-lumin}}
\end{center}
\end{figure}

The impact of nuclear deexcitation on the supernova evolution has been
studied in core-collapse supernova simulations of an 11.2~M$_\odot$
progenitor star \cite{Fischer14}.  Three simulations have been
performed which differ only in the treatment of the neutrino pair
de-excitation process.  In the first two simulations this process has
been implemented following the prescription of Ref. \cite{Fuller91}
and by adopting the Gaussian approximation model to the absorption
strength function, as described above. The third simulation served as
a control study, in which the neutrino-pair nucleear deexcitation has
been switched off.

The most important result of these simulations is that the global
quantities such as temperature and electron fraction profiles are the
same in all simulations. This implies that the neutrino-pair nuclear
deexcitation process has no impact on the global supernova evolution
\cite{Fischer14}. This is confirmed in Fig. \ref{fig:deexcite-lumin}
which shows that the luminosities of electron neutrinos, arising
mainly from electron capture, are about 4 orders of magnitude larger
than those of heavy-flavor neutrinos during the collapse phase.

Fig. \ref{fig:deexcite-lumin} also shows that the evolution of the
$\nu_e$ luminosities and average energies are the same in all three
simulations: the luminosity increases from about
$10^{51}$~erg~s$^{-1}$ to $10^{53}$~erg~s$^{-1}$ at $\nu_e$ trapping
shortly before core bounce caused by the increasing density which
enlarges the electron Fermi energy and the electron capture rates.
Relatedly, the average energies of $\nu_e$ neutrinos increase from
6~MeV to 10~MeV which, however are smaller than those of the neutrinos
directly produced by the capture process, reflecting the importance of
down-scattering by interaction with matter.

Nuclear deexcitation by neutrino-pair emission produces all
neutrino types, while electron capture is a pure source of electron
neutrinos. Indeed the de-excitation process is the dominating source
of $\nu_x$ neutrinos and, to a lesser extent, of electron antineutrinos
during collapse (see Fig. \ref{fig:deexcite-lumin}). 
Due to the charged-current contribution the production
of electron neutrino pairs is larger than that of the other two flavors.
At high densities
of order $10^{13}$ g~cm$^{-3}$ also nucleon-nucleon bremsstrahlung
becomes a significant source of neutrinos other than $\nu_e$
which causes the steep rise of the $\nu_{\mu,\tau}$ and ${\bar \nu_e}$
luminosities at times just before bounce in  Fig. \ref{fig:deexcite-lumin}. 
At densities below $10^6$ g~cm$^{-3}$ electron-positron annihilation
is another relevant neutrino-pair production process.
The average energies of the heavy-flavor neutrinos increase slightly
during the collapse (as temperature increases), but they are noticeably
smaller than the values of electron neutrinos produced by electron capture.
For densities $\rho > 5 \times 10^{11}$ g~cm$^{-3}$ neutrino-matter interactions
also become relevant for heavy-flavor neutrinos.

At core bounce where normal matter density is reached, heavy nuclei
dissociate into a state of homogeneous matter of nucleons and hence
the production of neutrino pairs from nuclear deexcitation disappears.
At shock formation and during the initial shock propagation
out of the stellar core, the infalling heavy nuclei that hit the
expanding shock wave also dissociate.
Consequently at the conditions behind the expanding shock front, weak
processes are determined by interactions with free neutrons and protons.
Hence, the inclusion of  heavy-nuclei de-excitations has no impact on the
supernova dynamics or  the neutrino signal after core bounce, e.g., in terms of
the energy loss   (as suggested in ref.~\cite{Fuller91}).
Although a small fraction of  heavy nuclei exist ahead of the expanding bounce
shock before  being dissociated
(see the discussion in the preceding subsection),
the conditions are such that other pair
processes dominate over the pair production from nuclear de-excitation.
Moreover,  the supernova dynamics is dominated by charged-current processes
on free nucleons behind the bounce shock in the dissociated regime.
Consequently, neutrino pair heavy-nuclei de-excitations has no impact in
the entire post-bounce period and the evolution of the neutrino  luminosities
and average energies in our three simulation become identical.

By studying the radial profiles of the collapsing core at different
times during collapse, Ref. \cite{Fischer14} gives a detailed account
of the evolution of the luminosities and average energies of the
various neutrino flavors and the competition of the different neutrino
sources.  It turns out that the two approaches which have been adopted
to describe nuclear deexcitation give qualitatively similar results,
they differ, however, in details. This is relevant if one is
interested in an accurate description of the heavy-flavor and electron
antineutrinos during collapse. An improved calculation of the
absorption strength function within a formalism like the Thermal QRPA
of Ref. \cite{Dzhioev13} might be useful to this end.

\section{Neutrino-nucleus reactions and supernova nucleosynthesis}

As discussed above, the continuous emission of neutrinos from the
protoneutron star drives a low-mass outflow (the neutrino-driven
wind). Due to the high temperature involved the matter is ejected as
free protons and neutrons. Upon reaching cooler regions further away
from the neutron star, the matter can assemble into nuclei.  The
outcome of this nucleosynthesis depends crucially on the
proton-to-neutron ratio and can either support the $\nu$p process (if
$Y_e >0.5$) or an r process ($Y_e < 0.5$). We will discuss both
processes in the two following subsections.  Furthermore neutrinos can
also contribute to the astrophysical production of certain nuclides,
in what is called the $\nu$ process, when they pass through the outer
shell of the star. The $\nu$ process is discussed in the third
subsection.

The proton-to-neutron ratio is mainly determined by the neutrino and
antineutrino charged-current reactions with the free neutrons and
protons \cite{Qian95}.  Assuming that the neutrino luminosities are
about equal for electron neutrinos and antineutrinos a detailed
analysis of these reactions shows that the matter is expected to be
proton rich if $4 \Delta > (\langle E_{\bar \nu_e} \rangle - \langle
E_{\nu_e} \rangle)$, where $\Delta=1.29$ MeV is the proton-neutron
mass difference and $\langle E_{\bar \nu_e} \rangle,\langle E_{\nu_e}
\rangle$ are the average energies of electron antineutrinos and
neutrinos, respectively \cite{Froehlich05}.  It is important that in
the kinematics of the neutrino-nucleon reactions corrections due to
the nuclear interaction are included. The dominant contribution is due
to mean-field corrections, induced by the astrophysical environment
and accounted for in the Equations of
State~\cite{Roberts12,Martinez12}.  Introducing the mean-field
potentials for protons, $U_p$, and neutrons, $U_n$, the energy of
neutrinos is reduced by $U_n - U_p$, while the energy of antineutrinos
is enlargened by the same amount. We note that these corrections act
like the $Q$ value in electron capture reactions on nuclei which grows
with increasing neutron excess throttling the capture rate. In fact,
for the neutron-rich environment in core-collapse supernovae,
e.g. shortly after bounce, $U_n-U_p >0$ and as a consequence, the
emissivity of neutrinos (antineutrinos) is increased (decreased) if
the mean-field corrections are taken into account.

Remembering that the charged-current neutrino reactions on nucleons
set the neutron-to-proton ratio of the matter ejected in the
neutrino-driven wind, the changes of the neutrino and antineutrino
energies due to the mean-field corrections have an essential impact on
the neutrino-driven wind nucleosynthesis: the early ejecta are
slightly neutron rich, while at later times the ejecta become proton
rich \cite{Martinez12}.

In Ref. \cite{Arcones08} it has been shown that in the outer layers of
the protoneutron star a substantial amount of light nuclei (deuterons,
$^3$H, $^3$He, and $^4$He) is present. The interaction of neutrinos
with these light nuclei has an impact on the average energies of
neutrinos and antineutrinos. The relevant neutrino cross sections on
the light nuclei have been calculated on the basis of ab-initio
few-body methods like the hyperspherical harmonics approach
\cite{Gazit08,Gazit07a,Gazit07b}.  These cross sections are more
accurate than previous calculations performed on less sophisticated
nuclear models (e.g. \cite{Woosley90,Kolbe93}), but do not differ too
dramatically.

\subsection{The $\nu p$ process}

\begin{figure}
  \begin{center}
    \includegraphics[width=0.8\linewidth]{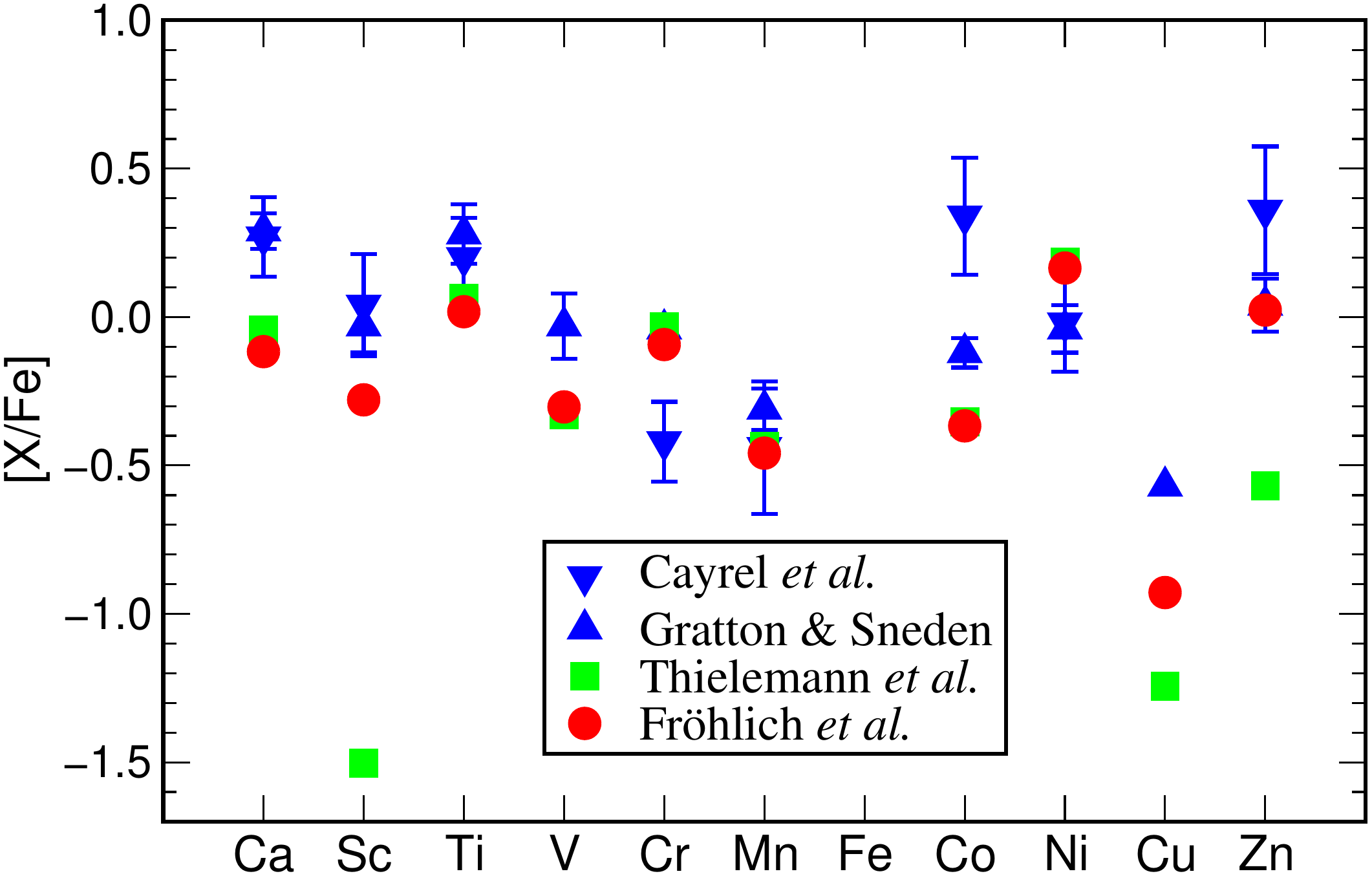}
    \caption{Comparison of elemental abundances in the mass range Ca
      to Zn produced in proton-rich environment (Fr\"ohlich et al.,
      \cite{Froehlich05}) with previous calculations (Thielemann et
      al., \cite{Thielemann96}) and observational data
      \cite{Cayrel04,Gratton91}  (adapted from
      \cite{Froehlich05}). \label{fig:abund-ca-ni}}
  \end{center}
\end{figure}

Supernova simulations indicate that during some period after bounce
the hot matter ejected from the surface of the freshly born neutron
star is proton-rich (i.e. $Y_e > 0.5$)
(e.g. \cite{Buras03,Liebendoerfer01,Janka03,Thompson05,gmp14}.  The
alpha rich freeze-out of such proton-rich matter favors the production
of $\alpha$ nuclei (mainly $^{56}$Ni) with some free protons left
\cite{Seitenzahl08}. We note that this freeze-out also results in
enhanced abundances of selected nuclei in the Ca-Fe mass range
bringing them into better agreement with observation
\cite{Froehlich05,Froehlich06,Pruet05} (see
Fig. \ref{fig:abund-ca-ni}).

Heavier nuclides can be synthesized from the freeze-out abundance
distribution by subsequent proton captures, competing with $\beta^+$
decays. We note that this reaction sequence is realized in the
so-called rapid-proton capture process (i.e. explosive hydrogen
burning on the surface of neutron stars in X-ray bursts)
\cite{Schatz98}.  However, in this scenario the mass flow to heavier
nuclides is strongly hampered by the increasing Coulomb barrier of the
produced elements and, in particular, by the so-called waiting point
nuclei. These are $\alpha$ nuclei like $^{56}$Ni, $^{64}$Ge,
$^{68}$Se...  which have relatively long $\beta$ half lives and for
which the proton capture is strongly hindered due to the small or
negative proton binding energies of the final nuclei (e.g. $^{57}$Cu,
$^{65}$As, $^{69}$Br, \ldots). In the case of unbound daughters it is
suggested that the mass flow might overcome the waiting point nuclei
by two-proton captures \cite{Wiescher01}.

\begin{figure}
\begin{center}
  \includegraphics[width=0.50\linewidth]{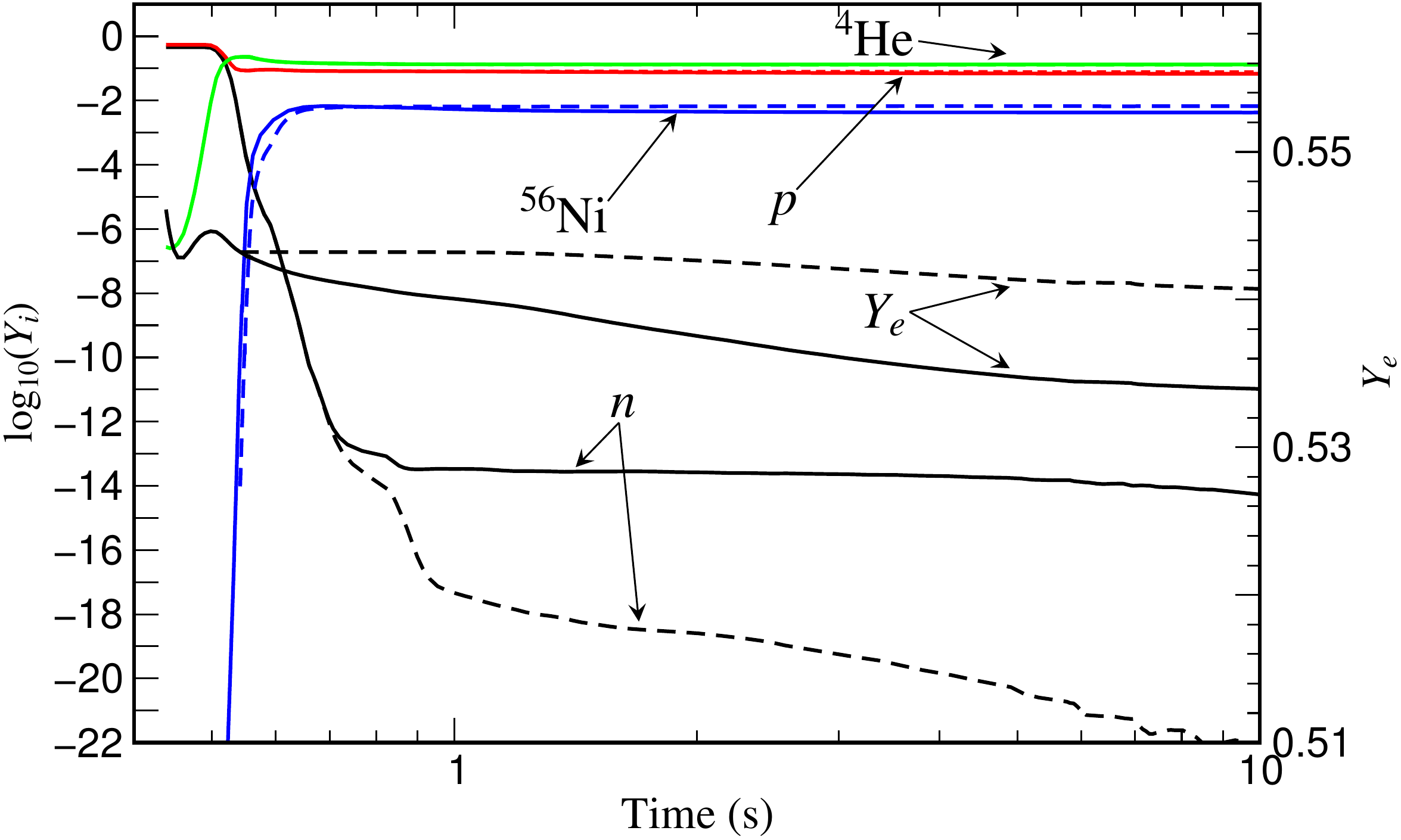}%
  \hspace{0.02\linewidth}%
  \includegraphics[width=0.46\linewidth]{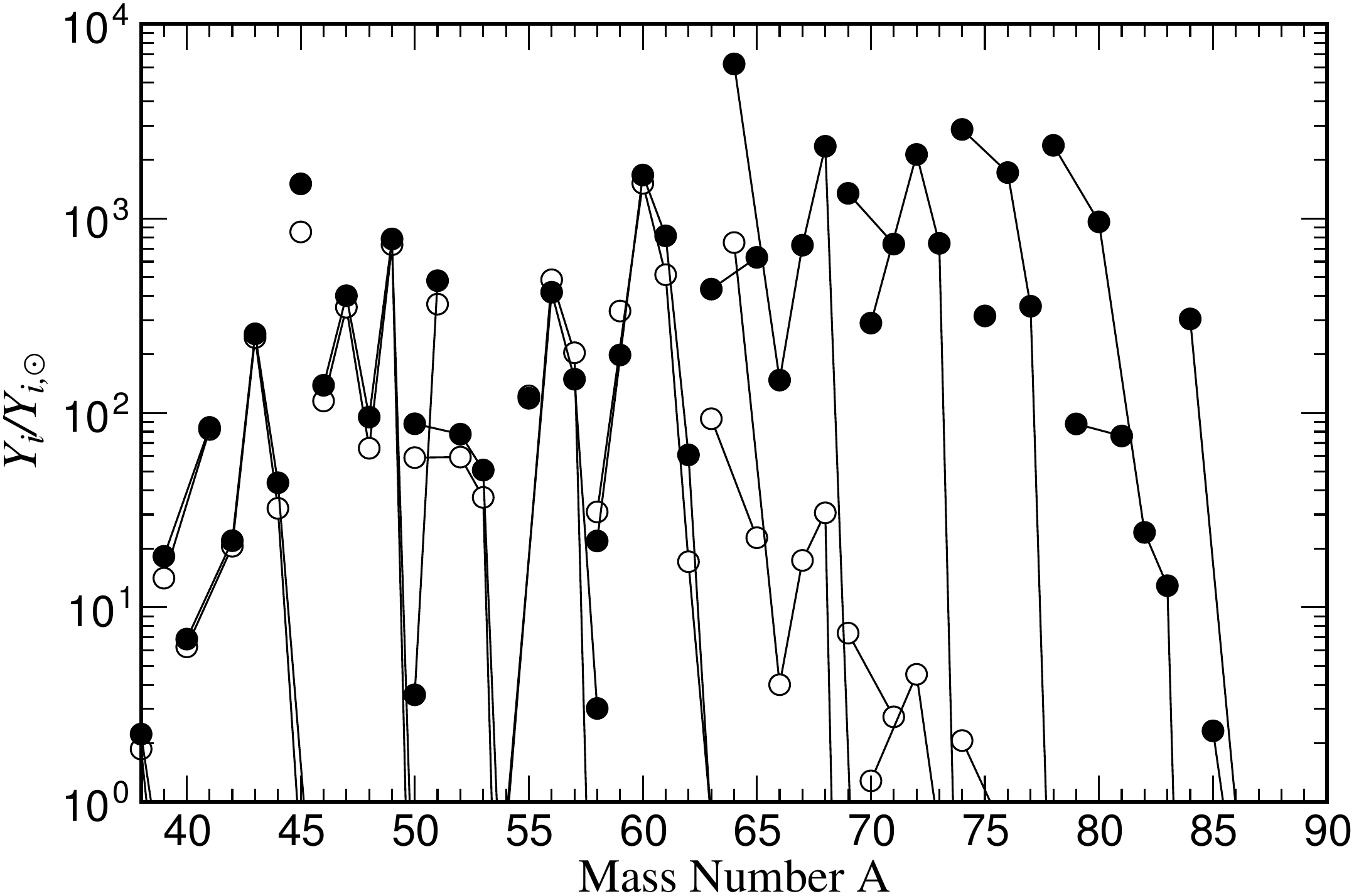}
  \caption{(left) Evolution of the abundances of neutrons, protons,
    alpha-particles, and $^{56}$Ni. The solid (dashed) lines display
    the nucleosynthesis results which include (omit) neutrino and
    antineutrino absorption interactions after nuclei are formed. The
    abscissa measures the time since the onset of the supernova
    explosion.  (right) Isotopic abundances (relative to solar)
    obtained within the supernova $\nu p$ process nucleosynthesis. The
    filled circles represent calculations where (anti)neutrino
    absorption reactions are included in the nucleosynthesis while for
    the open circles neutrino interactions are neglected.  The two
    panels base on a nucleosynthesis trajectory from the model B07
    from Ref.~\cite{Froehlich05} (from \cite{Froehlich06}). }
  \label{fig:nup-process}
\end{center}
\end{figure}

In contrast to the nucleosynthesis in X-ray bursts, the process occurs
in the supernova environment in the presence of extremely intense
neutrino fluxes which influences and alters the matter flow to heavier
nuclei substantially.  While the energy of supernova $\nu_e$ neutrinos
is too small to induce sizable reaction rates on $N \sim Z$ nuclei,
this is different for antineutrinos that are captured in a typical
time of a few seconds in the conditions of the hot neutrino bubble,
both on protons and nuclei, at the distances at which nuclei form
($\sim 1000$~km).  This time scale is much shorter than the beta-decay
half-life of the most abundant heavy nuclei (e.g. $^{56}$Ni,
$^{64}$Ge), which serve as obstacles in the mass flow to heavier
nuclei if neutrino interactions are not included.  As protons are more
abundant than heavy nuclei, antineutrino capture occurs predominantly
on protons, causing a  steady supply of free neutrons with a residual
density of $10^{14}$--$10^{15}$~cm$^{-3}$ for several seconds, when
temperatures are in the range 1--3~GK (see left panel in
Fig. \ref{fig:nup-process}).  The neutrons produced via antineutrino
absorption on protons can easily be captured by neutron-deficient
$N\sim Z$ nuclei (for example $^{64}$Ge), which have large neutron
capture cross sections. The amount of nuclei with $A>64$ produced is
then directly proportional to the number of antineutrinos captured.
While proton capture, $(p,\gamma)$, on $^{64}$Ge takes too long, the
$(n,p)$ reaction dominates (with a lifetime of 0.25~s at a temperature
of 2~GK), permitting the matter flow to continue to nuclei
heavier than $^{64}$Ge via subsequent proton capture up to the mass
range $A \sim 80-100$ (for a mass-flow diagram see, for example,
Ref.~\cite{Pruet.Hoffman.ea:2006}).  This nucleosynthesis process, operating in
proton-rich supernova environment, is called the $\nu p$-process
\cite{Froehlich06}.  The abundance distribution obtained by
$\nu p$-process nucleosynthesis is compared in the right panel of
Fig. \ref{fig:nup-process} to a simulation in which the effects of
(anti)neutrino captures are switched off.  Huther has explored the
influence of the various neutrino-induced neutral- and charged-current
reactions on the $\nu p$ process abundances. He finds that the impact
of the neutrino and antineutrino reactions on nuclei is negligible
\cite{Huther13}.

How far the massflow within the $\nu p$ process can proceed strongly
depends on the environment conditions, most noticeable on the $Y_e$
value of the matter. Obviously the larger $Y_e$, the larger the
abundance of free protons which can be transformed into neutrons by
antineutrino absorption, and the faster the bridging of the waiting
point nuclei like $^{64}$Ge by $(n,p)$ reactions. Other important
parameters are the location (radius) of the matter during the
formation of nuclei and the ejection velocity which both influence the
antineutrino fluence which the ejected matter experiences. A location
closer to the surface of the proto-neutron star and/or a slow ejection
velocity leads to an extended antineutrino exposure that allows for an
increased production of heavy
elements~\cite{Wanajo.Janka.Kubono:2011,Arcones.Froehlich.Martinez-Pinedo:2012}. The
sensitivity of the $\nu p$-process on parameters like the $Y_e$ value
has been explored in~\cite{Froehlich06,Pruet.Hoffman.ea:2006,Wanajo06}
These studies show that nuclei heavier than $A=64$ are only produced
for $Y_e>0.5$, showing a very strong dependence on $Y_e$ in the range
0.5--0.6.  A clear increase in the production of the light $p$-nuclei,
$^{92,94}$Mo and $^{96,98}$Ru, is observed as $Y_e$ gets larger. Thus
the $\nu p$ process offers an explanation for the production of these
light $p$-nuclei, which was unknown before.

\begin{figure}
  \begin{center}
    \includegraphics[width=0.50\textwidth]{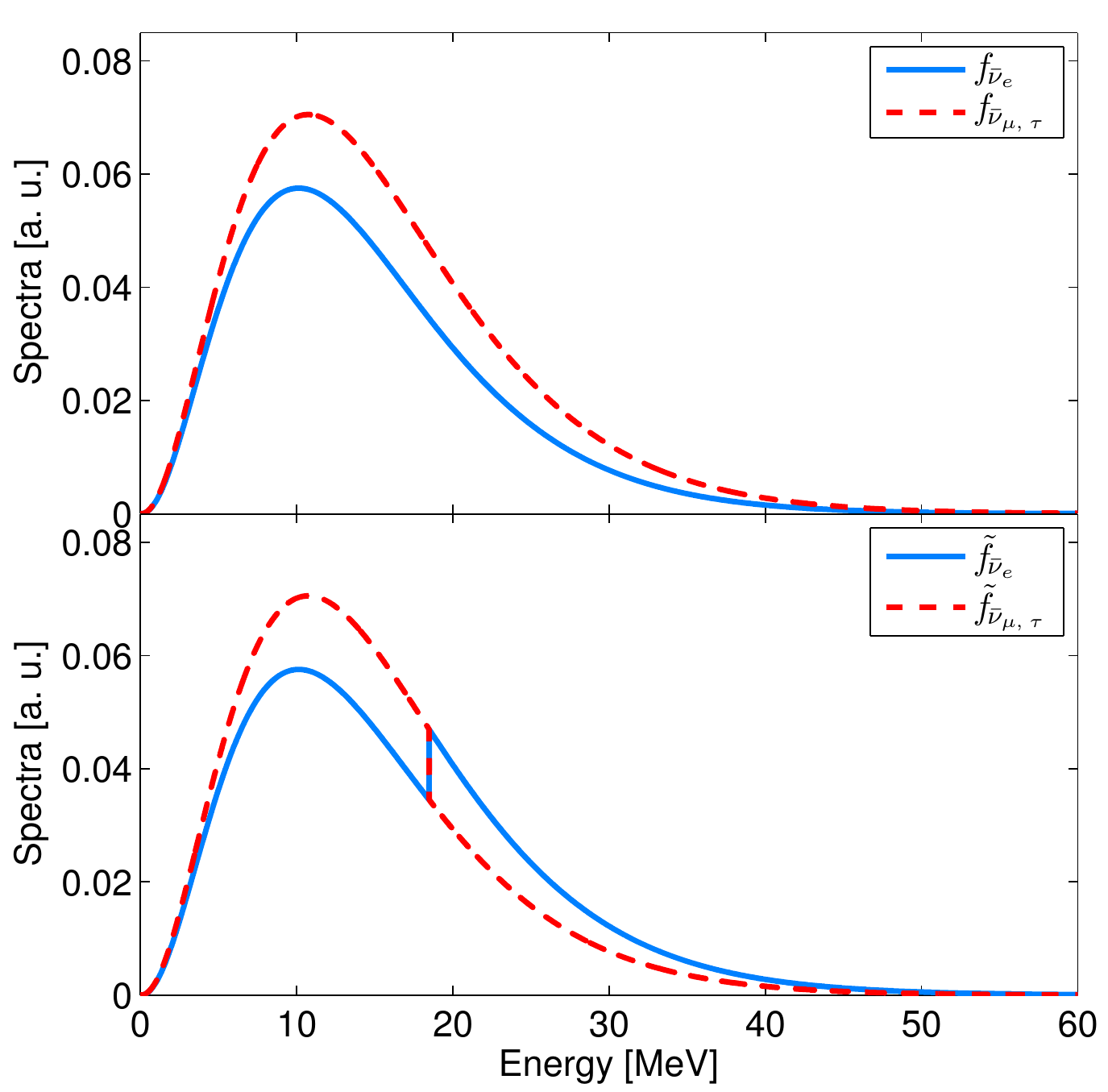}%
    \includegraphics[width=0.46\textwidth]{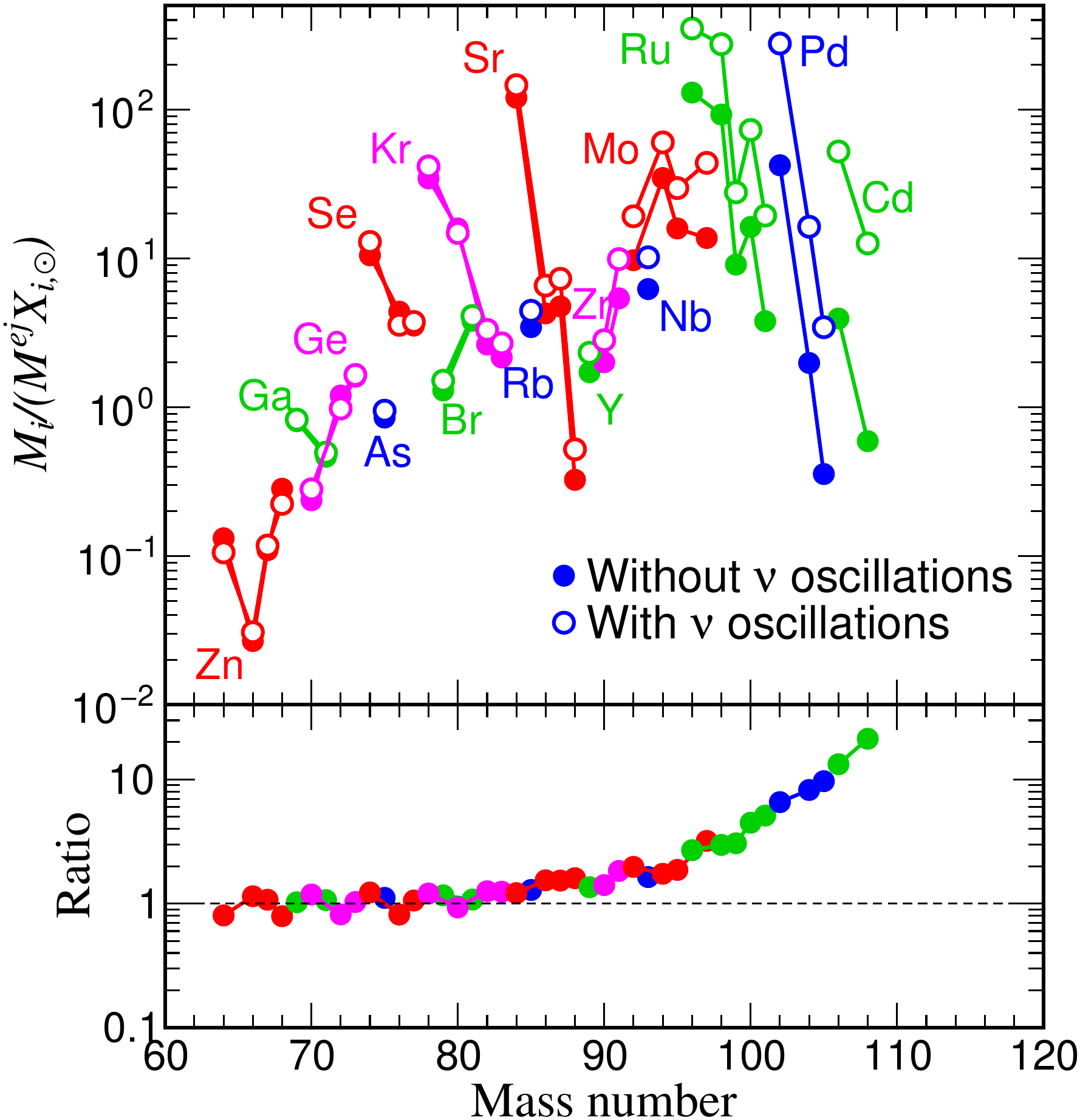}
    \caption{Left: (upper panel) ${\bar\nu_e}$ and ${\bar \nu_x}$
      spectra $f_{\bar\nu_e}$ and $f_{\bar \nu_x}$ based on the
      simulations of ref.~\cite{Buras06a} as given in table 1 of
      ref.~\cite{Pruet.Hoffman.ea:2006}. The ${\bar\nu_e}$ spectra
      have been arbitrarily normalized to one while the ${\bar \nu_x}$
      spectra are normalized to 1.3 to keep the relative ratio of the
      number luminosities. (lower panel): modified spectra $\tilde
      {f}_{\bar \nu_e}$ and $\tilde {f}_{\bar \nu_x}$, including the
      effects of collective neutrino oscillations, which induce a swap
      of the ${\bar \nu_e}$ and ${\bar \nu_x}$ spectra for energies
      above $E_s$.  Right: Abundances of $\nu p$ nucleosynthesis
      calculations with (open circles) and without (full circles)
      considering the effects of collective neutrino flavor
      oscillations (normalized to solar).  The lower panel shows the
      ratios for the two nucleosynthesis studies as function of mass
      number  (from \cite{Martinez11}). }
    \label{fig:collective}
  \end{center}
\end{figure}

The late-time abundance of free neutrons depends sensitively on the
$\bar{\nu}_e$ luminosities and spectra.  As has been shown recently
\cite{Dasgupta09,Duan10,Duan11}, the $\bar{\nu}_e$ spectrum is
affected by collective neutrino flavor oscillations which are expected
to occur in the high-neutrino-density environment surrounding the
neutron star \cite{Duan10} and can swap the ${\bar \nu_e}$ and ${\bar
  \nu_{\mu,\tau}}$ spectra above a certain split energy of order 15-20
MeV, where the split energy depends on the relative fluxes of $\bar
{\nu_e}$ and $\bar{\nu}_x$ and on the neutrino mass hirarchy
\cite{Fogli09}. Assuming such a swap scenario for the case of the
normal neutrino mass hierarchy (see left part of
Fig. \ref{fig:collective}), Ref.  \cite{Martinez11} has studied the
impact of collective neutrino flavor oscillations on the $\nu p$
process nucleosynthesis adopting anti-neutrino spectra from recent
supernova simulations \cite{Buras06a}.  As in general, $\bar{\nu}_x$
neutrinos have larger mean energies and a larger high-energy tail than
electron antineutrinos, the spectrum swap induced by collective
neutrino oscillations increases the neutron production rate by
antineutrino capture on protons during the $\nu p$
process. Ref. \cite{Martinez11} estimates this increase to vary by a
factor between 1.5 to 1.25 depending on a split energy between 0 and
25 MeV. Obviously the enhanced production of free neutrons boosts the
matter flow to heavier nuclides and results in larger $\nu p$ process
abundances. This is demonstrated in the right part of
Fig. \ref{fig:collective} which compares the $\nu p$-process
abundances calculated with a neutron production rate increased by a
factor of 1.4 due to collective neutrino oscillations and the standard
case without such oscillations. The production of the light p-nuclides
$^{92,94}$Mo and $^{96,98}$Ru is enhanced by a factor 2-3. The largest
production increase is observed for nuclides $A>96$ which is traced
back to a faster $(n,p)$ reaction time-scale for the $N=50$ nucleus
$^{96}$Pd which acts like a 'seed' for the production of nuclides with
$A>96$ \cite{Wanajo11}. The faster $(n,p)$ reactions also have an effect
on the abundances of waiting point nuclei like $^{64}$Ge or $^{68}$Se.
These nuclei are less of an obstacle in the mass flow in the
calculations which consider collective neutrino oscillations and hence
less matter is accumulated at the waiting points. More recently,
Wu~\emph{et al}~\cite{Wu.Qian.ea:2015} have performed a detailed analysis of the
impact of collective oscillations on neutrino-wind nucleosynthesis
accounting for the full temporal variation of neutrino spectra and
matter density profiles.

The $\nu p$ process constitutes a mass flow from seed nuclei, mainly
$^{56}$Ni, to heavier nuclides by $(p,\gamma)$ reactions, supplemented
by $(n,p)$ or $\beta^+$ reactions running through many nuclei at the
proton-deficient side of the nuclear chart. Many of the $\beta^+$
half-lives, for the ground states, are experimentally known.  However,
their potential modifications for the astrophysical environment with
finite temperature as well as the rates for the proton capture and
$(n,p)$ reactions have to be theoretically modelled, which is usually
performed on the basis of the statistical model. A particularly
important ingredient in such statistical model calculations are the
nuclear masses, which define the reaction $Q$ values and thus the
competition between reactions and their inverse. Nuclear mass
measurements have benefited tremendously in recent years by the
establishment of nuclear Penning traps or by storage ring experiments
at radioactive ion-beam facilities which also allowed the
determination of masses of $\nu p$-process nuclei with unprecedented
precision. With relevance to the $\nu p$ process such measurements
have been performed for nuclei in the mass range $A \sim 90$ at the
SHIPTRAP at GSI and JYFLTRAP in Jyv\"askyll\"a
\cite{Weber08,Haettner11}, for nuclei around $A=64$ at the CSRe
storage ring in Lanzhou \cite{Tu11}, for proton-deficient nuclei in
the mass range $A \sim 90$ at the Canadian Penning trap
\cite{Fallis11}. The impact of these mass measurements on the $\nu p$
process has been explored in Refs. \cite{Wanajo11,Weber08}. Studies
which investigated the sensitivity of the $\nu p$ process abundances
on the uncertainties in the $(n,p)$ rates have been reported
in~\cite{Wanajo.Janka.Kubono:2011,Froehlich12}. The results are
similar to those found for the effect of the collective neutrino
oscillations: faster $(n,p)$ rates spead up the mass flow to heavy
nuclides and hence increase the abundances for nuclides with $A>96$,
while they decrease the abundances for nuclides in the mass range $A
\sim 64-96$.

\subsection{The r-process}

More than 50 years ago the astrophysical r-process (short for rapid
neutron capture process) was introduced to explain the origin of
the transactinides and of half of the nuclides heavier than iron
\cite{BBFH,Cameron57}. The process is characterized by fast neutron
captures, in comparison to the competing $\beta$ decays, which, in
turn, requires astrophysical objects with extreme neutron densities
for the r process to operate \cite{Cowan91}. On the nuclear chart the
r-process path runs through nuclei with extreme neutron excess far off
the valley of stability.  Most of these nuclei have yet not been
produced in the laboratory and their properties, which are required
input to r-process simulations, have to be modelled theoretically. The
most relevant nuclear input are masses (reaction $Q$ values) and
$\beta$ half lives which influence the r-process path and the duration
of the process, respectively \cite{Cowan91}.  Particularly important
for the r-process mass flow to heavier nuclei are the waiting point
nuclei associated with the magic neutron numbers $N=82$ and 126 (and
possibly 184), which are also related to the peaks observed in the
r-process abundance distributions around mass numbers $A\sim 130$ and
195.

Decisive progress towards experimental information for r-process
nuclei is expected from current and future radioactive ion-beam
facilities like RIBF at RIKEN, FAIR at GSI and FRIB at NSCL/MSU. From
these facilities we expect not only the direct determination of
properties of r-process nuclei, but also stringent constraints to
improve the theoretical description of such neutron-rich nuclei which
are out of experimental reach even at these advanced facilities. We
have witnessed such indirect progress due to data for neutron-rich
nuclei for both, the global description of masses and the prediction
of half-lives. The advances in experimental mass measurements are
reviewed in Ref. \cite{Blaum06}; recent mass evaluations can be found
in \cite{Audi12,Audi03} (see also \cite{Pfeiffer12}). These data help
to improve the predictions of nuclear mass models like the
microscopic-macroscopic approach for many years pioneered by M\"oller
\cite{Moeller95,Moeller12} (see also \cite{Liu11}), the Extended
Thomas Fermi Model with Strutinski Integral \cite{Pearson96}, the
shell-model guided approach by Duflo and Zuker \cite{Duflo95} or
studies based on the Hartree-Fock-Bogoliubov model \cite{Goriely10}.
Half-life measurements of r-process nuclei at or in the vicinity of
the magic neutron numbers $N=50$ and 82 have been reported in
Refs. \cite{Hosmer05,Madurga12,Quinn12,Hosmer10} and
\cite{Kratz86,Pfeiffer01,Nishimura12}, respectively. First half-life
measurements towards r-process nuclei at the $N=126$ shell closure
have been performed at GSI \cite{Nieto14}, which are useful
constraints for theoretical half-life predictions in this until yet
unexplored part of the nuclear chart
(e.g. \cite{Moeller03,Suzuki12,Zhi13} and the calculations of Borzov
as quoted in \cite{Nieto14}).
 
One of the fundamental questions in astrophysics is: where does the
r-process operate? Although no conclusive answer can yet be given,
crucial clues come from abundance observations in so-called metal-poor
stars, i.e. stars with Fe/H ratios which are significantly (often by
orders of magnitude) smaller than the solar Fe/H ratio. While hydrogen
has mainly been produced by Big Bang nucleosynthesis, iron has
gradually been synthesized by both types of supernovae (core-collapse
and thermonuclear) during the history of the galaxy. The Fe/H
abundance ratio can thus serve as a proxy of age and stars with Fe/H
ratios much smaller than solar are supposed to be very old stars.  Due
to their abundance pattern, metal-poor stars can be classified roughly
into two categories: So-called r-II stars \cite{Christlieb04} exhibit
relative abundance patterns which agree nearly perfectly with the
solar r-process abundances for elements with charge numbers $Z>50$
\cite{Cowan06,Sneden08,Roederer12}.  Some other stars
\cite{Honda06,Roederer12a} are depleted in elements with $Z>50$, but
enriched in lighter elements like Sr, Y and Zr, with the star HD
122563 being a famous example \cite{Honda06}.  One often denotes the
process whose imprint is observed in stars like HD 122563 the 'weak
r-process', while the 'main r-process' is related to the robust
abundance pattern observed in the r-II stars.

\begin{figure}
  \begin{center}
    \includegraphics[width=0.8\linewidth]{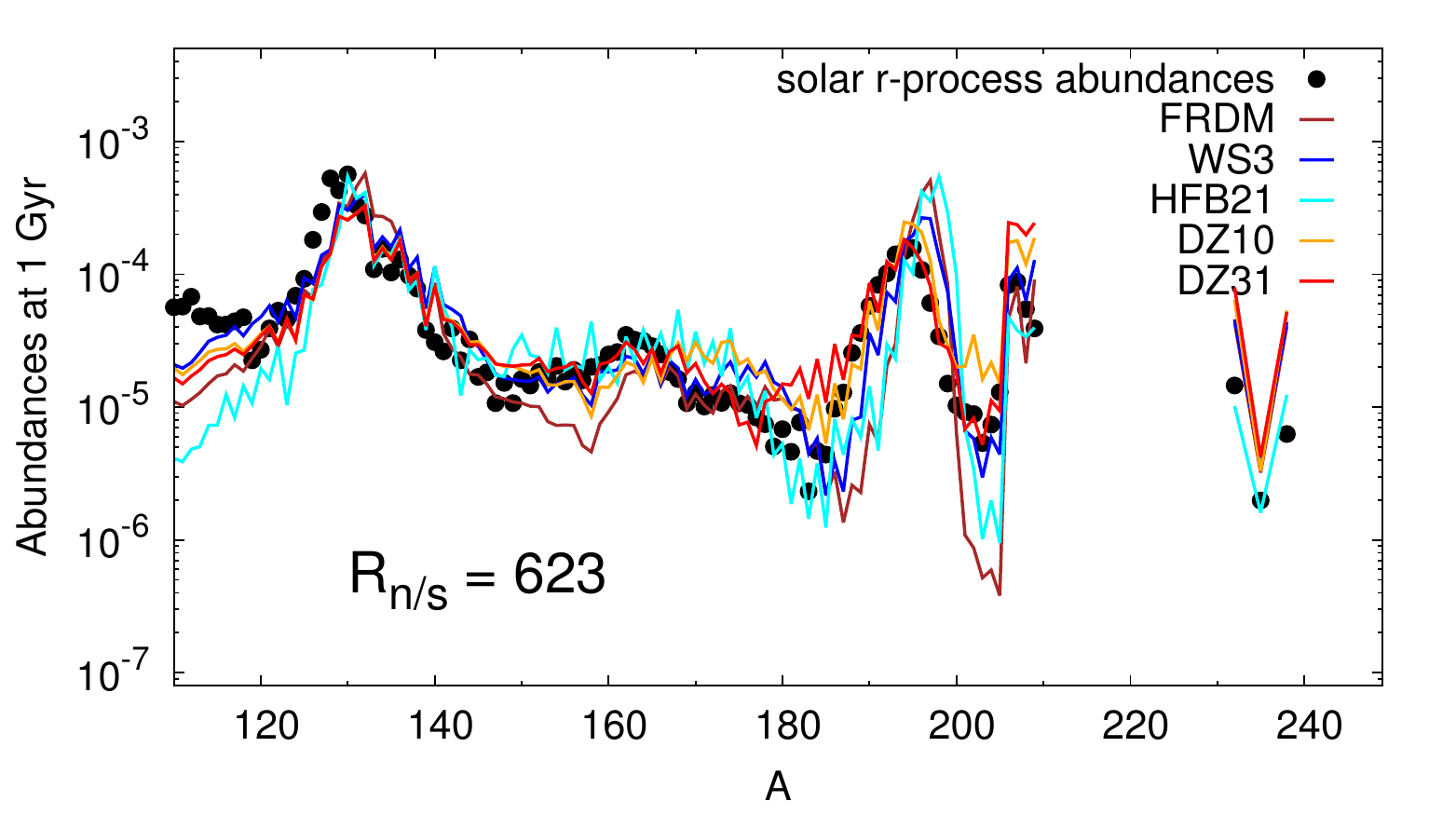}
    \caption{R-process abundances calculated for a neutron-star merger
      trajectory and different mass models (from
      \cite{Mendoza14}). \label{fig:rprocess-site1} }
  \end{center}
\end{figure}

It is tempting to identify potential astrophysical sites for the
r-process. The main r-process might be attributed to neutron star
mergers. Recent simulations \cite{Korobkin12,Bauswein13} show that the
extreme neutron richness of the matter dynamically ejected during the
merger ($Y_e \sim 0.05)$ leads to fast mass flow to nuclei up to mass
range $A \sim 280$ and the occurence of fission cycling. In fact, the
second r-process peak at $A \sim 130$ is produced by fission fragments
in this scenario. Hence fission rates and yield distributions,
including the production of free neutrons during the fission process
and by $\beta$-delayed neutron emission of fission yields, are crucial
ingredients in the simulations \cite{Goriely13,Winteler14}.
Fig. \ref{fig:rprocess-site1} shows the abundance distribution
obtained from r-process simulations based on neutron-star merger
trajectories of Ref. \cite{Bauswein13} and using the fission fragment
yields calculated on the basis of the ABLA code \cite{Gaimard91} which
has been carefully adjusted against experimental fission data
\cite{Kelic10}.  In fact the calculated abundance distributions
reproduce quite nicely the solar r-process pattern in a quite robust
way, i.e.  basically independent on the nuclear mass models and the
details of the merger trajectories \cite{Mendoza14}.  Despite these
intriguing results it is still an open question whether neutron-star
mergers have already operated with sufficient frequencies at the early
time of the galaxy to explain the observed r-process abundances in the
oldest
stars~\cite{Qian00,Argast04,Shen.Cooke.ea:2014,Voort.Quataert.ea:2015,Wehmeyer.Pignatari.Thielemann:2015}. Furthermore,
the impact of neutrinos in relativistic merger models requires further
investigation~\cite{Wanajo.Sekiguchi.ea:2014}.  Alternative sites for
the r-process which are expected to operate mainly at low
metallicities are jets from magnetorotational supernova
\cite{Winteler12} or an r-process in the He-shell of core-collapse
supernovae \cite{Banerjee11}. In the latter scenario, the required
neutrons are produced by charged-current reactions on $^4$He via
$^4\text{He}(\bar{\nu}_e,e^+ n)^3$H where the relevant cross sections
are given in \cite{Barnea07} based on an ab-initio calculation within
the hyperspherical harmonics approach \cite{Barnea00}.

\begin{figure}
  \begin{center}
    \includegraphics[width=0.8\linewidth]{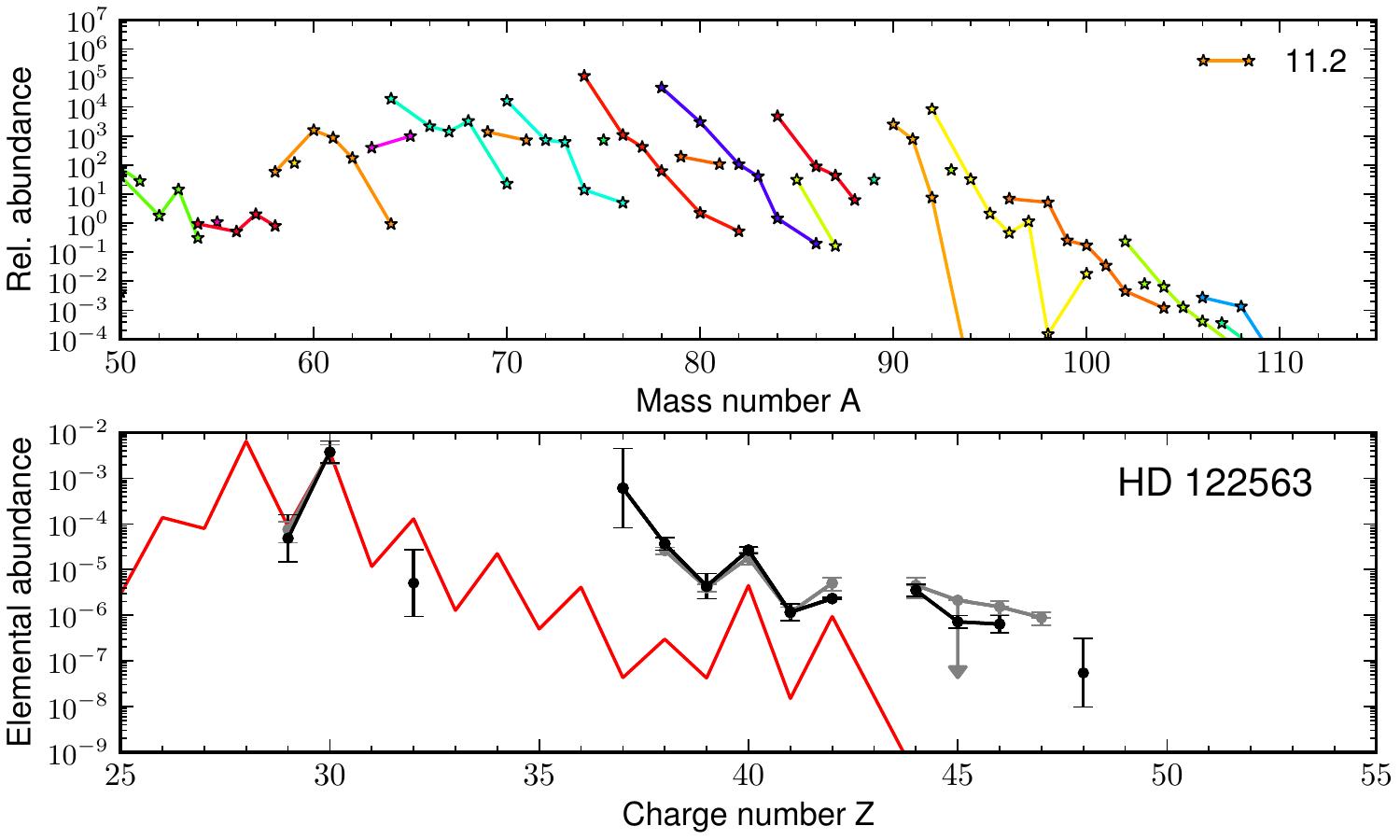}
    \caption{Mass-integrated abundances for the neutrino-driven wind
      of a core-collapse supernova  (adapted
      from~\cite{Martinez14}). \label{fig:rprocess-site2}}
  \end{center}
\end{figure}

The weak r-process has been suggested to operate in the
neutrino-driven wind developing in core-collapse supernovae after
bounce. This scenario has actually been the favorite site for the
r-process for many years \cite{Woosley94,Witti94}. To achieve the
neutron-to-seed ratio $n/s$ necessary to produce the third r-process
peak at $A \sim 195$ and the transactinides after freeze-out of
charged-particle reactions, $n/s > 150$, puts stringent constraints on
the wind properties. Such high ratios require either large entropies,
small $Y_e$ value or fast velocities of the ejected matter
\cite{Qian96,Hoffman97}. Recent supernova simulations imply that such
conditions are not met in the neutrino-driven scenario making the
operation of the 'main r-process' unlikely in this scenario. However,
the conditions might be sufficient to produce the light r-process
elements up to Sr, Y, and Zr \cite{Roberts10,Martinez14}.  This is
demonstrated in Fig. \ref{fig:rprocess-site2} which shows the
mass-integrated abundance distribution (upper panel) and elemental
abundances (lower panel) obtained from a core-collapse supernova
simulation of a $11.2 M_\odot$ progenitor star \cite{Martinez14} which
followed the post-bounce evolution up to 9~s after bounce.  The
calculation adopted an Equation of State which consistently provided
the mean-field corrections needed for the calculation of the
charged-current neutrino reactions on nucleons which determine the
$Y_e$ value of the ejected matter (as discussed above).  The ejected
matter was found to be neutron rich at early times ($Y_e \sim 0.48$
for the first 3 seconds after bounce). At later times the matter
becomes increasingly enriched in protons reaching $Y_e=0.58$ after
9s. The early neutron-rich ejecta produce isotopes of the elements Sr
to Mo ($Z=38-42$), while nuclei with $Z>42$ (or $A>92$) are produced
by the $\nu p$ process in the later proton-rich ejecta.  However, their
production is quite inefficient due to low antineutrino luminosities
(equivalently neutron production rates) at late
times. Ref. \cite{Martinez14} also finds relatively small amounts of
matter ejected at late times. Thus, the $\nu p$ process does not
contribute much to the mass-integrated abundances. It is very
interesting to note that the elemental abundances for the elements Sr,
Y, Zr, Nb, Mo reproduce the abundance pattern observed in the star HD
122563 quite well (see lower panel in
Fig. \ref{fig:rprocess-site2}). However, the calculation clearly
underproduces the observed abundances for the elements with $Z>42$,
including silver and palladium.  The production site of these elements
is yet an open question~\cite{Martinez14}. Active-sterile neutrino
oscillations, that assume a sterile neutrino mass $\sim 1$~eV, help to
achieve conditions suitable for the productions of elements in the
range $Z=42$-50~\cite{Wu14}

There have been several suggestions how neutrino-nucleus reactions
might alter the r-process abundance in the neutrino-driven wind
scenario (for a review see \cite{Langanke03a}). However, the impact of
the reactions has mainly been explored chosing parametrizations of the
wind which made it the site of the main r-process.  The studies
include r-process calculations including neutrino-nucleus reactions
\cite{Fuller95,McLaughlin95,Terasawa01a,Terasawa01b,Terasawa04}, the
exploration of the role of neutrino-induced fission
\cite{Qian02,Kolbe04,Kelic05} and potential fingerprints left by
neutrino-induced spallation on the final r-process abundances
\cite{Qian97,Haxton97}. The recent calculations by
Huther~\cite{Huther13} indicate that neutrino-nucleus reactions have
no noticeable impact on the abundance distribution of the weak
r-process~\cite{Martinez14}.  A potential exception is one of the
variants of the so-called '$\alpha$ effect', proposed in
Ref. \cite{Meyer95}.  In very strong neutrino fluxes neutral-current
reactions on $^4$He can trigger two-body reaction sequences
(e.g.$^4$He($\nu,\nu^\prime
p)^3$H$({^4}$He,$\gamma)^7$Li($^4$He,$\gamma$)$^{11}$B and
$^4$He($\nu,\nu^\prime
n)^3$He$({^4}$He,$\gamma)^7$iBe($^4$He,$\gamma$)$^{11}$C) which
compete with the three-body reaction $\alpha+\alpha+n \rightarrow {^9}$Be
which has been initially considered to bridge the mass gaps at $A=5$
and 8 in the neutrino-driven wind r-process \cite{Woosley94}. As a
consequence of the more effective bridging the mass flow to heavier
nuclei is faster, reducing in turn the neutron-to-seed ratio and thus
throttling the subsequent r-process.

\subsection{The $\nu$ process}

When neutrinos, produced in the hot supernova core, pass through the
outer shells of the star, they can induce nuclear reactions and in
this way contribute to the synthesis of elements (the
$\nu$-process)~\cite{Domogatski78,Woosley88}. As pointed out by
Woosley {\it et al.}~\cite{Woosley90}, the nuclides $^{11}$B and
$^{19}$F are produced by $(\nu,\nu' n)$ and $(\nu, \nu' p)$ reactions
on the quite abundant nuclei $^{12}$C and $^{20}$Ne. These reactions
are dominantly induced by $\nu_\mu$ and $\nu_\tau$ neutrinos and their
antiparticles due to their larger average energies than $\nu_e$ and
$\bar{\nu_e}$ neutrinos.  Also the synthesis of the odd-odd nuclides
$^{138}$La and $^{180}$Ta are attributed to the $\nu$
process. However, they are produced mainly by the charged-current
reaction $^{138}$Ba($\nu_e,e^-)$$^{138}$La and
$^{180}$Hf($\nu_e,e^-)$$^{180}$Ta \cite{Goriely01}.  Hence, the
$\nu$-process is potentially sensitive to the spectra and luminosity
of $\nu_e$ and $\nu_x$ neutrinos, which are the neutrino types not
observed from SN1987A.

Neutrino nucleosynthesis studies are quite evolved requiring
state-of-the-art stellar models with an extensive nuclear network
\cite{Woosley02,Heger05}. In the first step stellar evolution and
nucleosynthesis is followed from the initial hydrogen burning up to
the presupernova models. The post-supernova treatment then includes
the passage of a neutrino flux through the outer layers of the star
and of the shock wave which heats the material and also induces
noticeable nucleosynthesis, mainly by photodissociation (the $\gamma$
process \cite{Howard91}). Modelling of the shock heating is quite
essential as the associated $\gamma$ process destroys many of the
daughter nuclides previously produced by neutrino nucleosynthesis.

\begin{figure}
  \begin{center}
    \includegraphics[width=0.6\textwidth]{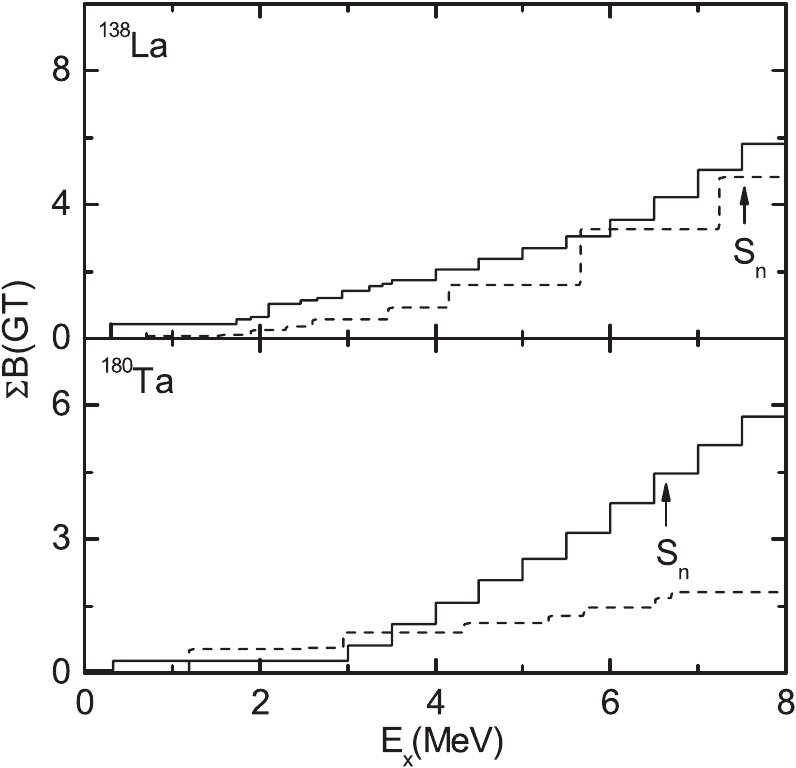}
    \caption{Summed B(GT) strength in $^{138}$La and $^{180}$Ta as a
      function of excitation energy up to the respective neutron
      emission thresholds. Solid lines: experimental data; dashed
      lines: RPA calculation \cite{Heger05}  (from
      \cite{Byelikov07}).\label{fig:lata-bgt}} 
\end{center}
\end{figure}

\begin{figure}
\begin{center}
  \includegraphics[width=0.52\linewidth]{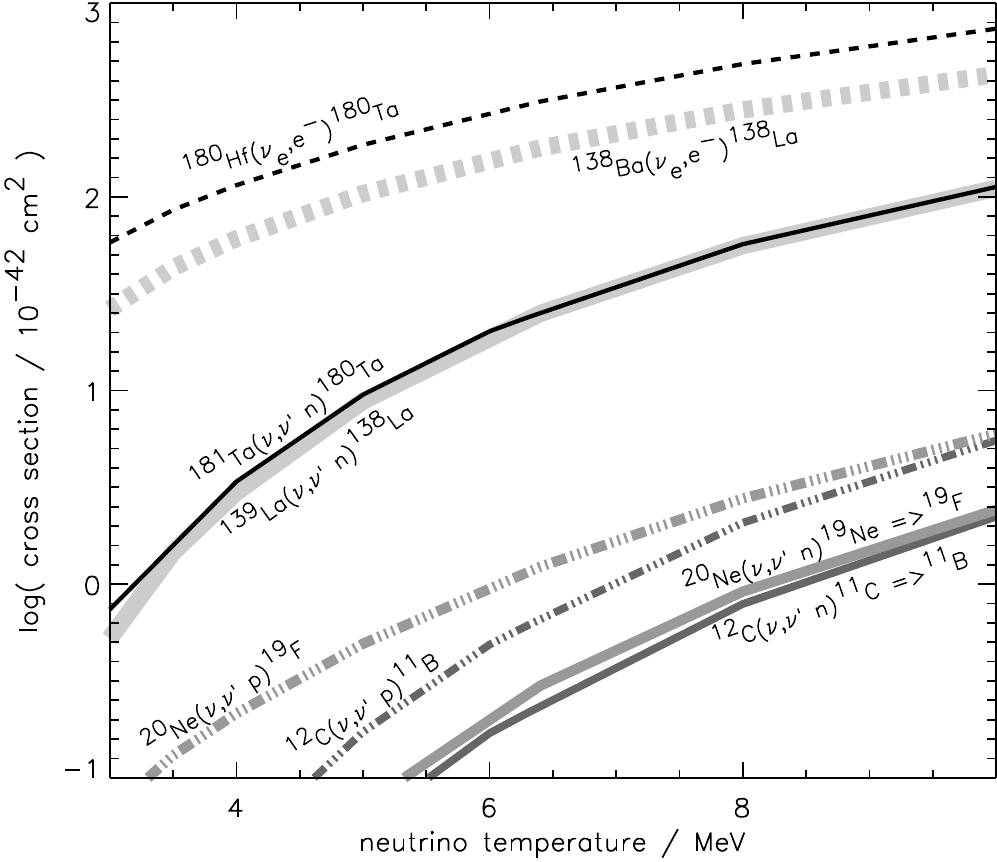}%
  \hspace{0.02\linewidth}%
  \includegraphics[width=0.45\linewidth]{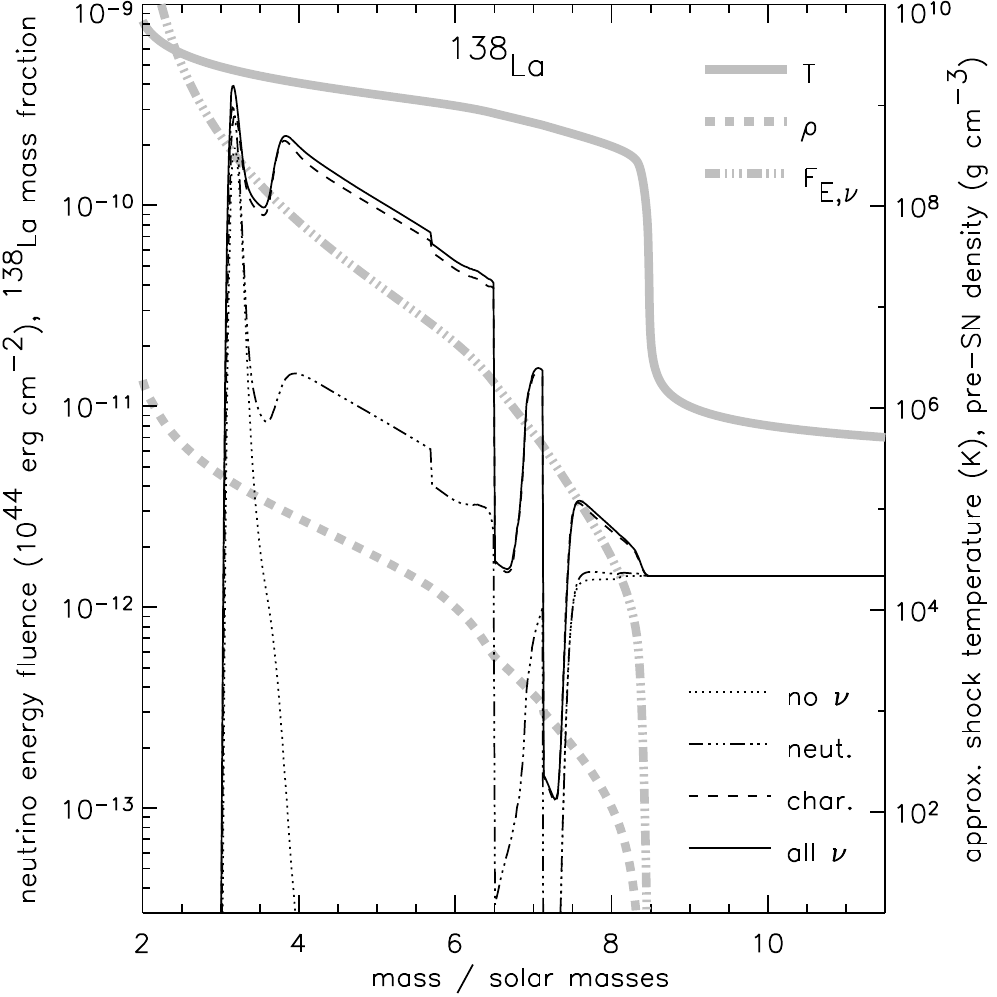}
  \caption{(Left) Electron neutrino charged-current cross sections on
    $^{138}$Ba for cascades up to two particle emission.  (right)
    Production of $^{138}$La in a $25\,$M$_\odot$ star.  The figure
    shows the production of $^{138}$La without neutrinos (dotted
    line), with the charged-current reaction $^{138}$Ba$(\nu_e,e^-)$
    (dashed), with the neutral current reaction
    $^{139}$La$(\nu,\nu'n)$ (dashed-dotted) and with both reactions
    (solid).  Additionally, the supernova shock temperature, the
    pre-supernova density, and the neutrino energy fluence
    $F_{\mathrm{E},\nu}$ assuming a total neutrino energy of
    $3\times10^{53}$~erg are given (from
    \cite{Heger05}). \label{fig:ba138}}
\end{center}
\end{figure}

The production of $^{138}$La and $^{180}$Ta by the $\nu$ process is
caused by charged-current $(\nu_e,e^-)$ reactions on $^{138}$Ba and
$^{180}$Hf, respectivley.  At the energies involved the cross sections
are dominated by the low-energy tail of the GT$_-$ strength
distribution. As large-scale shell model calculations are yet not
feasible for these relatively heavy nuclei, the cross sections had to
be calculated within the RPA model for both allowed and forbidden
transitions \cite{Woosley90,Heger05}.  As a major improvement the
GT$_-$ strengths on $^{138}$Ba and $^{180}$Hf below the particle
thresholds was measured at the RCNP in Osaka using the
$(^3\text{He},t)$ charge-exchange reaction \cite{Byelikov07}.  The
left panel of Fig. \ref{fig:lata-bgt} compares the running sum of the
measured B(GT$_-$) strength with the RPA prediction. Note that for the
$\nu$-process production of $^{138}$La and $^{180}$Hf one ist only
interested in the strength below the neutron threshold (indicated by
$S_n$ in Fig. \ref{fig:lata-bgt}) as the strength above $S_n$ connects
to states which will subsequently decay by neutron emission and hence
not contribute to the $^{138}$La and $^{180}$Hf production.  One finds
that the RPA prediction are slightly smaller than the B(GT$_-$) data
for $^{138}$Ba up to the neutron threshold, but underestimates the
$^{180}$Hf strength by nearly a factor of 3.  The inversion of the
B(GT$_-$) strength into cross section is complicated by the fact that
for both nuclei the proton decay channel opens below the neutron
channel. The respective corrections have been derived from branching
ratios obtained from the statistical model. Furthermore adding
forbidden contributions to the cross sections derived from RPA
calculations, one finds that the $^{138}$Ba($\nu_e,e^-)$$^{138}$La and
$^{180}$Hf($\nu_e,e^-)$$^{180}$Ta, calculated for a supernova spectrum
for $\nu_e$ neutrinos with temperature $T=4$ MeV and zero chemical
potential, are about 25$\%$ and 30$\%$ larger than estimated solely on
the basis of allowed transitions.  The cross sections are shown,
together with other relevant cross sections for the study of neutrino
nucleosynthesis in the left panel of Fig.  \ref{fig:ba138}.

Incorporated into stellar models \cite{Woosley02,Heger05} one finds
indeed that $^{138}$La is being produced by charged-current reactions
on $^{138}$Ba, This is demonstrated in Fig. \ref{fig:ba138} which
shows the $^{138}$La production in the oxygen/neon shell of a
25 $M_\odot$ star \cite{Heger05}.  The key is the enhancement of
$^{138}$Ba by an s-process prior to the supernova explosion.  A
$^{138}$La abundance is basically not existing prior to the passing of
the supernova neutrinos  and is the mainly produced by
the $(\nu_e,e^-)$ reaction on $^{138}$Ba (right panel).  In principle,
$^{138}$La can also be made by the neutral-current neutron-spallation
reaction $(\nu,\nu' n)$, but its contribution to the total $^{138}$La
abundance is insignificant (right panel).  The $\gamma$ process
produces also some $^{138}$La when the shock passages through the
oxygen/neon layer. But its contribution to the total $^{138}$La yield
is about an order of magnitude less than the charged-current
contribution (right panel). In fact, $^{138}$La appears to be
co-produced with nuclides like $^{16}$O and $^{24}$Mg in massive
stars. Using these nuclei as tracers for the contribution of massive
stars to the solar abundance \cite{Woosley02,Heger05,Byelikov07}, the
observed $^{138}$La abundance appears to be mainly due to the $\nu$
process.

A comparison of the $^{180}$Ta abundance calculated in stellar models
is complicated by the fact that, on the time scales of the $\nu$
process, $^{180}$Ta can exists in two states: the $J=9^-$ isomeric
state at excitation energy $E_x=75.3$ keV with a half life of more
than $10^{15}$ years and the ground state which decays with a half
live of 8.15 hours. Also the B(GT$_-$) data cannot distinguish between
the contributions to these two states. Furthermore in the finite
temperature astrophysical environment the two states couple via
excitation of states at intermediate energies
\cite{Belic99,Belic01}. To treat this coupling, Hayakawa {\it et al.}
have recently proposed a model \cite{Hayakawa10a,Hayakawa10b} in which
they distinguish between states build on the ground and isomeric
states (guided by data of Refs. \cite{Saitoh99,Dracoulis98}) and
assume these two band structures  couple only weakly to each
other.  They then followed the decoupling of the two-band structure
from thermal equilibrium to freeze-out in a time-dependent approach
assuming an exponential decrease of temperature with time. In their
calculation 39$\%$ of the $^{180}$Ta produced in the $\nu$ process
survives in the long-lived isomeric state.  A similar branching ratio
has been estimated earlier by Mohr {\it al.} \cite{Mohr07}.  Combining
these branching ratios with the total $^{180}$Ta $\nu$ process
abundance as given in \cite{Byelikov07} makes the $\nu$ process a
potential $^{180}$Ta production site.  We note, however, that also
s-process nucleosynthesis has been suggested to produce nearly 100\%
of the solar $^{180}$Ta abundance \cite{Wisshak01}.

\begin{figure}
  \begin{center}
    \includegraphics[width=0.40\linewidth]{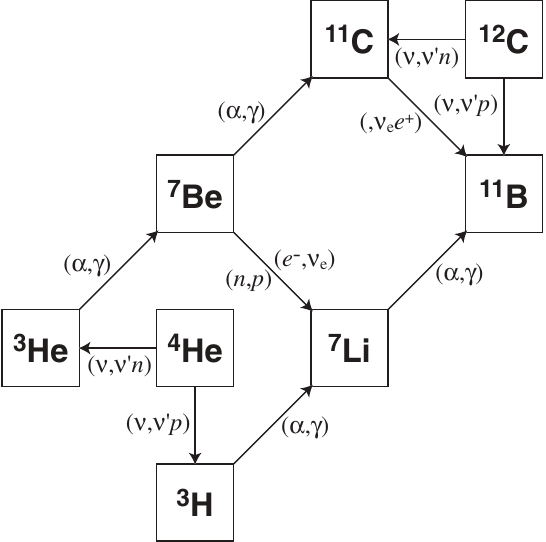}%
    \hspace{0.02\linewidth}%
    \includegraphics[width=0.56\linewidth]{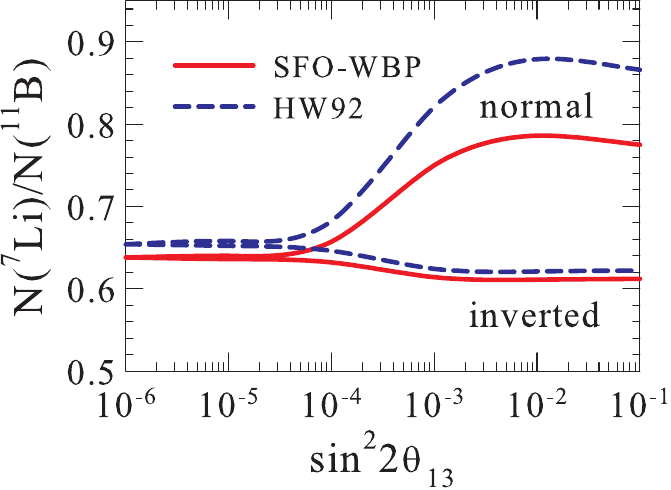}
    \caption{(left) Nuclear reaction flow path for the production of
      light elements induced by neutrino reactions on $^{12}$C during
      a supernova explosion.  (right) Dependence of the abundance
      ratio $^7$Li/$^{11}$B on the neutrino mixing angle $\theta_{13}$
      for normal and inverted neutrino mass hierarchies.  The
      calculations have been performed for two sets of neutrino cross
      sections and branching ratios  (from
      \cite{Suzuki13}).\label{fig:li7-b11}}
  \end{center}
\end{figure}

Following the work of Ref.~\cite{Woosley90}, several studies have
confirmed that the $\nu$ process can produce the nuclides $^{11}$B and
$^{19}$F in solar abundances by $(\nu,\nu' n)$ and $(\nu, \nu' p)$
reactions on $^{12}$C and $^{20}$Ne.  These studies considered
improved stellar simulations~\cite{Woosley02,Rauscher02,Heger05} and
more elaborate neutrino-induced spallation cross sections
(e.g.~\cite{Heger05,Suzuki06,Yoshida08}).  The shell model 
has been shown to reproduce the GT distributions for nuclei in the $p$
shell quite well \cite{Suzuki06}. Some $^7$Li is produced by
neutrino-induced reactions on $^4$He in the He layer of the star (the
$(\nu,\nu' p \alpha)$ reaction on $^{12}$C can in principle also
produce $^7$Li in the oxygen/carbon layers, but the respective
branching ratio is too small to be significant \cite{Kolbe03}).

Very recently Sieverding {\it et al.} have performed studies of 
neutrino-induced nucleosynthesis in supernovae for stars with solar
metallicity and initial main sequence masses between 15 and 40 $M_\odot$
\cite{Sieverding15}. These studies improve the previous
investigations i) by using a global set of partial differential
cross sections for neutrino-induced charged- and neutral-current
reactions on nuclei with charge numbers $Z < 78$ and ii) by
considering modern supernova neutrino spectra which are noticeably
shifted to lower energies (i.e. $T_{\nu_e} =2.8$ MeV, $T_{\bar {\nu_e}}
= T_{\nu_x} =4.0$ MeV assuming a Fermi-Dirac neutrino distribution
with zero chemical potential) compared to spectra previously adopted
and discussed above (e.g. \cite{Woosley90,Heger05}). Mainly due to the
change in neutrino spectra, Sieverding finds slightly smaller abundances for
$^7$Li, $^{11}$B, $^{138}$La and $^{180}$Ta, however, he confirms
the production of these nuclides by neutrino nucleosynthesis 
\cite{Sieverding15} (see Table \ref{tab:andre}).

\begin{table}
  \caption{Production factors relative to solar abundances normalized
    to $^{16}$O production. 
    The table shows results from calculations without neutrinos and  
    with neutrinos assuming recent neutrino energy spectra. The
    highest temperature values correspond to the  neutrino spectra  
    adopted in~\cite{Heger05}.}
  \begin{center}
 \begin{tabular}{ l c c c }
   \hline\hline
 Nucleus& no $\nu$ & $T_{\nu_e}=2.8$ MeV   & $T_{\nu_e}=4.0$ MeV \\
 & &$T_{\bar{\nu_e}}=4.0$ MeV &$T_{\bar{\nu_e}}=4.0$ MeV \\
 & &$T_{\nu_{\mu,\tau}}=4.0$ MeV &$T_{\nu_{\mu,\tau}}=6.0$ MeV \\ \hline
 \multicolumn{4}{l} {\bf 15 M$_\odot$ star} \\ \hline 
   $^{7}$Li& 0.001 & 0.28       & 2.54  \\
   $^{11}$B&0.007 & 1.43  & 6.13 \\
   $^{19}$F& 1.02 & 1.14  & 1.31 \\
   $^{138}$La &0.07 & 0.67  & 1.18 \\
   $^{180}$Ta& 0.07 & 1.14 & 1.81 \\ \hline
   \multicolumn{4}{l} {\bf 25 M$_\odot$ star} \\ \hline 
    $^{7}$Li& 0.0005 & 0.11 & 0.55  \\
   $^{11}$B&0.003 & 0.80  & 2.61 \\
   $^{19}$F& 0.06 & 0.24  & 0.43 \\
   $^{138}$La &0.03 & 0.63  & 1.14 \\
   $^{180}$Ta& 0.14 & 1.80 & 2.81 \\
   \hline \hline
\end{tabular}
 \label{tab:andre}
\end{center}
\end{table}

Interestingly neutrino-induced reactions noticeably contribute to the
production of $^{19}$F in more massive stars with $M > 20$~M$_\odot$,
while they are negligible in lower-mass stars \cite{Sieverding15}.
This is related to the fact that $^{19}$F is produced at two different
stellar sites: in the C-O layer by the reaction sequence
$^{18}$O(p,$\alpha$)$^{15}$N($\alpha,\gamma)$, which is not sensitive
to neutrino processes, and in the O-Ne layer by neutrino-induced
spallation from the abundant $^{20}$Ne.  As in lower-mass stars the
O-Ne layer is less massive (about 1~M$_\odot$ in the 15~M$_\odot$
star) than in more massive stars (about 3~M$_\odot$ in the 25~M$_\odot$ star) the relative importance of neutrino-induced reactions
for the total $^{19}$F abundance depends strongly on the progenitor
mass and the stellar evolution.

The studies of Ref. \cite{Sieverding15}  
find that neutrino-induced reactions,
either directly or indirectly by providing an enhanced abundance 
of light particles,
noticeably contribute to the production of the radioactive nuclides
$^{22}$Na and $^{26}$Al, which are both prime candidates for 
gamma-ray astronomy.
However, the studies do not find significant production of two 
other candidates, $^{44}$Ti and $^{60}$Fe,
due to neutrino-induced reactions.

The neutrino-induced reactions on $^4$He (in the He layer of the star)
and on $^{12}$C (in the oxygen/carbon layer) initiate a small nuclear
network, which produces a few light elements, in particular $^7$Li and
$^{11}$B (see left panel of Fig. \ref{fig:li7-b11}).  The main path
for the $^7$Li production is by
$^4$He($\nu,\nu'p$)$^3$H($\alpha,\gamma$)$^7$Li and
$^4$He($\nu,\nu'n$)$^3$He($\alpha,\gamma$)$^7$Be($e^-,\nu_e$)$^7$Li,
while the $^{11}$B production is mainly due to the neutral-current
$(\nu,\nu' n)$ and $(\nu, \nu' p)$ reactions on $^{12}$C.  However, in
both cases also charged-current reactions ($^4$He($\nu_e,e^- p$)$^3$He
and $^4$He($\nu_e,e^+ n$)$^3$H in the case of $^4$He, and
$^{12}$C($\nu_e, e^- p$)$^{11}$B and $^{12}$C($\nu_e, e^+ n$)$^{11}$C
for $^{12}$C) contribute to the production of $^7$Li and $^{11}$B,
respectively. This makes the abundance of these two nuclides sensitive
to neutrino oscillations. Due to the expected hierarchy of average
neutrino energies ($\langle E_{\nu_e} \rangle < \langle E_{\nu_x}
\rangle $, neutrino oscillations are expected to increase the average
$\nu_e$ energy and consequently also the charged-current cross section
induced by supernova neutrinos. as pointed out by Kajino and
collaborators, this makes the ratio of $^7$Li and $^{11}$B sensitive
to the $\theta_{13}$ mixing angle and to the mass hierarchy
\cite{Yoshida08,Mathews12,Cheoun12} as is demnostrated in the right
panel of Fig. \ref{fig:li7-b11}. Despite this intriguing sensitivity,
an accurate derivation of the $^7$Li/$^{11}$B abundance ratio requires
reliable stellar model calculations and neutrino and nuclear cross
sections (see \cite{Kajino14}, but must also consider the production
of the elements from other astrophysical sources; $^7$Li is, for
example, also produced by Big Bang nucleosynthesis
\cite{Ryan99}. 

Accounting for neutrino oscillations requires to consider the effect
of collective neutrino oscillations. This has been recently done in
ref.~\cite{Wu.Qian.ea:2015} based on neutrino spectra obtained in Boltzmann
transport simulations~\cite{fischer10}. Their results show that
collective neutrino oscillations increase the average energy of
$\nu_e$ neutrinos assuming inverted hierarchy. Hence, the production
of $^{138}$La and $^{180}$Ta, which are both generated by
charged-current $(\nu_e,e^-)$ reactions, is enhanced. $^7$Li and
$^{11}$B are produced in regions beyond the MSW resonance. Here the
situation is different for normal or inverted hierarchy. In the case
of normal hierarchy the increased average energy of $\nu_e$ neutrinos
results in an enhancement of the production of $^7$Li by the reaction
$^4\text{He}(\nu_e,e^-p)^3\text{He}(\alpha,\gamma)^7\text{Be}(e^+\nu_e)^7\text{Li}$. For
the case of inverted hierarchy the production of $^{11}$B is enhanced
by the reaction chain
$^4\text{He}(\bar{\nu}_e,e^+n)^3\text{H}(\alpha,\gamma)^7\text{Li}(\alpha,\gamma)^{11}\text{B}$.

Woosley {\it et al.} also suggested that some medium-mass nuclei, most
notably $^{51}$V and $^{55}$Mn, might be produced by neutral-current
$(\nu,\nu' p)$ reactions on $^{52}$Fe and $^{56}$Ni with subsequent
beta decays of the daughters $^{51}$Mn and $^{55}$Co, respectively
\cite{Woosley90}.  A recent calculation of the nucleosynthesis in a 25
$M_\odot$ star, based on the stellar evolution of the star performed by
Heger and for the first time taking the full set of neutrino-induced
partial cross sections (\cite{Huther13}, see above) into account,
indeed confirms this suggestion showing that a sizable part of the
abundances for $^{51}$V, $^{55}$Mn and $^{59}$Co can be attributed to
the $\nu$ process \cite{Sieverding14}.  However, the production of
these elements occurs close to the supernova core and therefore the
amount, which might be ejected, will depend on the matter fallback
during the onset of the supernova explosion requiring a more detailed
study.
  
\section{Detecting supernova neutrinos}
 
The observation of neutrinos from supernova SN1987A by the earthbound
detectors Kamiokande~\cite{Hirata.Kajita.ea:1987} and
IMB~\cite{Bionta87} has confirmed and advanced the understanding of
core-collapse supernovae. A similar boost is expected from the
observation of the next near-by supernova which is likely to test the
predictions of supernova models concerning the neutrino spectra for
the different flavors, including the noticeable neutrinoburst signal
in electron neutrinos, originating from electron capture on protons
set free by the shock, and the expected hierarchy in the average
neutrino energies between the different flavors ($\langle E_{\nu_e}
\rangle < \langle E_{\bar{\nu}_e} \rangle \lesssim \langle E_{\nu_x}
\rangle$) reflecting the differences in the interaction of the
neutrino flavors with the surrounding matter in the supernova core. To
reach this goal, several supernova detectors are operational, with
some more being in preparation or proposed.

Obviously translating the event rates of supernova neutrinos observed
in the detectors requires a detailed understanding of these detectors,
including the knowledge of the cross sections for the neutrino
interaction with the detector materials. For a recent overview on the
detectors and their scheme for observing supernova neutrinos the
reader is refered to Ref. \cite{Scholberg12}. In accordance with the
theme of the present review we will focus in the following on recent
advances in describing neutrino-induced cross sections for those
nuclei which serve as detector materials in current and future
supernova neutrino detectors. Much of the early work on this topic has
been described in \cite{Kolbe03}. We will briefly summarize these
studies and discuss in some details the more recent calculations.
 
Besides neutrino-induced inelastic reactions on nuclei - as we will
discuss in the following - supernova neutrinos can also be detected by
their interaction with electrons and by elastic scattering on nuclei
in the detector. The cross sections for the interaction of supernova
neutrinos with electrons is well understood \cite{Marciano03}, but
small. However, it has the advantage that the direction of the
neutrino can be reconstructed if the electron track after the scatter
can be reconstructed (like in Water Cerenkov detectors or liquid argon
time-projection chambers) \cite{Scholberg12}.  Elastic neutrino
scattering on nuclei must be detected by the nuclear recoil.  This can
be a viable scheme for protons \cite{Beacom02}, but for nuclei the
corresponding recoil energies are too small
\cite{Freedman77,Drukier84,Scholberg12}. The cross section for this
coherent process is well known and has been discussed above, see
eq.~(\ref{eq:elastic}). 

Table 2 in Ref. \cite{Scholberg12} lists the present and future
supernova neutrino detectors.  Their main material are liquid
scintillator (C$_n$H$_{2n}$), water, lead or liquid argon.  Hence we
will in the following discuss in turn reactions on $^{12}$C, $^{16}$O,
$^{40}$Ar and $^{208}$Pb. Finally we add some remarks on
neutrino-induced reactions for selected molybdenium and cadmium
isotopes motivated by the ability of the MOON and COBRA detectors to
potentially also observe supernova neutrinos.

\subsection{Carbon}

The nucleus $^{12}$C is an essential part of liquid scintillator
detectors like Borexino, KamLAND, MiniBooNe, LVD, Baksan. Electron
neutrinos and antineutrinos can interact with $^{12}$C by
charged-current reactions (i.e.  $^{12}$C($\nu_e,e^-$)$^{12}$N and
$^{12}$C($\bar{\nu}_e,e^+$)$^{12}$B), but for the relatively
low-energy supernova neutrinos with $\langle E_{\nu_e} \rangle \approx
10$ MeV and $\langle E_{\bar {\nu}_e} \rangle \approx 15$ MeV both
reactions are hindered due to the large $Q$ values with 17.34~MeV for
$\nu_e$ and 14.39 MeV for ${\bar \nu_e}$.  The reactions leading to
particle-bound states in the final nucleus can be tagged by the beta
decays of the $^{12}$N and $^{12}$B ground states.  The matrix element
governing the neutrino-induced excitations of either ground state can
be readily derived from the experimental value of the beta decay. We
note that in RPA calculations of the $^{12}$C($\nu_e,e^-$)$^{12}$N and
$^{12}$C($\bar{\nu}_e,e^+$)$^{12}$B inclusive cross sections the
ground state transition is adjusted to the beta-decay data. Then the
calculated $^{12}$C($\nu_e,e^-$)$^{12}$N cross section for electron
neutrinos, produced by muon decay at rest and with somewhat higher
energies than supernova $\nu_e$ neutrinos, agree quite well with the
data from the KARMEN and LSND collaborations (see
\cite{Kolbe03,Suzuki11}).

Recently shell model calculations have been performed for charged- and
neutral-current neutrino reactions on $^{12}$C using two different
residual interactions in a (0+2)$\hbar \omega$ model space
\cite{Suzuki06}.  Two results of these studies are quite
interesting. At first, the B(GT) strength is noticeably reduced if the
model space is extended from the pure $p$ shell (0$\hbar \omega$) to
the (0+2)$\hbar \omega$ requiring for the latter case nearly no
quenching of the GT strength anymore.  Furthermore, the various
neutrino-induced cross sections agree quite well with those obtained
by RPA calculations (e.g. \cite{Kolbe92}). Shell model calculations
performed in model spaces larger than the $p$ shell, but employing
different residual interactions than Ref. \cite{Suzuki06}, have also
been reported in \cite{Hayes00,Volpe00} again showing the importance
of correlations beyond the RPA approach to describe the
neutrino-induced transitions to the $T=1$ triad in the $A=12$ nuclei.

The detectors can observe neutral-current reactions, triggered by all
neutrino flavors, by detecting the $\gamma$ ray following the
$(\nu,\nu')$ excitation of the isospin $T=1$ state at excitation
energy $E_x=15.11$ MeV. This state is the Isobaric Analog State of the
$^{12}$B and $^{12}$N ground states. The Gamow-Teller matrix element
needed to calculate the $^{12}$C$(\nu,\nu')$$^{12}$C(15.11) cross
section can be derived from the beta-decay of the $^{12}$B and
$^{12}$N ground states after an appropriate isospin rotation or
directly from the measured M1 strength between these states determined
in inelastic electron scattering.

Due to the high thresholds for all neutrino reactions on $^{12}$C the
major detection channel is $\bar {\nu}_e$ absorption on free
protons for liquid scintillator detectors. The respective event rate is estimated to be about an order
of magnitude larger than the charged-current reactions on $^{12}$C and
about a factor of 5 larger for the neutral-current reactions
\cite{Ianni11} (although they can be induced by all neutrino flavors).

\subsection{Oxygen}

The double-magic nucleus $^{16}$O is a potential neutrino target in
water Cerenkov detectors like Superkamiokande. Like $^{12}$C it has
very large $Q$ values for charged-current reactions. Furthermore, the
closed-shell structure of the $^{16}$O ground state strongly
suppresses GT transitions (in the IPM limit the GT strength is
zero). These two effects strongly reduce the neutrino cross sections
on $^{16}$O for supernova neutrinos. In fact, charged-current
reactions have been estimated to give a negligible event rate in water
Cerenkov detectors for galactical supernovae with a distance of 10 kpc
or 30000 light years (the distance to the galactical center)
\cite{Haxton87,Haxton88a}. Due to their expected higher average
energies $\nu_x$ neutrinos might excite the spin dipole resonances in
$^{16}$O which will then decay by emission of protons or neutrons. If
this decay leaves the daughter nucleus, $^{15}$N or $^{15}$O, in
particle-bound excited states (i.e. at excitation energies $E_x <
10.7$ MeV in $^{15}$N and $E_x < 7.3$ MeV in $^{15}$O) the subsequent
$\gamma$ decay cascade of these excited states might be observable in
Superkamiokande, in particular as both daughter nuclei have no excited
state below 5 MeV. In Ref. \cite{Langanke96} this decay scheme has
been estimated to yield about 700 events in Superkamiokande for a
supernova in the galactical center. However, the estimate has been
made for a $\nu_x$ Fermi-Dirac neutrino spectrum with temperature
$T=8$ MeV (and zero chemical potential). More modern supernova
simulations indicate smaller average energies (temperatures) for
$\nu_x$ neutrinos, which would noticeably reduce the expected event
rate. The main observational channel of supernova $\nu_x$ neutrinos in
water Cerenkov detectors appears to be coherent elastic scattering on
protons \cite{Beacom02}.

\begin{figure}
  \begin{center}
    \includegraphics[width=0.48\linewidth]{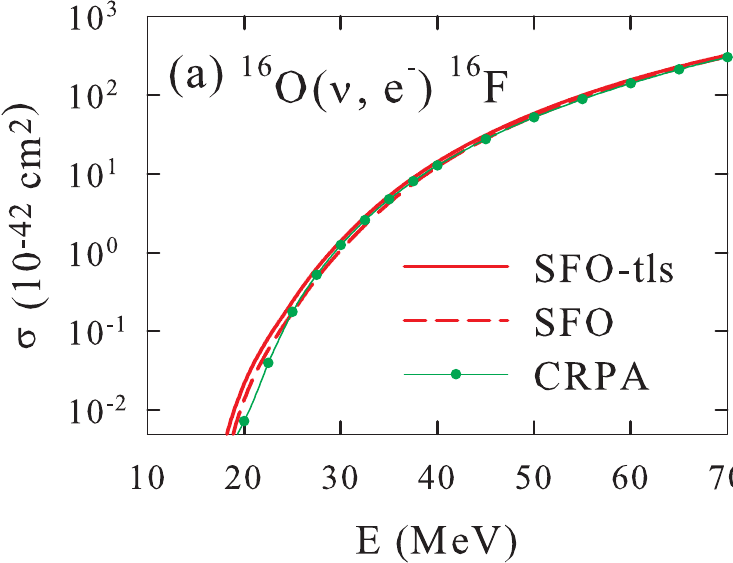}%
    \hspace{0.02\linewidth}%
    \includegraphics[width=0.48\linewidth]{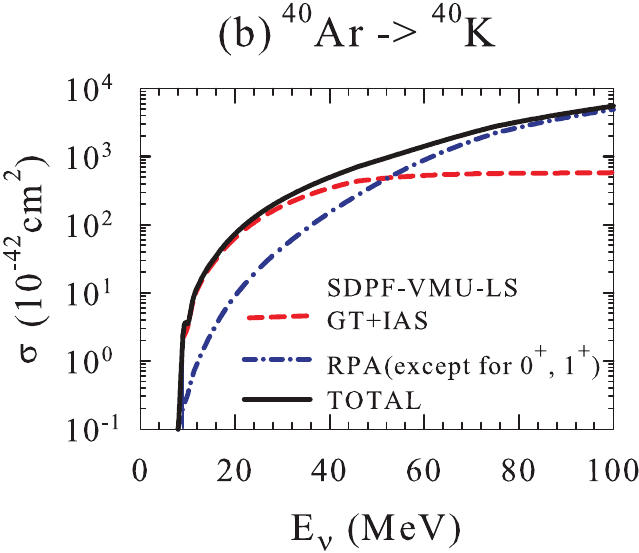}
    \caption{ (left) Comparison of $(\nu_e,e^-)$ cross sections on
      $^{16}$O calculated in shell model and RPA approaches (from
      \cite{Suzuki11}); (right) Contributions to the $(\nu_e,e^-)$
      cross sections on $^{40}$Ar from the IAS and from allowed
      (calculated in shell model) and forbidden (calculated by RPA)
      transitions (from \cite{Suzuki13a}). \label{fig:suzuki-cross}} 
  \end{center}
\end{figure}

As the Gamow-Teller contribution is very small, the charged- and
neutral-current cross sections on $^{16}$O are dominated by
spin-dipole transitions. These have recently been modelled in a shell
model calculation \cite{Suzuki11} considering the $p$ and $sd$ shells
and modifying the residual interaction to account for effects of the
tensor force \cite{Suzuki08}. This calculation finds the major dipole
strength for the $2^-$ channel with excitations around $E_x = 15$ and
22 MeV.  As is demonstrated in the left panel of
Fig. \ref{fig:suzuki-cross}, the calculated shell model
$^{16}$O($\nu_e,e^-$)$^{16}$F cross sections agrees very well with
results obtained within the Continuum RPA \cite{Kolbe02a} for neutrino
energies up to 100 MeV.

\subsection{Argon}

Liquid argon detectors are large time-projection chambers with a very
good energy resolution and full particle track reconstruction
\cite{Scholberg12}. For supernova neutrinos the main goal is to
observe $\nu_e$ neutrinos from $^{40}$Ar($\nu_e,e^-$)$^{40}$K$^*$ by
tagging on the gamma's from the deexcitation of the final states in
$^{40}$K \cite{Scholberg12}.  A similar detection scheme is possible
for the observation of $\bar {\nu}_e$ neutrinos. However, as we will
mention below, the respective cross section is reduced due to the
suppression of Gamow-Teller distributions. The ICARUS detector is a
currently operational liquid argon supernova neutrino detector
\cite{Bueno03,Gil04}.

The $^{40}$Ar($\nu_e,e^-$)$^{40}$K cross section for supernova
neutrinos is dominated by the Fermi contribution to the IAS in
$^{40}$K at $E_x=4.38$ MeV and the Gamow-Teller transitions.  The
first can readily be calculated applying the $(N-Z)$ sum rule for Fermi
transitions, while the second is constrained by the recent measurement
of the B(GT$_-$) strength of $^{40}$Ar by a $(p,n)$ charge-exchange
experiment \cite{Bhattacharya09}.  However, it has been found that the
GT strengths assigned to transitions to two low-lying states at
excitation energies $E_x = 2.289$ MeV and 2.730~MeV is relatively
large, hence significantly contributing to the charged-current cross
section, but it is noticably different than the B(GT) value derived
from the $\beta$ decay of the isospin-analog $^{40}$Ti
\cite{Bhattacharya98}. This difference has been recently investigated
in a $(^3\text{He},t)$ experiment, which has determined the relative strength
of the two transitions and found a value close to the $(p,n)$ experiment
and in clear contrast to the $\beta$-decay data (quoted in
\cite{Karakoc14}). A shell model calculation performed in the $pf$-$sd$
model space has nicely reproduced the B(GT) strength determined in the
$(p,n)$ experiment, including the relative transition strengths to the
two low-lying states \cite{Suzuki13a}.  Recently a shell model
calculation, including a residual interaction with consideration of
Coulomb contributions, hence breaking isospin symmetry, has explored
possible nuclear structure reasons for the differences between the
B(GT) values derived from the $(p,n)$ measurement and the $^{40}$Ti
$\beta$ decay \cite{Karakoc14}.

The authors of Ref. \cite{Karakoc14} recommend to use the Gamow-Teller
strength determined from the $(p,n)$ experiment to derive the
$^{40}$Ar($\nu_e,e^-$)$^{40}$K cross section for supernova neutrinos.
This has indirectly been done by Suzuki and Honma, as their shell
model Gamow-Teller strength distribution gives a very good account of
the one derived from the $(p,n)$ data \cite{Suzuki13a}.  The
calculation shows that there is additional GT strength outside of the
experimental energy window explored by the $(p,n)$ data. While theory
and experiment give a B(GT) strength of about 5 units up to excitation
energies of 8 MeV in $^{40}$K, the calculated total strength is 7.21
units.  The recent shell model calculation also gives a better account
of the data than obtained in a model space restricted to
2p-excitations from the $sd$ to the $pf$ shell \cite{Ormand95}.  Based
on a hybrid model approach (shell model GT strength distribution, the
transition to the IAS from the Fermi sum rule and forbidden transition
calculated within the RPA for multipolarities $\lambda = 1$-5 Suzuki
and Honma have calculated the $^{40}$Ar($\nu_e,e^-$)$^{40}$K cross
section for neutrino energies up to $E_\nu =100$ MeV (see right panel
of Fig. \ref{fig:suzuki-cross}). As expected, the cross section is
dominated by allowed contributions up to $E_\nu = 50$ MeV, which
covers the range of supernova $\nu_e$ energies. It is interesting to
note that the cross sections obtained within the hybrid model agrees
well with calculations where all multipole contributions have been
derived within the RPA \cite{Kolbe03} except at low neutrino energies
where the hybrid model cross sections are larger by about 20-40 \%. We
note again that at low neutrino energies the cross section is
sensitive to the detailed GT strength distribution which, in agreement
with recent data, is well described by the shell model. Hence the
results presented in Ref. \cite{Suzuki13} should be more reliable than
the RPA results of ref. \cite{Kolbe03}.  The neutrino-induced
charged-current reactions have also been studied within the Local
Density Approximation taking account of Pauli blocking and Fermi
motion \cite{Athar04}. The results agree reasonably well with those
from the hybrid model and the RPA approach for neutrino energies
$E_\nu > 20$ MeV.  Cross sections presented on the basis of a QRPA
calculation \cite{Cheoun11} deviate from the values obtained in the
hybrid model and the RPA study by about a factor of 3.

In the IPM, the GT$_+$ strength for $^{40}$Ar (with 18 protons and 22
neutrons) vanishes as all GT transitions, in which a proton is changed
into a neutron, are Pauli blocked. Cross-shell correlations might
introduce a non-vanishing GT$_+$ strength, but it is expected to be
small. Based on this assumption, the $^{40}$Ar($\bar
{\nu}_e,e^+$)$^{40}$Cl cross section has been calculated solely within
an RPA approach \cite{Kolbe03}. Due to the missing or highly
suppressed allowed contributions, the ($\bar {\nu}_e,e^+$) cross
section on $^{40}$Ar is much smaller than the ($\nu_e,e^-$) cross
section for supernova neutrino energies \cite{Kolbe03}, making the
observation of $\bar {\nu}_e$ neutrinos by liquid argon detectors a
challenge. It would be still desirable to put the low energy ($\bar
{\nu}_e,e^+$) cross section on a more reliable basis by calculating
the allowed and forbidden transitions to low-lying states within shell
model calculations which take cross-shell correlations properly into
account. Such studies have been proven to be up to the task in
studying the M1 strength in $^{38}$Ar \cite{Lisetzki} or the isotope
shift in calcium \cite{Caurier01}.

\subsection{Lead}

The Fermi and Ikeda sum rules both scale with neutron excess $(N-Z)$.
As the charged-current response induced by supernova $\nu_e$ neutrinos
(with average energies around 12 MeV) are dominated by Fermi and GT
transitions, the charged-current cross sections on lead is expected to
be significantly larger than on other materials like iron or carbon.
This makes $^{208}$Pb an attractive target for supernova neutrino
detectors like the HALO detector in Canada \cite{Duba08}. The main detection schemes
are the ($\nu_e,e^-$) and $(\nu_x,\nu_x'$) reactions where in both
cases the final nuclei, $^{208}$Bi and $^{208}$Pb, respectively are
left in excited states which mainly decay by particle emission. Due to
the relatively low excitation energies involved and the large Coulomb
barriers associated in the proton channel, the particle decay occurs
mainly by emission of (one or two) neutrons. In HALO these neutrons
can be detected from capture on $^3$He, after they have been moderated
in polypropylene \cite{Duba08,Scholberg12}.  Hence the calculation of
the partial neutrino-induced neutron spallation cross sections on
$^{208}$Pb are of relevance to interpret supernova neutrinos detected
by HALO.

There have been several evaluations of the $(\nu_e,e^-$) cross
sections on $^{208}$Pb, which is the dominating isotope of natural
lead.  Unfortunately the calculation of the allowed contributions to
the cross sections within the shell model is yet computationally not
feasible. The Fermi contribution to the cross section can be derived
from the transition to the known IAS in $^{208}$Bi and the Fermi sum
rule $B(F) = (N-Z)$. All studies have calculated the
contributions from the other multipoles on the basis of RPA variants
which reproduce the strength and the position (around $E_x = 15$ MeV
in $^{208}$Bi) of the Gamow-Teller distribution as well as of the
spin-dipole resonances \cite{Kolbe01a,Volpe02,Engel03} The B(GT$_-$)
strength in $^{208}$Pb is fixed by the Ikeda sum rule, however, it has
been renormalized in Refs. \cite{Kolbe01a,Engel03} (by $(0.7)^2$ and
$(0.8)^2$, respectively) to account for the quenching of the GT
strength at low excitation energies, while Ref. \cite{Volpe02} used
the unquenched $g_A$ value.  Despite these differences the
calculations give quite similar cross sections (within $20-30 \%$) as
function of neutrino energy.  The results are, however, noticeable
smaller (by a factor 3) than those obtained in the first study of the
$(\nu_e,e^-)$ reaction on $^{208}$Pb \cite{Fuller99} which is due to
the differences in the nuclear model, the treatment of the multipole
operator expansion and the Coulomb correction for the electron in the
outgoing channel \cite{Kolbe01a,Volpe02}. A QRPA calculation, on top
of a Hartree-Fock study with BCS pairing, has been presented in
Ref. \cite{Lazauskas07}. The obtained results are about a factor of 2
larger than in Refs. \cite{Kolbe01a,Engel03} (also for $^{56}$Fe),
likely caused by the neglect of renormalization of the GT contribution
to the cross section.  Suzuki and Sagawa presented $(\nu_e,e^-$) cross
sections which have been obtained using GT data from a $(p,n)$
experiment performed at Osaka \cite{Krasznahorkay01} and adjusting
their Hartree-Fock + Tamm-Dancoff approach for the first-forbidden
response to the peaks of the spin-dipole resonances \cite{Suzuki03}.
Furthermore, the spreading and quenching of the GT response has been
considered by coupling to 2p-2h configurations.  Although the
calculations have been performed assuming a muon-decay-at-rest rather
than a supernova neutrino spectrum, and are thus not directly relevant
to the present discussion, the result, however, is satisfyingly:
Suzuki and Sagawa obtain ($3.2 \times 10^{-39}$~cm$^2$) for the
$(\nu_e,e^-)$ cross sections \cite{Suzuki03}, in close agreement to
the RPA result of Ref. \cite{Kolbe01a} ($3.62 \times 10^{-39}$ cm$^2$).

Jachowicz and co-workers \cite{Jachowicz03,Jachowicz08} have
investigated the influence of different neutrino spectra and of
neutrino oscillations on the neutrino signal in lead-based supernova
detectors.  Total charged-current $(\nu_e,e^-)$ cross sections for the
chain of lead isotopes for supernova neutrinos, assuming a Fermi-Dirac
distribution with temperature $T=4$ MeV and zero chemical potential,
are reported in Refs. \cite{Kolbe01,Paar13} based on RPA
calculations yielding quite similar cross sections.  RPA-based studies
have been performed for the neutral-current $(\nu,\nu'$) cross
sections on $^{208}$Pb \cite{Kolbe01a,Engel03}. Again the results
agree quite well (within 10-15~\%).

It might not be surprising that the RPA sudies of
Refs. \cite{Kolbe01a,Volpe02,Engel03} predict similar cross sections,
as they are constrained by similar indirect data like the positions
and strengths of the leading responses. However, the calculations
differ noticeably in their treatment of the quenching of the GT
strength. Here a recent measurement of the Gamow-Teller and
spin-dipole responses by a $({\vec p},{\vec n}$) charge-exchange
reaction with polarized protons and neutrons has determined the GT$_-$
strength up to excitation energies of nearly 50 MeV. The data yield a
GT$_-$ strength of about 85 units up to an excitation energy of 25
MeV, which includes the GT resonance at about 15 MeV.  This value is
significantly smaller than the Ikeda sum rule, $3(N-Z) = 132$,
corresponding to a quenching of the low-lying GT$_-$ strength by a
factor of $(0.8)^2$ as assumed in Ref.~\cite{Engel03}. Ref.
\cite{Kolbe01} used too strong a quenching, hence the cross sections
might be slightly too low, while Ref. \cite{Volpe02} did not consider
any quenching and might have overestimated the GT contribution to the
cross section.  The GT and spin-dipole data show another known
shortcoming of RPA calculations as they often do not describe the
observed fragmentation and spreading of the resonant strength well due
to missing correlations. As we have discussed above this is not too
relevant for the total cross sections if the neutrino energies are
sufficiently large. However, these shortcomings are quite relevant if
one wants to describe the neutron spallation cross sections as the GT
resonance resides quite closely to the threshold of the $2n$ decay
channel which opens at $E_x = 14.98$ MeV.  Such studies have been
performed for example in Refs. \cite{Kolbe01a,Engel03} where the
latter simply assumes that all states with excitation energies between
100 keV above the $1n$ threshold at $E_x=6.9$~MeV to 2.2~MeV above the
$2n$ threshold decay by emission of one neutron, while states at
higher excitation energies decay by multiple neutron
emission. Ref. \cite{Kolbe01a} treats the various decay modes of
excited states in $^{208}$Bi as a cascade of decays followed on the
basis of the statistical model. Ref. \cite{Kolbe01a} uses the
information from the statistical decays to calculate neutron spectra
for various supernova neutrino spectra.  These results, however,
should be effected by the underestimation of the spreading and
fragmentation of the resonant responses (see the discussions in
Refs. \cite{Kolbe01a,Kolbe03}).  These references discuss whether the
observation of the neutron count rate in a lead supernova detector
might be a suitable tool to study neutrino oscillations as proposed in
\cite{Fuller99}.

\subsection{Molybdenium and cadmium}

There have been recently several studies of neutral-
and charged-current reactions on molybdenium and cadmium isotopes
induced by supernova neutrinos. These investigations have been
motivated to explore the sensitivity of the 
Cadmium Zinc Telluride 0-Neutrino Double-Beta Research Apparatus
(COBRA) experiment \cite{Zuber01} 
and the Molybdenium Observatory of Neutrinos (MOON) \cite{Nomachi05}
to the observation of supernova neutrinos. The COBRA and MOON detectors
are both designed and planned for the detection of neutrino-less 
double beta decay. If 
this mode is successfully observed it will establish the Majorana nature
of the neutrino and, even more importantly, will establish violation
of lepton number conservation (e.g. \cite{Rodejohann11}.
The COBRA detector is made out of CdZnTe crystals which serve as both 
source and detection material. Its potential to detect neutrino-less 
double beta decay rests mainly on the two nuclides $^{116}$Cd and $^{130}$Te.
The MOON detector is a tracking-calorimeter detector which is highly
enriched in $^{100}$Mo \cite{Ejiri13}. 
Besides searching for neutrino-less double beta decay the MOON detector 
is also anticipated as solar neutrino detector.

Motivated by the relevance for double-beta decay, GT strength
distributions for $^{100}$Mo and $^{116}$Cd have been determined by
charge-exchange reactions. Using a ($^{3}$He,t) experiment at RCNP
Thies {\it et al} showed that the GT$_+$ strength on $^{100}$Mo is
basically concentrated in a single transition to the $^{100}$Nb ground
state \cite{Thies12}, likely corresponding to the transformation of a
$g_{9/2}$.  proton into a $g_{7/2}$ neutron. The GT strength between
these two ground states has independently been deduced from the
$^{100}$Tc electron capture branching ratios, obtained at the IGISOL
facility \cite{Garcia12}.  We note that a similar domination of the GT
strength by a single transition (to the $1^+$ state at $E_x=0.69$ MeV)
has been observed in the GT$_+$ strength of $^{96}$Mo, deduced by
$(d,^2\text{He})$ and $(^3\text{He},t)$ charge-exchange experiments.
For $^{116}$Cd the GT$_-$ strength has been determined recently using
the $(p,n)$ reaction \cite{Sasano07,Sasano12}.  The centroid of the
strength associated with the GT resonance is located at an excitation
energy around $E_x = 14$ MeV, with some additional strength forming a
shoulder in the GT distribution around $E_x=8$ MeV. The GT$_-$
strength on $^{116}$Cd had previously been determined by a ($^3$He,t)
experiment \cite{Akimune97} which turned out, however, to be erraneous
due to the use of a natural cadmium target \cite{Akimune08}. The GT
strength between the $^{116}$Cd and $^{116}$In ground states has also
been determined from the electron capture decay branch of
$^{116}$In~\cite{Wrede13} yielding a transition strength in agreement
with the value deduced from the $(p,n)$ charge-exchange
experiment~\cite{Sasano12}.

The experimental GT strength distributions are valuable constraints 
for calculations discussing the MOON and COBRA detectors
as devices to observe supernova neutrino. A detailed study of
neutrino- and antineutrino scattering off $^{116}$Cd has recently been
reported in Ref. \cite{Almosly14} performed within the 
charge-changing mode of the QRPA. The authors performed detailed studies
of the GT$_-$ strength distribution for
several Skyrme interactions supplemented by finite-range 
(isovector and isoscalar) pairing interactions and compare
it to the results obtained with the Bonn one-boson
exchange potential. (Similar studies based on
the Bonn potential have been previously reported in \cite{Almosly13}.
These authors adjusted the parameters of the force for each multipole
individually.) Only the Bonn potential
was found to reproduce the giant GT resonance at 14 MeV and its satellite 
around 8 MeV, while the calculations performed with the Skyrme forces
missed these energies typically by 1-2 MeV \cite{Almosly14}.
These  differences in the position of the
strong GT transitions translate into different predictions
for the $(\nu_e,e^-)$ cross section on $^{116}$Cd for supernova neutrinos,
which, as calculated in Ref. \cite{Almosly14}, 
is dominated by GT transitions. Using the cross section obtained with
the Bonn potential as benchmark,  the Skyrme interactions yield results
which can deviate upto a factor of 2; however, most Skyrme interactions
predict cross sections which agree within $50 \%$. For the 
$({\bar \nu_e}, e^+)$ cross section on $^{116}$Cd for supernova
antineutrinos the calculations predict results which agree within a 
factor of 2 among each other, with the Skyrme forces typically yielding a
somewhat larger value than the Bonn interaction. The largest 
contribution to the cross sections arizes from
first-forbidden transitions to the $J=1^-$ states
in $^{116}$Ag.  The partial cross sections reported in Ref. \cite{Almosly13}
for the (differently parametrized) Bonn potential are very similar, 
except for the contribution of the $J=0^+$ multipole 
to the  $(\nu_e,e^-)$ cross section which is noticeably
larger in the study of \cite{Almosly13} than found in Ref. \cite{Almosly14}, in
particular compared to the results for the Skyrme interactions. 

The formalism of charge-changing QRPA has also been adopted to calculate
the cross sections for neutral-current reactions on stable 
cadmium and molybdenium
isotopes induced by supernova neutrinos of all flavor. 
For the cadmium isotopes the inelastic cross sections for supernova
$\nu_e$ neutrinos  are dominated
by the contributions from the $J=1^+$ multipole (GT$_0$ strength)
\cite{Almosly15}. For the odd-A isotopes $^{111,113}$Cd Almosly {\it et al.}
find a surprisingly large vector contribution to excited $J=0^+$ states
\cite{Almosly15}.    

Also the calculation of neutral-current reactions on the stable
molybdenium isotopes show a somewhat larger cross sections for the
odd-A nuclei $^{95,97}$Mo compared to their even-even neighbor
isotopes~\cite{Ydrefors12}. The difference amounts to about $40 \%$
for neutrinos with energies typical for supernova $\nu_e$'s and
decreases with neutrino energy. The neutral-current response of
supernova neutrinos on odd-A and even-even molybdenium isotopes has
also been studied in Refs.  \cite{Ydrefors11,Balasi11}. Interestingly
the authors find a relatively large contribution of vector transitions
to excited $0^+$ states for the odd isotopes $^{95,97}$Mo
\cite{Ydrefors11}, which is less pronounced for the even-even isotopes
\cite{Balasi11}.  The charged-current reactions on the odd isotopes
$^{95,97}$Mo for supernova neutrinos has been explored in
Ref. \cite{Ydrefors13}.  The authors find the ($\nu_e,e^-$) cross
section dominated by allowed Fermi (to the IAS) and GT transitions,
while the $({\bar \nu_e},e^+$) cross section has some forbidden
contributions arizing from $1^-$ and $2^-$ multipoles, besides the
dominating GT contribution.

Engel {\it et al.} \cite{Engel00} have studied the charged-current
cross sections on $^{100}$Mo, replacing in the dominating $1^+$
channel the QRPA results with the GT strength measured by the
$(^3\text{He},t)$ charge-exchange reaction~\cite{Akimune97}. These authors
discuss how the MOON detector might distinguish between charged- and
neutral-current events, assuming that neutral-current reactions will
excite the molybdenium isotopes to states which mainly decay by
emitting neutrons and successive $\gamma$-rays \cite{Engel00}. Hence
it appears useful if the studies reported in
Refs. \cite{Ydrefors11,Balasi11} would be extended to calculations of
the partial cross sections with particle emission in the final
channel.

\section{Summary}

Neutrinos have been identified as one of the essential players in
core-collapse supernovae. This insight - predicted by models - has been
unambiguously verified by the observations from supernova SN1987A. 
The undisputably most important role of neutrinos for the supernova dynamics
concerns the energy balance: due to their tiny cross sections with the
surrounding matter neutrino emission efficiently cools the inner core
of the collapsing core and, after bounce, carries away an overwhelmingly
large portion of the gravitational binding energy released in the explosion.
Even despite the tiny cross sections, the 
tremendously large number of neutrinos involved makes them an important 
means of energy transport during the explosion, reviving the stalled
shock front. 

Neutrinos also play a crucial role for the explosive
supernova nucleosynthesis as their interaction with the hot matter
surrounding the protoneutron star (mainly free nucleons, but also light
nuclear clusters) determines the ratio of protons and neutrons
available for synthesis of heavier nuclei in the $\nu p$ process
or potentially the r-process. While (anti)neutrino absorption on protons
is at the very heart of the $\nu p$ process, the role of neutrinos
during the mass flow to heavy nuclides in the astrophysical r-process
(if it occurs in the core-collapse supernova environment) is likely
to be small. However, particle spallation induced by neutrino reactions
on abundant nuclei in the outer layers of the star has been identified
as the synthesis mechanism of selected isotopes like $^{11}$B, $^{138}$La
and $^{180}$Ta in a process called neutrino nucleosynthesis.

In several of these aspects of supernova dynamics and nucleosynthesis,
neutrinos interact with free nucleons. The respective cross sections
are well defined based on the framework of the electroweak
theory. However, in the dense and hot supernova environment these
interactions are modified by correlations which is a topic of ongoing
research interest.  Neutrino interactions with nuclei can occur during
collapse as well as after bounce during the various nucleosynthesis
processes.  During collapse, an important neutrino process is coherent
scattering on nuclei which leads to neutrino trapping in the final
phase of the collapse. Except for small corrections induced by isospin
breaking and additional Gamow-Teller contributions (at finite
temperature and for odd-$A$ and odd-odd nuclei), the coherent cross
sections are determined by the Fermi transition to the Isobaric Analog
State and hence well under control. Other neutrino processes during
collapse include absorption on nuclei and inelastic neutrino
scattering with nuclei. As electron capture on nuclei is the other
very relevant weak-interaction process during collapse, mainly $\nu_e$
neutrinos are present in the core. They can be absorbed on neutrons in
nuclei. As this is the inverse reaction of electron capture, the
respective cross section is derived applying the principle of detailed
balance.  Hence, its accuracy has benefitted from the improved
description of electron capture on nuclei made possible in recent
years by more precise data from advanced charge-exchange experiments
and from progress in nuclear modelling, mainly due to the application
of shell model techniques.

The shell model has also been the many-body technique which has been
applied in the first derivation of inelastic neutrino-nucleus
scattering cross sections at finite supernova temperatures. The
calculations have been validated against precision data derived from
inelastic electron scattering.  As these are taken for nuclear ground
states only, the validation is restricted to cross sections at
temperature $T=0$. However, neutrino-induced deexcitation of nuclear
states, thermally populated at finite temperatures, are found to
strongly increase the inelastic cross sections at low neutrino
energies. Although the shell model is expected to describe also the
(Gamow-Teller) transitions on excited states adequately well, a full
state-by-state summation to obtain the inelastic cross section at
supernova temperatures is computationally unfeasible and some
approximation has to be applied. However, it has been quite gratifying
that calculations performed within the Thermal QRPA approach confirmed
the strong increase of the cross sections at low neutrino energies at
finite temperature.  This method derives the cross sections
consistently from a thermal ensemble and accounts for effects induced
by the melting of nucleon pairs at finite temperature. It is important
that both approaches - the TQRPA and the shell model - predict very
similar cross sections at higher neutrino energies as supernova
simulations have indicated that inelastic scattering of such
high-energy neutrinos off nuclei leads to down-scattering of the
neutrinos and hence strongly reduces the tail of the neutrino spectra
produced in the neutrino-burst just after bounce. As a consequence,
this change in spectra reduces the expected event rates for the future
observation of neutrinos from the burst phase by supernova neutrino
detectors. Inelastic neutrino-nucleus scattering is also a mean of
energy exchange between neutrinos and matter during collapse.  But
simulations have indicated that this process only adds rather mildly
to neutrino thermalization, which is dominated by inelastic neutrino
scattering on electrons.
  
Neutrino types other than electron  neutrinos are produced during infall by 
deexcitation of thermally populated nuclear states by neutrino-pair emission. 
However, for the supernova dynamics this process has been recently shown as unimportant
as the cross section for $\nu_e$ production by electron capture is several
orders of magnitude larger and dominates the energy balance of the core during collapse.
This finding is based on a reasonable, but quite schematic description
of the neutrino-pair deexcitation process, but is unlikely to change if more accurate
cross sections, for example derived on the basis of the Thermal QRPA model, become available.

Neutrino-induced reactions have been proposed to play roles in various supernova
nucleosynthesis processes; e.g. in the $\nu p$ and r-process nucleosynthesis and
obviously in the process called neutrino nucleosynthesis. To allow for systematic and
detailed investigations into the importance of neutrino-nucleus reactions for these
various processes, a global set of partial differential cross sections for neutrino-induced
charged- and neutral-current reactions on nuclei (up to charge number $Z =78$) has been derived.
Based on the observation that for the moderate supernova neutrino energies 
the initial neutrino-nucleus excitation function is dominated by the contributions
of giant resonances, the calculation of the respective cross sections
has been performed within the Random Phase Approximation, which is constructed to give a fair account
of the energy and the total transition strength of these collective resonances. 
In these initial processes the neutrino excites the nucleus to levels which
subsequently decay by particle or $\gamma$ emission.
The decay of the excited nuclear levels is followed within the Statistical Model
considering a cascade of multi-particle decays.  
   
Inserting these neutrino-nucleus cross sections into supernova
nucleosynthesis simulations indicates that neutrino reactions on
nuclei have only limited impact on the calculated abundances. For the
$\nu p$ process ${\bar \nu_e}$ absorption on protons bound in nuclei
can in principle compete with the absorption on free protons, but is
kinematically suppressed due to the relatively large $Q$ values of the
neutron-deficient nuclei encountered in the $\nu p$ process mass
flow. Current supernova simulations indicate only a `weak
r-process' which produces nuclides up to the second r-process peak
(mass numbers $A < 130$). In this scenario neutrino-nucleus reactions
are found unimportant.  Uncertainties related to supernovae as
r-process sites are introduced due to the yet insufficiently
understood roles of collective neutrino oscillations and of sterile
neutrinos as well as to the undetermined mass hierarchy of neutrinos.

More than two decades ago, neutrino-induced spallation of abundant
nuclei in the outer layers of a supernova has been identified as a
major nucleosynthesis source of selected nuclides.  This finding has
been confirmed over the years using improved stellar models, more
accurate neutrino-nucleus cross sections and, last, but not least,
refined spectra for the various types of supernova neutrinos. Although
the predictions for these spectra shifted systematically to lower
average energies (or temperatures if one describes the spectra as
usual by a Fermi-Dirac distribution with zero chemical potential)
stellar models still imply that the nuclides $^7$Li, $^{11}$B,
$^{138}$La and $^{180}$Ta are produced by neutrino nucleosynthesis
nearly in solar abundances.  As $^{11}$B is the product of
neutral-current reactions on $^{12}$C induced by $\nu_\mu$ and
$\nu_\tau$ neutrinos and their anti-particles and $^{138}$La and
$^{18}$Ta are made by $(\nu_e,e^-)$ reactions on $^{138}$Ba and
$^{180}$Hf, respectively, neutrino nucleosynthesis can serve as a
thermometer for the spectra of those neutrino types, which have not
been observed from supernova SN1987A.

The observation of neutrinos from a future close-by supernova,
preferably from our galaxy, is the ultimate goal of the various
neutrino detectors either operational or under construction. If the
supernova is indeed close enough, e.g. from the galactic center, these
detectors have the ability to test the hierarchy of neutrino energies
predicted by supernova models. Furthermore they will have the time
resolution to probe the evolution of neutrino luminosities during
collapse and explosion, including the prominent burst of electron
neutrinos expected just after bounce when heavy nuclei are dissociated
into free nucleons and the electrons, which have survived the
collapse, are captured on free protons. To translate the observed
neutrino event rates from a supernova requires the knowledge of the
respective neutrino-induced reactions on the detector nuclei. These
cross sections have been derived on the basis of nuclear models
currently available to treat the related many-body problem. Hence for
lead or cadmium targets only cross sections calculated within the
frame work of the RPA are available, while for lighter target nuclei
more sophisticated nuclear models have been applied. As a consequence
for the lighter nuclei a more accurate description of the detailed
nuclear response has been considered than for the heavier nuclei,
which is of relevance if particle decays of levels in the daughter
nucleus might serve as detection scheme. Furthermore some more
fundamental questions like the quenching of higher multipoles, in
relation to the one established for shell model calculations of the
Gamow-Teller strength when compared to data, or the role of two-body
currents has to be resolved.

Even if a few questions concerning the role of neutrino-induced
reactions on nuclei for the supernova dynamics and nucleosynthesis
remain insufficiently answered, the field has witnessed significant
progress in recent years. The decisive push in this field in specific,
and, of course, to our general understanding of supernovae globally
would come from the observation of a nearby supernova by $\gamma$'s,
neutrinos and the composition of the ejecta. Perhaps Nature is so kind
and we do not have to wait for too long. The observers are ready.

\section*{Acknowledgments}

The research presented in this review has strongly benefited from
collaborations with A. Arcones, A.~Bauswein, E. Caurier, T.~Fischer,
C. Fr\"ohlich, A. Heger, L. Huther, H.-Th. Janka, A. Juodagalvis,
E. Kolbe, M. Liebend\"orfer, B. M\"uller, J. J. Mendoza-Temis, P. von
Neumann-Cosel, F. Nowacki, Y.-Z.~Qian, A.~Richter, K. Sieja,
A. Sieverding, F.-K.~Thielemann, P. Vogel, S. E. Woosley, M.-R. Wu,
and Q. Zhi. This work was partly supported by the Deutsche
Forschungsgemeinschaft through contract SFB~634, the Helmholtz
International Center for FAIR within the framework of the LOEWE
program launched by the state of Hesse, and the Helmholtz Association
through the Nuclear Astrophysics Virtual Institute (VH-VI-417).


\end{document}